\documentclass[12pt]{article}
\usepackage{amsmath}
\usepackage{graphicx}
\usepackage{enumerate}
\usepackage{booktabs}       
\usepackage{natbib} 
\usepackage{makecell}
\usepackage{longtable}
\usepackage{array} 
\usepackage{subcaption}
\newcommand{\blind}{1}

\usepackage{amsmath,amssymb,amsfonts,amsthm,bbm}

\usepackage{mathtools}   
\usepackage{stmaryrd} 

\usepackage{enumitem}   
\usepackage{romannum}

\usepackage[titletoc,title]{appendix}  

\usepackage[table,xcdraw,dvipsnames]{xcolor}   
\usepackage{hyperref}
\newcommand\myshade{85}
\colorlet{mylinkcolor}{YellowOrange}
\colorlet{mycitecolor}{Aquamarine}
\colorlet{myurlcolor}{violet}

\hypersetup{
	linkcolor  = mylinkcolor!\myshade!black,
	citecolor  = mycitecolor!\myshade!black,
	urlcolor   = myurlcolor!\myshade!black,
	colorlinks = true,
}

\usepackage{caption}
\usepackage{subcaption}

\usepackage[normalem]{ulem}
\usepackage{setspace}

\usepackage{graphicx}
\usepackage{comment}
\graphicspath{{figs/}}

\renewcommand{\hat}{\widehat}
\renewcommand{\tilde}{\widetilde}

\newcommand{\bfm}[1]{\ensuremath{\boldsymbol{#1}}} 

\def\bp{\bfm p}

\def\bx{\bfm x}   \def\bX{\bfm X}  \def\XX{\mathbb{X}}
\def\by{\bfm y}     
     \def\ZZ{\mathbb{Z}}

\def\calL{{\cal  L}}

\def\calT{{\cal  T}}

\def\calX{{\cal  X}}

\newcommand{\bfsym}[1]{\ensuremath{\boldsymbol{#1}}}

 \def\bgamma{\bfsym \gamma}             
 \def\bdelta{\bfsym {\delta}}

 \def\bnu{\bfsym {\nu}}
 \def\btheta{\bfsym {\theta}}

\providecommand{\angles}[1]{\left\langle #1 \right\rangle}
\providecommand{\paren}[1]{\left( #1 \right)}

\providecommand{\braces}[1]{\left\{ #1 \right\}}

\usepackage{mathtools}
\DeclarePairedDelimiterX{\infdivx}[2]{(}{)}{%
  #1 \; \delimsize\| \; #2%
}

\DeclareMathOperator*{\argmin}{argmin}
\DeclareMathOperator*{\argmax}{argmax}

\DeclareMathOperator{\Tr}{Tr}

\newcommand*\xbar[1]{%
  \hbox{%
    \vbox{%
      \hrule height 0.4pt 
      \kern0.5ex
      \hbox{%
        \kern-0em
        \ensuremath{#1}%
        \kern-0em
      }%
    }%
  }%
} 
 \newtheorem{definition}{Definition}
 \newtheorem{assumption}[definition]{Assumption}
 \newtheorem{lemma}[definition]{Lemma}
 
 \newtheorem{theorem}[definition]{Theorem}
 \newtheorem{corollary}[definition]{Corollary}

 \newtheorem{remark}{Remark}
\newtheorem{replemma}{Lemma}  
\newenvironment{restatelemma}[2][]{%
  \renewcommand\thereplemma{\ref{#2}}%
  \begin{replemma}[#1]%
}{%
  \end{replemma}%
}

\definecolor{royalpurple}{rgb}{0.47, 0.32, 0.66}
\definecolor{greenfresh}{HTML}{00897B}
\definecolor{bluefresh}{HTML}{1E88E5}
\definecolor{redfresh}{HTML}{E53935}

\definecolor{royalpurple}{rgb}{0.47, 0.32, 0.66}

\def\beq{\begin{equation}}
\def\eeq{\end{equation}}

\def\bet{\begin{theorem}}
\def\eet{\end{theorem}}

\def\bel{\begin{lemma}}
\def\eel{\end{lemma}}

\usepackage[framemethod=TikZ]{mdframed} 

\usepackage{fancybox}

\usepackage[linesnumbered,ruled,vlined]{algorithm2e}

\SetCommentSty{mycommfont}

\SetKwInput{KwInput}{Input}                
\SetKwInput{KwOutput}{Output}              
\SetKwInput{KwInitialize}{Initialize}    

\usepackage{listings}             
\begin{document}
\pagenumbering{arabic}

\def\spacingset#1{\renewcommand{\baselinestretch}%
{#1}\small\normalsize} \spacingset{1}

%
%
%

\def\TITLE{Transfer Learning for Contextual Joint Assortment-Pricing under Cross-Market Heterogeneity}

\if1\blind
{
\title{\bf \TITLE}
\author{
Elynn Chen$^\sharp$ \hspace{6ex}
Xi Chen $^\diamond$ \hspace{6ex}
Yi Zhang$^\dag$ \\
\medskip
\normalsize \vspace{-2ex}
$^{\sharp \; \diamond}$New York University, Stern School of Business \\ 
\normalsize
$^{\dag}$ Tsinghua University, School of Economics and Management
}
\maketitle
} \fi

\if0\blind
{
\bigskip
\bigskip
\bigskip
\begin{center}
{\LARGE\bf \TITLE}
\end{center}
\medskip
} \fi

\bigskip
\begin{abstract}
\spacingset{1}
We study transfer learning for contextual joint assortment-pricing under a multinomial logit choice model with bandit feedback. A seller operates across multiple related markets and observes only posted prices and realized purchases. While data from source markets can accelerate learning in a target market, cross-market differences in customer preferences may introduce systematic bias if pooled indiscriminately.

We model heterogeneity through a structured utility shift, where markets share a common contextual utility structure but differ along a sparse set of latent preference coordinates. Building on this, we develop Transfer Joint Assortment-Pricing (TJAP), a bias-aware framework that combines aggregate-then-debias estimation with a UCB-style policy. TJAP constructs two-radius confidence bounds that separately capture statistical uncertainty and transfer-induced bias, uniformly over continuous prices.

We establish matching minimax regret bounds of order
$
\tilde{O}\!\left(d\sqrt{\frac{T}{1+H}} + s_0\sqrt{T}\right),
$
revealing a transparent variance-bias tradeoff: transfer accelerates learning along shared preference directions, while heterogeneous components impose an irreducible adaptation cost. Numerical experiments corroborate the theory, showing that TJAP outperforms both target-only learning and naive pooling while remaining robust to cross-market differences.
\end{abstract}

\noindent%


\spacingset{1.2}

\addtolength{\textheight}{.1in}%


\section{Introduction}  \label{sec:intro}

Digital platforms and retailers increasingly operate across multiple related markets. Firms routinely launch the same product lines in different cities, expand sequentially into new regions, or run repeated market-level experiments to refine pricing and assortment decisions. Such multi-market deployments and repeated experimentation are central to modern data-driven revenue management (e.g., \cite{ban2021personalized}; \cite{bastani2022meta}). In each new deployment, managers face a fundamental tension: decisions must be made quickly under substantial uncertainty, yet the firm often possesses extensive historical data from prior markets that may be similar but not identical. Leveraging prior data or past experiments can significantly improve learning efficiency (e.g., \cite{bu2020online};\cite{bastani2021predicting}), but cross-market heterogeneity in customer preferences can limit its effectiveness. In particular, naive pooling across markets may introduce systematic bias and lead to distorted decisions when markets differ in economically meaningful ways.

This challenge is especially acute in {\em joint assortment-pricing problems}, where firms must simultaneously determine which products to offer and what prices to charge. In many retail and platform settings, these decisions are intrinsically coupled: assortments impose discrete combinatorial constraints, prices are continuous decision variables, and demand depends nonlinearly on both through substitution effects. Moreover, firms typically observe only bandit feedback -- posted prices and realized purchases -- rather than full demand curves. As a result, errors in learning customer preferences propagate through both pricing and assortment decisions, amplifying their impact on revenue outcomes.

While incorporating data from related markets can substantially reduce estimation uncertainty, doing so without explicitly accounting for heterogeneity risks introducing persistent bias. This leads to a fundamental question:
\begin{quote}
    \emph{``When and how can data from related markets safely accelerate contextual joint assortment-pricing under discrete-choice demand with bandit feedback?''}
\end{quote}
A large literature in revenue management studies learning and optimization under discrete-choice demand models, particularly the multinomial logit (MNL) model, developing algorithms and performance guarantees for dynamic pricing, assortment selection, and their contextual extensions \citep{keskin2014dynamic,javanmard2020multi,agrawal2017thompson,chen2020dynamic,oh2021multinomial}. These works, however, typically treat each market independently and do not leverage cross-market information in a principled way.  
In parallel, the statistics and machine learning literature has established that transfer learning and multitask learning can substantially improve sample efficiency when tasks share structure \citep{bastani2021predicting,li2022transfer,tian2023transfer,bastani2022meta,xu2024robust,xu2025multitask}.
Yet these approaches largely focus on linear or full-feedback settings and do not accommodate the discrete-choice structure and joint pricing–assortment decisions central to revenue management.

As a result, despite the prevalence of multi-market deployments in practice, there is currently {\em no framework that enables safe and theoretically grounded transfer for contextual joint assortment-pricing under discrete-choice demand with bandit feedback.}

\subsection*{Structured Transfer in Joint Assortment-Pricing}

We study a multi-market setting under a contextual multinomial logit model, consisting of one target market and multiple source markets. Rather than assuming markets are identical or arbitrarily different, we model cross-market heterogeneity through a {\em structured preference shift}: markets share a common contextual utility structure, while deviations relative to the target market are confined to a sparse set of preference coordinates.

This formulation reflects a realistic operational perspective. In practice, markets often differ along a limited number of salient dimensions, such as price sensitivity for specific product categories or affinity toward certain attributes, while remaining broadly aligned elsewhere. The sparsity structure captures such localized heterogeneity while preserving sufficient structure for statistical efficiency. 

Importantly, these heterogeneous coordinates are not known a priori, but are learned from data in a data-driven manner through the estimation procedure. The sparsity structure therefore serves as a statistical regularity that enables the model to identify and adapt to these differences while maintaining efficiency in high dimensions. It yields a natural separation between {\it shared preference directions}, where source data can safely reduce uncertainty, and
{\it shifted coordinates}, where target-specific adaptation is unavoidable.

\subsection*{A Bias-Aware Transfer and Decision Framework}

Building on this structure, we develop {\em Transfer Joint Assortment-Pricing (TJAP)}, a unified learning-and-decision framework for contextual joint assortment-pricing across heterogeneous markets. 
The design of TJAP follows a simple principle:
\begin{quote}
    {\em ``Information shared across markets should be pooled to reduce variance;
deviations specific to the target market must be isolated to prevent bias.''}
\end{quote}
To implement this principle, TJAP integrates three components.

First, we propose an {\em aggregate-then-debias estimation procedure}. Source-market data are pooled to estimate shared preference components, thereby reducing estimation variance. Because source markets may differ from the target along a sparse set of coordinates, the pooled estimator is then refined using target-market data via an $\ell_1$-regularized debiasing step that corrects for sparse deviations.

Second, we develop a {\em bias-aware optimistic decision rule}. We construct frequentist UCB-style confidence bounds that incorporate a two-radius structure: one radius captures statistical uncertainty shrinking with pooled information, while the other accounts for residual transfer bias due to heterogeneity. These bounds are designed to hold uniformly over continuous prices, enabling their integration into the joint assortment-pricing problem.

Third, we incorporate {\em episodic information-geometry control} to stabilize learning under adaptive decisions. Because pricing and assortment choices influence the informativeness of future observations, TJAP freezes the information geometry within each episode and invokes targeted exploration only when necessary to ensure identifiability in the target market.

Together, these components yield a unified framework that balances {\em variance reduction through transfer} with {\em bias control through target adaptation}.

\subsection*{Contributions}

This paper makes four primary contributions.

\textbf{First}, we introduce a structured transfer-learning formulation for contextual joint assortment-pricing under multinomial logit demand. We model cross-market heterogeneity through sparse preference shifts, in which markets share a common contextual utility structure while differing along a small number of latent preference coordinates. This formulation captures realistic multi-market variation while enabling principled information sharing across markets.

\textbf{Second}, we develop a bias-aware learning-and-decision framework, Transfer Joint Assortment-Pricing (TJAP), that explicitly separates variance reduction from bias control. The framework integrates an aggregate–then–debias estimation procedure with a two-radius, price-uniform optimistic policy and episodic information-geometry control. This design enables effective reuse of source-market data while preventing negative transfer arising from structural heterogeneity, and accommodates the joint optimization of discrete assortments and continuous prices under bandit feedback.

\textbf{Third}, we establish finite-time regret guarantees that reveal a transparent variance-bias decomposition in transfer learning. The regret scales as
$$
\tilde{O}\!\left(d\sqrt{\frac{T}{1+H}} + s_0\sqrt{T}\right),
$$
where $H$ is the number of source markets, $d$ is the feature dimension, and $s_0$ measures the sparsity of cross-market preference shifts. The first term captures variance reduction from transfer, while the second term reflects an irreducible adaptation cost along heterogeneous coordinates.

\textbf{Fourth}, we establish a matching minimax lower bound, showing that this variance–bias tradeoff is fundamental. In particular, no policy can improve the dependence on $H$ or eliminate the $s_0$-driven term over the structured preference-shift class. Together, these results precisely characterize the value of transfer: when heterogeneity is sparse, transfer yields substantial gains; when heterogeneity is diffuse, these gains saturate; and bias-aware correction is essential for safe data reuse. \emph{This yields the first characterization of transfer limits in joint assortment-pricing.}

Beyond formal guarantees, our results provide operational insights for multi-market revenue management. The variance-bias decomposition clarifies when historical markets should be pooled and when caution is required. When cross-market differences are localized, leveraging auxiliary markets can significantly accelerate learning and improve both pricing and assortment decisions. In contrast, indiscriminate aggregation without bias correction can lead to systematically distorted decisions under structural heterogeneity. Our numerical experiments corroborate these insights, showing that TJAP consistently outperforms both target-only learning and naive pooling while remaining robust to cross-market differences.

\subsection{Related Work and Our Distinction}\label{sec:related-work}

Our paper lies at the intersection of online revenue management under discrete-choice demand, bandit learning for structured decision problems, and transfer learning across related environments. We connect these \emph{literatures} in a setting that is both practically important and technically challenging: \emph{contextual joint assortment-pricing under multinomial logit demand with bandit feedback and cross-market heterogeneity}.

\paragraph{Online pricing and assortment under discrete-choice demand.}
Choice-based revenue management studies pricing and assortment decisions under random-utility models, most prominently the multinomial logit (MNL) model for its tractability and substitution structure~\citep{talluri2006theory,kok2008assortment}. 
In static settings, joint assortment-pricing under MNL admits tractable structure and has been extended to richer behavioral models~\citep{wang2012offline,aouad2018approximability,aouad2021assortment,gao2021assortment,najafi2025pricing}. 
Related work also considers assortment and pricing without explicit demand prediction or under alternative feedback models~\citep{chen2023model,chen2025assortment}.

In online settings, most studies treat assortment learning and pricing learning separately. 
For MNL assortment bandits, both Bayesian and UCB-style approaches establish finite-time regret guarantees in non-contextual and contextual formulations~\citep{agrawal2017thompson,agrawal2018mnlbanditDP,cheung2017thompson,chen2020dynamic,Oh2019TSforMNL,oh2021multinomial,ou2018ijcai,perivier2022dynamic}. 
Dynamic pricing under demand uncertainty, including high-dimensional and personalized settings, has also been extensively studied~\citep{keskin2014dynamic,ban2021personalized,den2022dynamic,chen2025dynamic}. 
However, online learning for \emph{joint} assortment-pricing remains largely confined to single-market formulations~\citep{miao2021dynamic,chen2021nonparametric,erginbas2025online,kim2025censored,chen2019joint,jia2024online,feng2023principro}, and does not address principled cross-market information reuse.

\paragraph{Bandits and structured exploration.}
Contextual bandits provide the methodological foundation for learning with covariates, with sharp regret guarantees for linear and generalized linear reward models using self-normalized confidence bounds~\citep{dani2008stochastic,rusmevichientong2010linearly,abbasi2011improved,filippi2010parametric,Li2017GeneralizedLinBandits,chen2025stochastic}. 
Combinatorial bandits extend these tools to constrained action spaces~\citep{chen2013combinatorial,kveton2015cascading}. 
Within discrete-choice bandits, MNL-based methods exploit logit structure to handle combinatorial feasibility while achieving tight regret bounds~\citep{agrawal2018mnlbanditDP,chen2020dynamic,oh2021multinomial}. 

Our approach builds on the frequentist UCB paradigm, which has proven effective in balancing exploration and exploitation in classical, linear, and generalized linear bandits~\citep{auer2002using,lattimore2020bandit,chu2011contextual,Li2017GeneralizedLinBandits,oh2021multinomial,erginbas2025online}. 
We extend this framework to joint assortment-pricing by developing confidence bounds that are \emph{uniform over prices} and remain valid under discrete-choice substitution effects. Moreover, our optimism explicitly accounts for transfer-induced bias, which is not present in existing assortment-only or pricing-only bandit models.

\paragraph{Transfer, multitask, and offline-to-online learning across markets.}
A parallel literature studies transfer and multitask learning under structured heterogeneity. 
In supervised settings, debiasing and robust procedures exploit sparse task differences or distributional shifts~\citep{bastani2021predicting,li2022transfer,tian2023transfer,liu2023unified,huang2025optimal,zeng2024llm,chai2026low}. 
In sequential decision problems, multitask and meta-learning frameworks analyze information sharing across tasks~\citep{bastani2022meta,xu2025multitask,huang2025optimal,repasky2024multi,kim2025collaborative,chen2025transfer,chai2025transfer,chai2025deep,zhou2025prior,chai2026optimistic}. 
Related work in pricing studies offline-to-online reuse and meta dynamic pricing across experiments~\citep{bu2020online,bu2025feature,bastani2022meta,han2025meta,zhang2025contextual,zhang2025transfer}. 
These approaches primarily focus on pricing-only or linear-demand models and do not accommodate joint assortment interactions under discrete-choice demand, nor do they explicitly address bias arising from cross-market heterogeneity.

\paragraph{Our Distinction.}
To our knowledge, this paper is the first to study \emph{safe transfer across heterogeneous markets in contextual joint assortment-pricing under multinomial logit demand with bandit feedback}. Our contribution is not simply to combine ingredients from these literatures, but to develop a framework tailored to the joint pricing-assortment setting. Specifically, we contribute:
(i) a structured multi-market model with sparse utility shifts that makes cross-market similarity statistically exploitable without assuming identical markets;
(ii) a bias-aware aggregate-then-debias estimation strategy that separates shared learning from target-specific adaptation;
(iii) a frequentist, UCB-style optimistic policy with confidence bounds that are uniform over prices and explicitly account for transfer-induced bias; and
(iv) matching minimax regret bounds that identify the fundamental variance–bias tradeoff governing transfer in this setting.

Together, these results provide a unified characterization of when transfer improves learning, when its benefits saturate, and why bias-aware correction is essential for reliable data reuse in joint assortment-pricing.

\paragraph{Organization.}
Section~\ref{sec:problem-formulation} formulates the multi-market contextual joint assortment-pricing problem. 
Section~\ref{sec:algo} presents the proposed algorithm.
Section~\ref{sec:theory} establishes theoretical guarantees.
Section~\ref{sec:exp} reports numerical experiments.
Section~\ref{sec:proof-sketch} presents the key steps of the proofs.
Section~\ref{sec:conc} concludes and discusses future research directions.

\section{Multi-Market Joint Assortment-Pricing Problem}\label{sec:problem-formulation}

We consider a seller operating across multiple related markets, such as geographic locations, store formats, or platform environments. The seller’s primary objective is to optimize revenue in a {\it designated target market}, indexed by superscript $(0)$, which may correspond to a newly launched or strategically important market with limited historical data. In addition, the seller has access to data from $H$ {\it source markets}, indexed by $(h)$ for $h \in [H]$, which represent previously operated or concurrently running markets with richer data histories. While decisions are made only in the target market, data from source markets may be leveraged to accelerate learning, provided cross-market differences are properly accounted for.

Time is discrete and indexed by $t = 1, \ldots, T$. In each period, customers arrive in every market together with observable contextual information. For market $h$, let $\bx^{(h)}_{it} \in \mathbb{R}^d$ denote the covariate vector associated with product $i$ at time $t$, capturing observable product attributes, customer characteristics, or their interactions. These covariates are observed prior to decision making.

The seller makes decisions only in the target market. At each time $t$, after observing $\{\bx^{(0)}_{it}\}_{i \in [N]}$, the seller selects an assortment $S^{(0)}_t \subseteq [N]$ satisfying $|S^{(0)}_t| \le K$, and posts prices $p^{(0)}_{it}$ for each $i \in S^{(0)}_t$. Products not included in the assortment are unavailable for purchase. In contrast, source markets generate data according to their own assortment and pricing policies, which are not controlled by the seller in our formulation. This asymmetric structure reflects settings in which firms deploy new pricing and assortment policies in a focal market while relying on historical data from related markets.

After observing the offered assortment and prices, the customer in each market either purchases one of the available products or chooses the outside option. The seller observes realized purchases and revenues but does not observe latent utilities or counterfactual demand. Learning therefore proceeds under bandit feedback.

The objective is to design a policy that maximizes cumulative expected revenue in the target market over horizon $T$. Performance is evaluated relative to a clairvoyant benchmark that knows the true target-market demand parameters and, in each period, selects the revenue-maximizing assortment and prices given the realized context.

This multi-market formulation makes data reuse intrinsic to the learning problem. The central challenge is to leverage information from related markets to reduce estimation variance while remaining robust to systematic differences in customer preferences across markets.

\subsection{Choice Model and Revenue Structure}

We model customer demand in each market using a contextual multinomial logit (MNL) specification. For market $h \in \{0\} \cup [H]$, product $i$, and period $t$, the latent utility takes the form
\begin{equation}\label{eq:utility-func}
f^{(h)}_{it}
= \langle \bx^{(h)}_{it}, \btheta^{(h)} \rangle
- \langle \bx^{(h)}_{it}, \bgamma^{(h)} \rangle p^{(h)}_{it}
+ \varepsilon^{(h)}_{it},
\end{equation}
where $\bx^{(h)}_{it} \in \mathbb{R}^d$ is the observed feature vector,
$\btheta^{(h)}$ captures baseline preference for product attributes,
$\bgamma^{(h)}$ captures price sensitivity,
and $\varepsilon^{(h)}_{it}$ are i.i.d. standard Gumbel shocks.

Under this model, when assortment $S^{(h)}_t$ and prices $\bp^{(h)}_t$ are offered in market $h$, the probability that a customer in market $h$ purchases product $i \in S^{(h)}_t$ is
\begin{equation}\label{eq:MNL-chocie-model}
q^{(h)}_t(i \mid S^{(h)}_t, \bp^{(h)}_t)
= \frac{\exp(v^{(h)}_{it})}{1+\sum_{\ell\in S^{(h)}_t}\exp(v^{(h)}_{\ell t})},
\quad
v^{(h)}_{it}
:= \langle \bx^{(h)}_{it}, \btheta^{(h)} \rangle
- \langle \bx^{(h)}_{it}, \bgamma^{(h)} \rangle p^{(h)}_{it},
\end{equation}
where the outside option has normalized utility zero. 

Given assortment $S$ and price vector $\bp$, the expected revenue in market $h$ at time $t$ is
\begin{equation}\label{eq:revenue}
R^{(h)}_t(S,\bp)
:= \sum_{i\in S} p_i \, q^{(h)}_t(i \mid S,\bp).
\end{equation}
In our setting, the seller optimizes $R^{(0)}_t(S,\bp)$ in the target market, while source markets follow the same structural demand model and generate data for learning.

A useful structural property of the MNL model is that optimal prices are uniformly bounded. Under positive price sensitivity, the revenue-maximizing price for any product is finite and depends only on primitive problem constants. Accordingly, without loss of generality, we restrict attention to prices in a compact interval $[0,\bar P]$ for some finite $\bar P$ (see Lemma \ref{lem:bounded-opt-price} in the Appendix).

Together, the shared MNL structure and bounded price domain define a common revenue landscape across markets, while allowing market-specific parameters $\bnu^{(h)} := (\btheta^{(h)}, \bgamma^{(h)})$ to capture heterogeneity in baseline preferences and price sensitivity.

\subsection{Learning Objective and Regret}

The preference parameters $\bnu^{(h)}$ are unknown in all markets. While source-market observations provide auxiliary information, the seller’s objective is entirely target-market–centric: to maximize cumulative expected revenue in the target market $h = 0$.

Let $(S^{(0)}_t, \bp^{(0)}_t)$ denote the assortment-price pair selected in the target market at time $t$. The expected revenue in period $t$ is $R^{(0)}_t(S^{(0)}_t, \bp^{(0)}_t)$, as defined in \eqref{eq:revenue}. Learning proceeds sequentially under bandit feedback, as only realized purchases and revenues are observed.

To evaluate performance, we compare any policy to a clairvoyant benchmark that knows the true target-market parameter $\bnu^{(0)}$. Given realized contexts at time $t$, the benchmark selects
\begin{equation*}
(S^{\ast}_t, \bp^{\ast}_t)
\in \argmax_{S\in\mathcal{S}_K,\; \bp\in[0,\bar P]^N}
R^{(0)}_t(S,\bp),
\end{equation*}
where $\mathcal{S}_K := \{S\subseteq[N]: |S|\le K\}$. The regret of a policy $\pi$ over horizon $T$ is therefore
$$
\text{Regret}(T;\pi)
:= \sum_{t=1}^T
\Big(
R^{(0)}_t(S^{\ast}_t, \bp^{\ast}_t)
-
R^{(0)}_t(S^{(0)}_t, \bp^{(0)}_t)
\Big).
$$
This benchmark isolates the cost of demand learning in the target market. Source-market data influence regret only indirectly through their effect on parameter estimation and decision quality in the target market.

\subsection{Structured Cross-Market Heterogeneity (Utility Shift Model)} \label{ssec:similarity_assumption}

We model cross-market heterogeneity through a \emph{structured preference shift}. Markets share a common contextual utility structure, while differences relative to the target market are confined to a limited set of economically meaningful dimensions. 
This reflects environments in which markets may vary in sensitivity to particular attributes or price responsiveness, yet exhibit similar substitution patterns across products.

\begin{assumption}[Structured Preference Shift]\label{assump:task-simi}
There exists an index set $\mathcal{S}_* \subseteq [2d]$ with $|\mathcal{S}_*| \le s_0$ such that, for every source market $h \in [H]$,
\begin{equation*}
(\bnu^{(0)} - \bnu^{(h)})_j = 0
\quad
\text{for all } j \notin \mathcal{S}_*.
\end{equation*}
That is, discrepancies between the target market and each source market are supported on a common subset of at most $s_0$ coordinates.
\end{assumption}

The sparsity level $s_0$ quantifies the degree of cross-market similarity. When $s_0$ is small, markets differ only along a limited number of preference dimensions, enabling substantial information sharing. When $s_0$ is large, heterogeneity becomes more diffuse and transfer becomes less effective.

The requirement of a common support across source markets captures settings in which structural differences arise from a stable set of market-specific factors, rather than arbitrary idiosyncratic shifts. As we show later, this structure yields a transparent variance-bias tradeoff: source data reduce estimation variance along shared directions, while adaptation along shifted coordinates incurs a cost that depends on $s_0$.

\section{Algorithmic Framework: Transfer Joint Assortment-Pricing}\label{sec:algo}

We develop \emph{Transfer Joint Assortment-Pricing} (TJAP), a unified learning-and-decision framework for contextual joint assortment-pricing across multiple markets. The goal is to leverage data from related source markets to accelerate learning in a target market, while remaining robust to cross-market heterogeneity.
At a high level, TJAP is designed around a central principle:
\begin{quote}
    \emph{``Shared information should be pooled to reduce variance, while market-specific deviations must be isolated to control transfer bias.''}
\end{quote}
This principle reflects the variance–bias tradeoff inherent in transfer learning under heterogeneous markets.

\paragraph{High-level overview.}
TJAP alternates between learning shared structure across markets and adapting to target-specific differences. In each episode, the algorithm first aggregates source-market data to estimate common preference components, and then refines this estimate using target-market observations to correct for sparse deviations. Given this bias-aware estimate, it selects assortments and prices by optimizing an optimistic revenue objective that accounts for both statistical uncertainty and potential transfer bias. To ensure stable learning under adaptive decisions, the algorithm updates estimates episodically and invokes targeted exploration only when the target-market data are insufficient to identify heterogeneous components.

Concretely, TJAP consists of three components:

\textit{(i) Aggregate-then-debias estimation.} 
Source-market data are first pooled to estimate shared preference components, and the resulting estimator is then adjusted using target-market observations to correct for sparse deviations.

\textit{(ii) Optimistic decision making with price-uniform confidence bounds.} 
To balance exploration and exploitation, TJAP adopts a frequentist UCB-style optimism principle tailored to joint assortment-pricing. Unlike standard bandit settings with finite action sets, prices are continuous and enter revenue nonlinearly. We therefore construct utility-level confidence bounds that are uniform over prices and incorporate them into an optimistic revenue objective.

\textit{(iii) Episodic information-geometry control.}
Because adaptive assortment and pricing decisions affect the informativeness of future data, TJAP stabilizes learning through an episodic design that freezes the confidence geometry within each episode. Source-market data tighten the geometry, while a target-market Fisher information criterion ensures identifiability, with randomized exploration invoked only when necessary.

\begin{figure}[!ht]
\centering
\includegraphics[width=\textwidth]{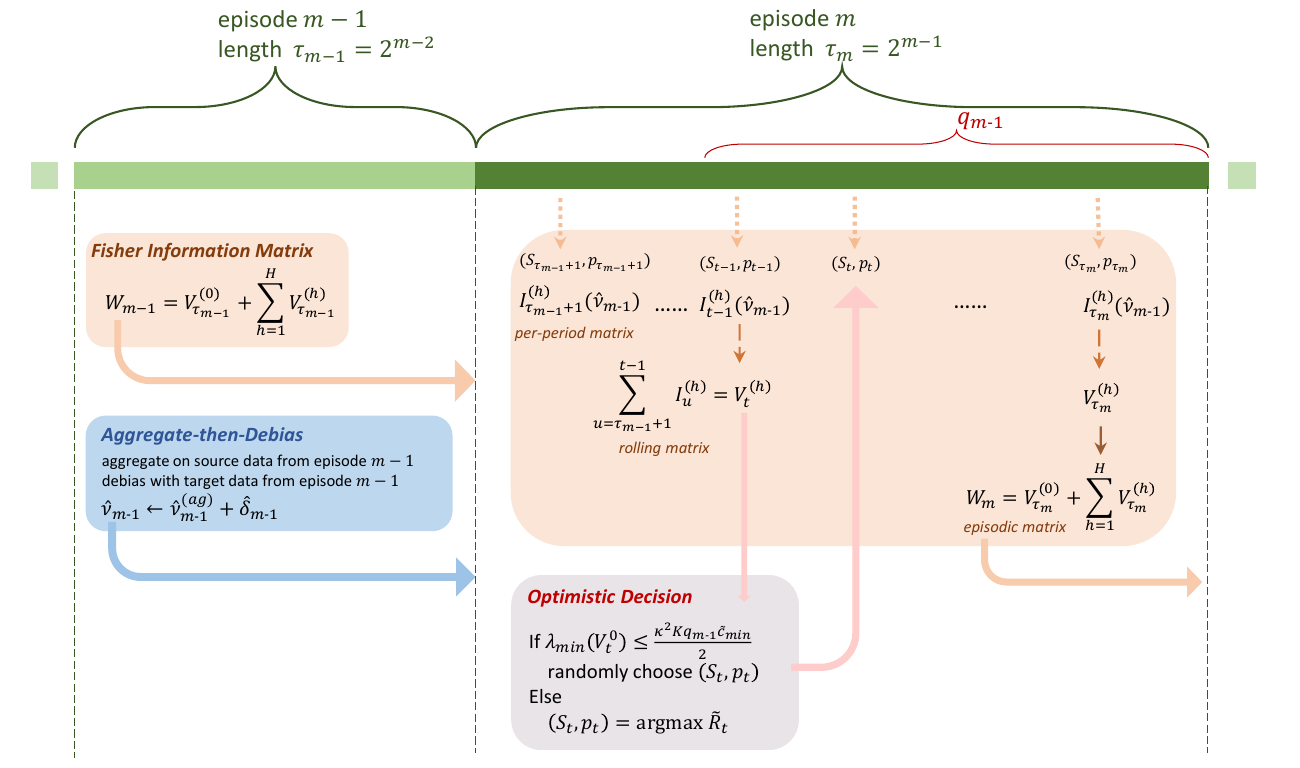}
\caption{Schematic illustration of Algorithm~\ref{algo:trans-assort-price-o2o} over two episodes. At the start of episode $m$, the estimate $\hat\bnu_{m-1}$ is computed from episode $m\!-\!1$ data via the aggregate-then-debias procedure~(Section~\ref{ssec:agg_then_debias}). During episode $m$, Fisher information is accumulated via per-period increments $I^{(h)}_t(\hat\bnu_{m-1})$, updating the rolling matrices $V^{(h)}_t$ and yielding the pooled matrix $W_{m}$~(Section~\ref{ssec:info_geometry}). The geometry from the previous episode, $W_{m-1}$, is used to construct the UCB bonus throughout episode $m$~(Section~\ref{ssec:opt_decision_rule}). The policy follows optimistic selection by maximizing $\tilde R_t(\cdot)$~(Section~\ref{ssec:opt_decision_rule}). A forced-exploration is considered only in the final $q_{m-1}$ periods, when the target curvature~(via $V^{(0)}_t$)~is insufficient.}
\label{fig:tjap-cwf-schematic}
\end{figure}

Together, these components implement a unified variance–bias strategy: source markets reduce estimation variance along shared directions, while target-market adjustments control transfer-induced bias. Section~\ref{ssec:agg_then_debias}-~\ref{ssec:info_geometry} develop each component in detail; complete pseudo-code is provided in Algorithm~\ref{algo:trans-assort-price-o2o} and illustrated in Figure~\ref{fig:tjap-cwf-schematic}. 

\begin{algorithm}[!ht]
\caption{TJAP-CWF}
\label{algo:trans-assort-price-o2o}
\KwInput{Streaming data $\{\{ \bx_{it}^{(h)} \}_{i \in [N]}, S_t^{(h)}, \bp_t^{(h)}, \by_t^{(h)} \}_{t\geq 1}$ for $h \in \{0\}\cup[H]$}
\KwInitialize{$V^{(0)}_0 \leftarrow \mathbf{0}_{2d \times 2d}$; $V^{(h)}_0 \leftarrow \mathbf{0}_{2d \times 2d}$ for all $h\in[H]$}

\For{$t \in [2d]$}{
Randomly choose $S_t^{(0)} \in \mathcal{S}_K$ and $\bp_t^{(0)} \sim \mathrm{Uniform}([0,\bar P]^N)$\;
$V^{(0)}_t \leftarrow V^{(0)}_{t-1} + \frac{1}{K^2}\sum_{i \in S_t^{(0)}} \tilde{\bx}_{it}^{(0)}(\tilde{\bx}_{it}^{(0)})^\top$\;
}

\For{each episode $m = 2, 3, \ldots$}{
Compute $\tau_m \leftarrow 2^{m-1}$, $\mathcal{T}_m \leftarrow \{\tau_{m-1}+1,\ldots,\tau_m\}$\;

\tcp{(i) aggregate on source data from episode $m-1$}
$\hat{\bnu}_{m-1}^{(ag)} \leftarrow \argmin_{\bnu \in \mathbb{R}^{2d}} \frac{1}{H|\mathcal{T}_{m-1}|}\sum_{h\in [H]}\sum_{t\in \mathcal{T}_{m-1}} \ell_t^{(h)}(\bnu)$\;

\tcp{(ii) debias with target data from episode $m-1$}
$\hat{\bdelta}_{m-1} \leftarrow \argmin_{\bdelta \in \mathbb{R}^{2d}}\Big( \frac{1}{|\mathcal{T}_{m-1}^{(0)}|}\sum_{t\in \mathcal{T}_{m-1}^{(0)}} \ell_t^{(0)}(\hat{\bnu}_{m-1}^{(ag)}+\bdelta) + \lambda_{m-1} \|\bdelta\|_1 \Big)$\;
Set $\hat{\bnu}_{m-1} \leftarrow \hat{\bnu}_{m-1}^{(ag)} + \hat{\bdelta}_{m-1}$ \;

Set $W_{m-1} \leftarrow V^{(0)}_{\tau_{m-1}} + \sum_{h=1}^H V^{(h)}_{\tau_{m-1}}$\;
$V^{(0)}_{\tau_{m-1}} \leftarrow \mathbf{0}_{2d \times 2d}$; $V^{(h)}_{\tau_{m-1}} \leftarrow \mathbf{0}_{2d \times 2d}, \forall h\in[H]$\;

\For{each period $t \in \mathcal{T}_m$}{
\If{$\tau_m - t \leq q_{m-1}$ \textbf{and} $\lambda_{\min}(V^{(0)}_t) \leq \dfrac{\kappa^2 K q_{m-1} \tilde{C}_{\rm min}}{2}$}{
Randomly choose $S_t^{(0)} \in \mathcal{S}_K$ and $\bp_t^{(0)} \sim \mathrm{Uniform}([0,\bar P]^N)$\;
}
\Else{
Offer $(S_t^{(0)},\bp_t^{(0)}) = \argmax\limits_{S \in \mathcal{S}_K, \ \bp \in \mathcal{P}^N} \ \tilde{R}_t(S,\bp;\alpha_{m-1},\beta_{m-1},W_{m-1})$\;
}

\For{$h\in\{0\}\cup[H]$}{
$V^{(h)}_{t+1} \leftarrow V^{(h)}_{t}+
\sum_{i \in S_t^{(h)}} q_{it}^{(h)}(\hat{\bnu}_{m-1})\,\tilde{\bx}_{it}^{(h)}(\tilde{\bx}_{it}^{(h)})^\top
- \sum_{i \in S_t^{(h)}}\sum_{j \in S_t^{(h)}} q_{it}^{(h)}(\hat{\bnu}_{m-1}) q_{jt}^{(h)}(\hat{\bnu}_{m-1})\,\tilde{\bx}_{it}^{(h)}(\tilde{\bx}_{jt}^{(h)})^\top$\;
}
}
}
\end{algorithm}

For clarity of exposition, we first present the framework under homogeneous covariates, so that cross-market heterogeneity arises solely through the structured utility shift (Assumption~\ref{assump:Homo-Cov}). The extension to covariate shifts is discussed in Section~\ref{ssec:heterogeneous}.

\paragraph{Episodic updates.}
TJAP operates in episodes of geometrically increasing length.
Let $\tau_1 := 1$ and define episode endpoints $\tau_m := 2^{m-1}$ for $m = 2,3,\ldots$.
Episode $m$ consists of time indices
$$
\mathcal{T}_m
:=
\{\tau_{m-1}+1,\ldots,\min\{\tau_m,T\}\}.
$$
The horizon $[T]$ is therefore partitioned into $M = \lceil \log_2(T+1) \rceil$ episodes, so parameter updates occur only $\mathcal{O}(\log T)$ times \citep{auer2008near,zhang2025transfer}.

At the beginning of episode $m$, parameter estimates and confidence sets are updated using data collected in episode $m-1$. Within each episode, these quantities are held fixed. This freezing of the information geometry stabilizes learning under adaptive pricing and assortment decisions while ensuring that estimation error contracts geometrically over time.

For each market $h \in \{0\}\cup[H]$, let
$
\mathcal{T}^{(h)}_{m-1}
:=
\mathcal{T}_{m-1}
$
denote the set of periods in episode $m-1$.
Source-market data $\{\mathcal{T}^{(h)}_{m-1}\}_{h=1}^H$ are pooled to estimate shared preference components, while target-market data $\mathcal{T}^{(0)}_{m-1}$ are used for debiasing and identifiability control.

\subsection{Aggregate-then-Debias Estimation}\label{ssec:agg_then_debias}

The structured preference shift in Section \ref{ssec:similarity_assumption} suggests separating estimation into shared and market-specific components. In each episode, TJAP first aggregates source-market data to estimate the common preference structure, thereby reducing estimation variance along directions that are stable across markets. Because source markets may differ from the target along a sparse set of coordinates, the aggregate estimator may be biased relative to the target parameter.
To correct this bias, we refine the aggregate center using only target-market data. This second step estimates the sparse discrepancy between the pooled source center and the target preference parameter, yielding an estimator that balances cross-market variance reduction with target-specific adaptation.

For each market $h$, let
\begin{equation*}
\mathcal{D}^{(h)}_{m-1}
:=\big\{
(\bx^{(h)}_{it},\, S^{(h)}_t,\, \bp^{(h)}_t,\, \by^{(h)}_t)
:\; t\in \mathcal{T}^{(h)}_{m-1}
\big\}
\end{equation*}
denote the data collected in episode $m-1$, where $\bx^{(h)}_{it}\in\mathbb{R}^d$ denotes the observed covariate vector associated with product $i$ in market $h$, capturing product attributes, customer characteristics, or their interactions;
$S^{(h)}_t\subseteq[N]$ is the assortment offered in market $h$ at time $t$;
$\bp^{(h)}_t=(p^{(h)}_{it})_{i\in S^{(h)}_t}$ is the vector of prices posted for the offered products;
and $\by^{(h)}_t\in\{0,1\}^{|S^{(h)}_t|+1}$ is the realized purchase indicator, including the outside option, with exactly one entry equal to one.
TJAP uses source-market data $\{\mathcal{D}^{(h)}_{m-1}\}_{h=1}^H$ to estimate shared preference components and target-market data $\mathcal{D}^{(0)}_{m-1}$ to correct for sparse deviations.

Writing the systematic utility in augmented form,
$$
v^{(0)}_{it}(p;\bnu)
=
\langle \tilde{\bx}^{(0)}_{it}(p), \bnu \rangle,
$$
where $\tilde{\bx}^{(0)}_{it}(p) = (\bx^{(0)}_{it}, -p \bx^{(0)}_{it}) \in \mathbb{R}^{2d}$, the negative log-likelihood for a single observation in market $h$ is
$$
\ell^{(h)}_t(\bnu)
= -\sum_{i\in S^{(h)}_t\cup\{0\}} y^{(h)}_{it}
\log\!\left(
\frac{\exp(\langle \tilde \bx^{(h)}_{it}(p^{(h)}_{it}),\bnu\rangle)}
{1+\sum_{j\in S^{(h)}_t}\exp(\langle \tilde \bx^{(h)}_{jt}(p^{(h)}_{jt}),\bnu\rangle)}
\right).
$$

\paragraph{Step I. Aggregation (variance reduction).}
At the beginning of episode $m$, TJAP first computes an {\it aggregate estimator} using only source-market data:
\begin{equation}\label{eq:agg_estimator}
\widehat{\bnu}^{\mathrm{ag}}_{m-1}\in
\argmin_{\bnu\in\mathbb{R}^{2d}}
\frac{1}{\sum_{h=1}^H |\mathcal{T}_{m-1}|}
\sum_{h=1}^H \sum_{t\in \mathcal{T}_{m-1}} \ell^{(h)}_t(\bnu).
\end{equation}
Because source markets share preference structure outside the sparse shift set, pooling reduces variance along shared coordinates.
In particular, the variance of $\widehat{\bnu}^{\mathrm{ag}}_{m-1}$ contracts at a rate proportional to the total amount of source data, which scales with $H$.

To align the debiasing step with the pooled estimator in~\eqref{eq:agg_estimator}, we define the
\emph{ground-truth aggregate center} as the population minimizer of the same pooled objective.
Let $\mathcal{T}_{m-1}$ denote the set of decision epochs within episode $m-1$, and define the episode-$m\!-\!1$ population pooled objective
\begin{equation}\label{eq:def-pop-pool-obj-homo}
L^{\mathrm{ag}}_{m-1}(\bnu)
:=
\frac{1}{H}\sum_{h=1}^H \; \mathbb{E}\!\left[
\frac{1}{|\mathcal{T}_{m-1}|}\sum_{t\in \mathcal{T}_{m-1}} \ell^{(h)}_t(\bnu)
\right],
\end{equation}
where the expectation is taken under the data-generating process of each source market in episode $m-1$.
We then define
\begin{equation}\label{eq:def-nu-ag-pop-homo}
\bnu^{\mathrm{ag}}_{m-1}
\in \argmin_{\bnu\in\mathbb{R}^{2d}} L^{\mathrm{ag}}_{m-1}(\bnu).
\end{equation}
This population center is the limit point that the pooled estimator in~\eqref{eq:agg_estimator} targets.

\paragraph{Step II. Debiasing (Target Adaptation).}
To correct the (potential) mismatch between the source pooled center and the target parameter, TJAP refines the aggregate center using target-market data only. Let
\begin{equation}\label{eq:def-delta-star-pop}
\bdelta^\ast_{m-1} := \bnu^{(0)}-\bnu^{\mathrm{ag}}_{m-1}
\end{equation}
denote the discrepancy between the \emph{population pooled source center} and the true target parameter.
Under Assumption~\ref{assump:task-simi}, $\bdelta^\ast_{m-1}$ is supported on at most $s_0$ coordinates.

TJAP estimates $\bdelta^\ast_{m-1}$ by solving an $\ell_1$-regularized likelihood problem on the target data:
$$
\widehat{\bdelta}_{m-1}\in
\argmin_{\bdelta\in\mathbb{R}^{2d}}
\left\{
\frac{1}{|\mathcal{T}_{m-1}|}
\sum_{t\in \mathcal{T}_{m-1}}
\ell^{(0)}_t\big(\widehat{\bnu}^{\mathrm{ag}}_{m-1}+\bdelta\big)+
\lambda_{m-1} \|\bdelta\|_1
\right\},
$$
where $\lambda_{m-1}>0$ is a regularization parameter.
The final estimator used in episode $m$ is
$$
\widehat{\bnu}_{m-1} := \widehat{\bnu}^{\mathrm{ag}}_{m-1}+\widehat{\bdelta}_{m-1}.
$$

\subsection{Optimistic Decision Rule with Price-Uniform Confidence Bounds} \label{ssec:opt_decision_rule}

Having constructed a bias-aware estimator of the target preference parameter, we now translate statistical uncertainty into pricing and assortment decisions. TJAP adopts a frequentist UCB-style optimism principle to balance exploration and exploitation.

Unlike standard bandit settings with finite action sets, joint assortment-pricing involves combinatorial assortment choices and continuous price decisions that enter revenue nonlinearly. We therefore construct confidence bounds at the utility level that are uniform over prices and embed them into an optimistic revenue objective.

To quantify statistical uncertainty in a way that is compatible with adaptive decisions, TJAP maintains an episodic information matrix $W_{m-1}$ defined in \eqref{eq:def-Wm}. 
The matrix $W_{m-1}$ aggregates information from all markets over episode $m-1$.
Throughout episode $m$, the parameter estimate $\widehat{\bnu}_{m-1}$ and the episodic information matrix $W_{m-1}$ are fixed.

\paragraph{Two-radius optimistic utility.} At episode $m$, TJAP constructs an intermediate optimistic utility
$$
\bar v_{it}(p)
:= \langle \tilde{\bx}^{(0)}_{it}(p),\widehat{\bnu}_{m-1}\rangle
+ u_{it}(p),
$$
where the bonus $u_{it}(p)$ has a {\it two-radius form}
\begin{equation}
\label{eq:two-radius-bonus}
u_{it}(p)=
\underbrace{\alpha_{m-1}\,\|\tilde{\bx}_{it}^{(0)}(p)\|_{W_{m-1}^{-1}}}_{\text{variance term}}+
\underbrace{\beta_{m-1}\,\|\tilde{\bx}_{it}^{(0)}(p)\|_{\infty}}_{\text{transfer-bias term}},
\end{equation}
where $\|\cdot\|_{W_{m-1}^{-1}}$ and $\|\cdot\|_\infty$ denote the Mahalanobis norm and infinity norm, respectively.  
The first term reflects statistical uncertainty and shrinks as pooled information accumulates across markets. The second term accounts for residual transfer bias arising from sparse cross-market shifts.
They arise directly from the estimation error decomposition in Section \ref{ssec:agg_then_debias}.
Let $\widehat{\bnu}_{m-1}$ denote the current estimator and $\bnu^{(0)}$ the true parameter. Under self-normalized concentration, we have
$$
\big|
\tilde{\bx}_{it}^{(0)}(p)^\top
\big(
\widehat{\bnu}_{m-1}
-
\bnu^{(0)}
\big)
\big|
\le
\alpha_{m-1}
\big\|
\tilde{\bx}_{it}^{(0)}(p)
\big\|_{W_{m-1}^{-1}},
$$
so the Mahalanobis norm naturally scales uncertainty by the inverse information matrix, upweighting poorly explored directions.
In addition, under the sparse utility-shift model, the residual
$
r_{m-1}
:=
\widehat{\bnu}_{m-1}
-
\bnu^{(0)}
$
satisfies $\| r_{m-1} \|_1 \le \beta_{m-1}$.
By Hölder’s inequality,
$$
\big|
\tilde{\bx}_{it}^{(0)}(p)^\top r_{m-1}
\big|
\le
\|
\tilde{\bx}_{it}^{(0)}(p)
\|_\infty
\| r_{m-1} \|_1
\le
\beta_{m-1}
\|
\tilde{\bx}_{it}^{(0)}(p)
\|_\infty,
$$
which explains the $\ell_\infty$-based transfer-bias radius.

\paragraph{Price-uniform envelope.} The intermediate bound $\bar v_{it}(p)$ may not preserve the monotonicity of utility in price implied by positive price sensitivity (Assumption~\ref{assump:price-sensitivity}). To restore this structure while maintaining optimism, we leverage the structural property $v'_{it}(p)\le -L_0<0$ and tighten the bound via a monotone-Lipschitz envelope:
\begin{equation} \label{eq:optimistic-utility} 
\tilde v_{it}(p)
:= \min_{p'\le p}
\big\{\bar v_{it}(p') - L_0 (p-p')\big\},
\qquad p\in[0,\bar P],
\end{equation}
where $L_0$ is the uniform lower bound on price sensitivity. By construction (Lemma~\ref{lem:lipschitz-envelope}), $\tilde v_{it}(p)$ is decreasing, $L_0$-Lipschitz, and satisfies
$$
v^{(0)}_{it}(p) \;\le\; \tilde v_{it}(p) \;\le\; \bar v_{it}(p)
\quad\text{for all }p\in[0,\bar P].
$$

\paragraph{Optimistic revenue maximization.} For assortment $S$ and price vector $p$, we define the optimistic revenue
\begin{equation*}
\tilde{R}_t(S,\bp;\alpha_{m-1},\beta_{m-1},W_{m-1}):=
\frac{\sum_{i\in S} p_{i}\exp\!\big(\tilde{v}_{it}(p_i)\big)}
{1+\sum_{j\in S}\exp\!\big(\tilde{v}_{jt}(p_j)\big)}.
\end{equation*}
At each period $t$ in episode $m$, TJAP selects
$$
(S_t^{(0)},\bp_t^{(0)})
\in \argmax_{S\in\mathcal{S}_K,\; \bp\in[0,\bar P]^N}
\widetilde R_t(S,\bp;\alpha_{m-1},\beta_{m-1},W_{m-1}).
$$
Because $\tilde v_{it}(\cdot)$ is decreasing and Lipschitz, the optimization over prices admits the same fixed-point characterization as in classical MNL joint pricing-assortment problems \citep{wang2012offline}. In particular, for any fixed assortment $S$, the optimal prices can be computed efficiently by solving a one-dimensional fixed-point equation, and the overall maximization remains tractable.

Lemma~\ref{lem:price-optimism} further shows that this preserves optimism at the optimized revenue level:
$$
R_t(S_t^*,\bp_t^*)
\le
\tilde R_t(S_t^{(0)},\bp_t^{(0)};\alpha_{m-1},\beta_{m-1},W_{m-1}).
$$
This construction yields a valid optimistic upper bound on the true target-market revenue uniformly over prices and assortments. In Section~\ref{sec:theory}, we show that the resulting instantaneous regret decomposes into a variance term governed by $\alpha_{m-1}$ and a transfer-bias term governed by $\beta_{m-1}$.

\subsection{Information Geometry and Episodic Control}\label{ssec:info_geometry}

The optimistic decision rule in Section \ref{ssec:opt_decision_rule} relies on confidence bounds that must remain valid under adaptively chosen assortments and prices. Because the informativeness of observations depends endogenously on past actions, naive online updates of confidence geometry can lead to ill-conditioned matrices and invalidate self-normalized concentration arguments. 

TJAP addresses this issue through {\it episodic information-geometry control}.
Within each episode, both the parameter estimate and the information matrix used in the confidence radius are frozen. This separation decouples confidence construction from within-episode adaptivity, allowing standard self-normalized concentration arguments to apply.

The Fisher information matrix measures the informativeness of the data and controls the self-normalized confidence radius that enters our optimism bonus. In implementation, we maintain three Fisher information matrices.

\paragraph{Episodic information matrix.} 
Fix an episode $m$. For each market $h\in\{0\}\cup[H]$, a time $s$, and a parameter vector $\bnu\in\mathbb{R}^{2d}$.
Conditioning on the realized context and actions $(\bx^{(h)}_s, S^{(h)}_s, \bp^{(h)}_s)$, the {\it per-period Fisher information matrix} $I^{(h)}_s(\bnu)$ of the MNL model at parameter $\bnu$ is defined as the conditional expectation of the score outer product.
The explicit formula is:
\begin{equation*}
I_s^{(h)}(\bnu)
=\ \sum_{i\in S_s^{(h)}} q_{is}^{(h)}(\bnu)\,\tilde{\bx}_{is}^{(h)}\tilde{\bx}_{is}^{(h)\top}
\ -\ \sum_{i\in S_s^{(h)}}\sum_{j\in S_s^{(h)}} q_{is}^{(h)}(\bnu)\,q_{js}^{(h)}(\bnu)\,\tilde{\bx}_{is}^{(h)}\tilde{\bx}_{js}^{(h)\top},
\end{equation*}
where we use the shorthand $\tilde{\bx}_{is}^{(h)}:=\tilde{\bx}_{is}^{(h)}(p_{is}^{(h)})$, and $q_{is}^{(h)}(\bnu)$ is the purchase probability given in~\eqref{eq:MNL-chocie-model}. 

\paragraph{Rolling Fisher information within an episode.} 
At a time $t\in\mathcal{T}_m$ in the current episode, we define the {\it rolling Fisher information matrix} by 
$$
V^{(h)}_t
\;:=\;
\sum_{s=\tau_{m-1}+1}^{t}
I^{(h)}_s\!\left(\widehat{\bnu}_{m-1}\right),
$$
where the per-period Fisher information matrix is evaluated at the fixed estimate $\widehat{\bnu}_{m-1}$.

The matrix $V^{(h)}_t$ accumulates curvature information generated within the current episode and is reset to zero at the beginning of the next episode.
Among these matrices, the target-market Fisher information $V^{(0)}_t$ plays a special role: it tracks the extent to which the algorithm has explored directions that are informative for identifying the target-market parameters under the current policy.

\paragraph{Episodic information matrix and geometry freezing.}
At the end of episode $m-1$, TJAP aggregates curvature information from all markets to form the episodic information matrix
\begin{equation}\label{eq:def-Wm}
W_{m-1}\ :=\ V^{(0)}_{\tau_{m-1}}\ +\ \sum_{h=1}^H V^{(h)}_{\tau_{m-1}}.
\end{equation}
which defines the confidence geometry used throughout episode $m$ (Section~\ref{ssec:opt_decision_rule}). The matrix $W_{m-1}$ remains fixed within the episode and enters only through the variance term of the optimistic objective. This \emph{geometry freezing} decouples confidence construction from within-episode adaptive decisions, ensuring that estimation error can be controlled uniformly over $\mathcal{T}_m$ through self-normalized martingale concentration. At the same time, source-market data contribute directly to $W_{m-1}$, tightening the information geometry and thereby shrinking the variance radius.

\paragraph{Identifiability gate and forced exploration.} While pooled curvature tightens confidence regions, consistent estimation of the sparse target deviation requires sufficient curvature in the target-market likelihood. TJAP therefore monitors the minimum eigenvalue of the rolling target Fisher matrix $V^{(0)}_t$.

If, near the end of an episode, the target curvature falls below a prescribed threshold, i.e. $\lambda_{\min}\!\left(V^{(0)}_t\right) \;\ge\; c_0$ fails for a prescribed constant $c_0>0$, the algorithm enters a short {\it forced-exploration phase} in the target market. During this phase, assortments and prices are selected from a fixed exploratory distribution designed to inject information along all coordinates. The duration of this phase is chosen so that, with high probability, the target Fisher matrix satisfies the required eigenvalue condition (Lemma~\ref{lem:topup-chernoff}).

Crucially, this exploration is {\it episodic and gate-controlled}: if the optimistic policy already generates sufficient target-market curvature, no additional exploration is invoked. As shown in Section \ref{sec:theory}, the total number of forced rounds across all episodes is logarithmic in $T$, and their contribution to regret is asymptotically negligible.

\paragraph{Forced-exploration length.} 
Let $q_{m-1}$ denote the length of the forced-exploration window available at the end of episode $m$. We choose $q_{m-1}$ to be the smallest integer such that, under the exploration policy (uniformly random assortments and prices), the target Fisher information satisfies the eigenvalue condition in Lemma~\ref{lem:topup-chernoff} with failure budget $\eta_{m-1}$. Equivalently, $q_{m-1}$ is selected so that with probability at least $1-\eta_{m-1}$, the forced exploration increases $\lambda_{\min}(V^{(0)}_t)$ above the prescribed threshold.
 In our analysis, this choice implies $\sum_m q_m = \tilde{\mathcal{O}}(d)$, so forced exploration contributes only a lower-order term to regret.

\paragraph{Role in the analysis.} 
The episodic information-geometry control serves two purposes in the analysis. First, it guarantees that the variance radius in the optimistic utility remains well defined and contracts at a rate governed by the aggregated information in $W_{m-1}$. Second, by enforcing target-market curvature through the identifiability gate, it ensures that the debiasing step satisfies a uniform restricted eigenvalue condition, which is essential for controlling the transfer-bias radius.

Together, episodic freezing and gate-controlled exploration allow TJAP to combine aggressive cross-market variance reduction with stable and identifiable target-market learning under adaptive joint assortment-pricing decisions.

\subsection{Extension to Heterogeneous Covariates}\label{ssec:heterogeneous}

Thus far, we have assumed homogeneous covariate distributions across markets, so that cross-market heterogeneity arises solely through preference shifts. When covariate distributions differ across markets, naive pooling may distort the information geometry and overstate the effective sample size contributed by source markets.
To address this, TJAP is modified through the
construction of the aggregate center in~\eqref{eq:agg_estimator}, and the Mahalanobis term inside the variance bonus in~\eqref{eq:two-radius-bonus}.

\paragraph{Reweighted episodic information geometry.}
Under heterogeneous covariates, the key change is to reweight the pooled Fisher information.
Concretely, we replace the unweighted pooled information matrix in~\eqref{eq:def-Wm} by
\begin{equation}\label{eq:def-Wm-hetero}
W_{m-1}
:= V^{(0)}_{\tau_{m-1}} + \sum_{h=1}^{H} \omega_{h,m-1}\, V^{(h)}_{\tau_{m-1}},
\end{equation}
where $\omega_{h,m-1}\ge 0$ controls the contribution of source market $h$ to the pooled geometry.
In the homogeneous setting, $\omega_{h,m-1} \equiv 1$, and \eqref{eq:def-Wm} is recovered.

The variance bonus in \eqref{eq:two-radius-bonus} continues to use the Mahalanobis norm induced by this reweighted matrix. By downweighting misaligned sources, the confidence geometry more accurately reflects the curvature of the target-market likelihood.

\paragraph{Reweighting the aggregate center.}
To maintain consistency between the pooled center and the information geometry, the aggregate estimator in~\eqref{eq:agg_estimator} can be modified to
\begin{equation}\label{eq:agg_estimator_hetero}
\widehat{\bnu}^{\mathrm{(ag)}}_{m-1}
\in \argmin_{\bnu\in\mathbb{R}^{2d}}
\;\;\frac{1}{\sum_{h=1}^{H}\omega_{h,m-1}}
\sum_{h=1}^{H}\omega_{h,m-1}\cdot \frac{1}{|\mathcal{T}_{m-1}|}\sum_{t\in \mathcal{T}_{m-1}} \ell^{(h)}_{t}(\bnu).
\end{equation}
Thus both curvature and pooled estimation are aligned with the effective target geometry.

\paragraph{Choice of weights.}
We propose two lightweight choices that fit seamlessly into TJAP.

\begin{itemize}
\item[(i)] \emph{Sample-level importance reweighting.} If a density ratio $w_h(x)=\frac{dP_x^{(0)}}{dP_x^{(h)}}(x)$ is available or can be reliably estimated, the rolling Fisher matrix may be replaced by an importance-weighted version
$$
\widetilde V_{\tau_{m-1}}^{(h)} := \sum_{t\in T_{m-1}} w_h(\bx_t^{(h)})\, I_t^{(h)}(\hat\bnu_{m-1}),
$$
so that, in expectation, source curvature $\mathbb{E}[\widetilde V_{\tau_{m-1}}^{(h)}]$ aligns with the target geometry. In this case, we use $W_{m-1}=V_{\tau_{m-1}}^{(0)}+\sum_{h=1}^H \widetilde V_{\tau_{m-1}}^{(h)}$ and no additional market-level downweighting is required.

\item[(ii)] \emph{Market-level information weights.} When density-ratio estimation is impractical, a more robust alternative is to retain $V_{\tau_{m-1}}^{(h)}$ and set $\omega_{h,m-1}$ based on a covariate-mismatch score,
\[
\omega_{h,m-1}
=\frac{1}{1+\chi^2(\widehat P_x^{(0)}\|\widehat P_x^{(h)})}
\qquad\text{or}\qquad
\omega_{h,m-1}=\min\Bigl\{1,\frac{c}{\widehat\rho_h}\Bigr\},
\]
where $\widehat\rho_h$ is an estimated density-ratio bound and $c>0$ caps the influence of any single source.
These choices preserve positive definiteness of $W_{m-1}$ while preventing heavily shifted sources from dominating the pooled geometry.
\end{itemize}
The remainder of the algorithm, including the debiasing step and the optimistic decision rule, remains unchanged. The extension therefore preserves the variance-bias structure of TJAP, with covariate shift handled through controlled reweighting of source curvature.

\section{Theoretical Results and Insights}\label{sec:theory}

We now establish finite-time regret guarantees for TJAP under the structural conditions introduced in Section \ref{sec:problem-formulation}. Our analysis characterizes how transfer from $H$ source markets accelerates learning in the target market, while explicitly quantifying the residual cost of cross-market preference shifts.

We begin by formalizing standard regularity conditions that ensure well-behaved pricing and estimation under the MNL model.

\begin{assumption}[Positive price sensitivity] \label{assump:price-sensitivity}
There exists $L_0 > 0$ such that
$$
\min_{i \in [N],\, t \in [T]} \langle \bx_{it}, \bgamma \rangle \ge L_0.
$$
\end{assumption}
This condition ensures that utility is strictly decreasing in price and guarantees the existence of finite revenue-maximizing prices.

\begin{assumption}[Bounded parameter space] \label{assump:param-space}
We assume that $\|\bx_{it}\|_\infty \leq 1, \forall i \in [N], t \in [T]$, and $\|(\btheta, \bgamma)\| \leq 1$.
\end{assumption}
This normalization ensures scale-free regret bounds and bounded curvature of the log-likelihood.

\begin{assumption}[Non-degeneracy]\label{assump:fisher-invertible}
There exist constants $\kappa > 0$ and $r > 0$ such that for all feasible assortments $S \in \mathcal{S}_K$, prices $\bp \in [0,\bar P]^N$, and parameters $\bnu$ satisfying $\|\bnu - \bnu^{(0)}\|_2 \le r$,
$$
\min_{i\in S\cup\{0\}} q_t\!\big(i \,\big|\, S,\bp,\bnu\big) \;\ge\; \kappa.
$$
\end{assumption}
Assumption~\ref{assump:fisher-invertible} ensures that choice probabilities are uniformly bounded away from zero in a neighborhood of the true parameter, yielding well-conditioned Fisher information and valid confidence sets. It is standard in MNL bandits literature~\citep{Oh2019TSforMNL,chen2020dynamic}.

To isolate transfer effects from distributional differences in observed features, we impose:

\begin{assumption}[Homogeneous Covariates with Bounded Eigenvalues] \label{assump:Homo-Cov}
For each $h\in \{0\}\cup [H]$, covariates $ \bx_{it}^{(h)}$ are drawn i.i.d. across items, rounds and markets from a fixed, but a priori unknown, distribution $ \mathcal{P}_x $, supported on a bounded set $ \mathcal{X} \subset \mathbb{R}^d $, with mean zero and covariance matrix $\Sigma = \mathbb{E}[\bx_{it}^{(h)} (\bx_{it}^{(h)})^\top]$. The covariance matrix $\Sigma$ satisfies
$$
0 < C_{\min} \le \lambda_{\min}(\Sigma) \le \lambda_{\max}(\Sigma) \le C_{\max} < \infty,
$$
where $\lambda_{\min}(\Sigma)$ and $\lambda_{\max}$ denote the minimum and maximum eigenvalues of $\Sigma$ respectively. 
\end{assumption}

Assumption~\ref{assump:Homo-Cov} plays two important roles. 
(i) The boundedness of the covariate support ensures that regret bounds are scale-free, a standard technical condition in online learning literature~\citep{javanmard2019dynamic,erginbas2025online}. This requirement is not restrictive, as it is satisfied by many practically relevant distributions, such as truncated Gaussian or uniform distributions. 
(ii) The assumption that covariates share a common distribution across all markets ($h \in \{0\} \cup [H]$) serves to isolate cross-market heterogeneity to \emph{model~(utility) shift} rather than \emph{distributional~(covariate) shift}. This modeling choice is particularly reasonable when the underlying populations are comparable in their observed characteristics, while differences across markets manifest primarily in latent preference parameters. By decoupling feature distributional similarity from preference heterogeneity, the assumption provides a clean setting to analyze the impact of transfer.
This condition can be relaxed to allow covariate shift, as discussed in Section~\ref{ssec:heterogeneous}.

Under these conditions, we first characterize the statistical convergence of the target parameter estimator, and then establish regret guarantees and fundamental limits of transfer learning in this setting.

\subsection{Statistical Error of the Aggregate-Then-Debias Estimator}

We first establish the statistical accuracy of the aggregate-then-debias estimator, and then translate this result into regret guarantees that quantify the benefit and limitation of transfer.

Let $\widehat{\bnu}_{m-1}$ denote the estimator constructed at the beginning of episode $m$, based on source-market aggregation and target-only $\ell_1$-debiasing. Let $\bnu^{(0)}$ denote the true target parameter.
The following theorem provides a high-probability bound that decomposes estimation error into a variance component, reduced through cross-market pooling, and a bias component, driven by sparse preference shifts.

\begin{theorem}[Statistical Error Bound]\label{thm:stats_error_bound}
Under Assumptions \ref{assump:task-simi}–\ref{assump:Homo-Cov}, there exist constants $C_v, C_b > 0$, depending only on $L_0, C_{\min}, C_{\max}, \kappa, r$, such that with probability at least $1 - \eta$, for every episode $m$,
$$
\|\widehat{\bnu}_{m-1} - \bnu^{(0)}\|_2
\;\le\;
C_v
\sqrt{\frac{d \log((1+H)T/\eta)}{(1+H)\,\tau_{m-1}}}
\;+\;
C_b
\, s_0
\sqrt{\frac{\log(dT/\eta)}{\tau_{m-1}}}.
$$
\end{theorem}

The bound separates two sources of estimation error: firstly, the variance term, 
$
\mathcal{O}\!\left(
\sqrt{d\,\tau_{m-1}^{-1}(1+H)^{-1}}
\right),
$
decreases as additional source markets contribute curvature.
and the transfer-bias term:
$
\mathcal{O}\!\left(
s_0\,\tau_{m-1}^{-1/2}
\right),
$
reflects the cost of estimating sparse target-specific deviations and does not benefit from additional sources.
Thus, pooling $H$ source markets effectively increases the sample size for shared preference coordinates by a factor of $1+H$, while estimation along shifted coordinates depends solely on target data.

\subsection{Regret Upper Bound}

We now quantify the learning performance of TJAP under the structural conditions introduced above. 
The tuning parameters in TJAP directly reflect the two-layer estimation structure introduced in Section~\ref{sec:algo}. Specifically:
\begin{itemize}
\item $\alpha_m$ controls the {\it variance radius} in the optimistic utility and arises from self-normalized concentration in the pooled information geometry. It scales with the effective curvature accumulated across the target and source markets.
\item $\lambda_m$ is the regularization parameter in the $\ell_1$-debiasing step and determines the accuracy with which sparse target-specific deviations are estimated from target-market data.
\item $\beta_m$ is the resulting {\it transfer-bias radius}, proportional to $s_0 \lambda_m$, and captures the residual error induced by cross-market preference shifts.
\end{itemize}
The forced-exploration length $q_m$ ensures that the target Fisher information satisfies a restricted eigenvalue condition, which is necessary for stable sparse recovery.

With these choices, the estimation error admits a variance-bias decomposition that propagates to regret. The following result provides a finite-time regret bound that makes explicit how transfer from $H$ source markets accelerates learning through pooled curvature while accounting for residual error induced by cross-market preference shifts.

\begin{theorem}[Regret Upper Bound]\label{thm:upper-linear-o2o}
Suppose Assumptions~\ref{assump:task-simi}–\ref{assump:Homo-Cov} hold. 
Run Algorithm~\ref{algo:trans-assort-price-o2o} with episodic parameters
\[
\alpha_{m}=c_\alpha\,\sqrt{\,2d\log\!\Bigl(1+\frac{\mathrm{tr}(W_{m})}{2d\,\lambda_0}\Bigr)+\log\!\frac{2}{\eta_{m}}\,},\quad
\lambda_{m}=c_\lambda\,\sqrt{\frac{\log(2d/\eta_{m})}{|\mathcal{T}_{m}|}},\quad
\beta_{m}=\frac{c_\beta\,s_0\,\lambda_{m}}{\phi_{m}^2},
\]
and choose the forced-exploration schedule $q_m$ according to Lemma~\ref{lem:topup-chernoff} with per-episode failure budget $\eta_{m}$, where $q_m$ is the (episode-dependent) exploration-window length defined in Section \ref{ssec:info_geometry}, $\{\eta_m\}_{m\ge1}$ satisfies $\sum_m \eta_m \le T^{-2}$, 
and $\phi_{m}^2$ is the restricted strong convexity constant in~\eqref{eq:RE-target}.

Then there exist constants $C_1,C_2,C_3>0$, depending only on
$L_0,\bar P,C_{\rm min},C_{\rm max},\kappa,r$, such that the cumulative expected
regret up to horizon $T$ satisfies
\begin{equation}\label{eqn:regret-upper}
\operatorname{Regret}(T;\pi)\le
K\,\bar P\Bigg[  C_1\,
d\,\sqrt{\frac{T\log\!\big((1+H)T^3\big)}{1+H}}\,
\;+\;
C_2\,s_0\,\sqrt{T\,\log\!\big(4dT^2\big)}
\;+\;
C_3\,d\,\log^2\!\big(4dT^2\big) \Bigg].
\end{equation}
\end{theorem}

The regret bound exhibits a transparent variance-bias decomposition. The leading term scales as
$
\mathcal{O}\!\left(
d\,T^{1/2}\,(1+H)^{-1/2}
\right),
$
reflecting variance reduction from pooling $H$ source markets. Under homogeneous covariates, source data contribute curvature in the same geometry as the target, effectively increasing the information available for shared preference directions by a factor of $1+H$. Consequently, the statistical uncertainty contracts at rate $1/\sqrt{1+H}$.

The second term,
$
\mathcal{O}\!\left(
s_0 \sqrt{T}
\right),
$
captures residual transfer bias due to sparse cross-market preference shifts. This term depends only on the sparsity level $s_0$ and does not improve with additional source markets. Intuitively, deviations confined to shifted coordinates must be learned from target data alone.

The remaining logarithmic term arises from episodic identifiability control and forced exploration; it is lower order relative to the main $\sqrt{T}$ contributions.
In particular, when $s_0$ is small relative to $d$, transfer yields substantial gains, while when heterogeneity is diffuse ($s_0$ large), the improvement from additional source markets diminishes.

\subsection{Regret Lower Bound}

The following theorem characterizes the fundamental limits of transfer learning under the structured preference-shift model.

\begin{theorem}[Minimax lower bound]\label{thm:lower-linear-o2o}
Under the utility model~\eqref{eq:utility-func} with Assumptions~\ref{assump:task-simi}–\ref{assump:Homo-Cov},   
for any $d \ge 1$, $K \in [d]$, and sparsity level $s_0 \in \{0,1,\dots,\min\{K,d\}\}$, there exists a constant $c_0 > 0$, depending only on $L_0, \bar P, C_{\min}, C_{\max}$, such that for all horizons $T$,
\begin{equation}\label{eq:LB-final-s0}
\inf_{\pi}\;\sup_{\text{instances}}\ \operatorname{Regret}(T;\pi)
\ \ge\ 
c_0\Bigg[\sqrt{\frac{K\,(d-s_0)\,T}{\,1+H\,}}+s_0\,\sqrt{K\,T}
\Bigg],
\end{equation}
where the infimum is taken over all policies $\pi$, and the supremum is over all problem instances satisfying Assumptions~\ref{assump:task-simi}–\ref{assump:Homo-Cov}.
\end{theorem}

The lower bound mirrors the two components in Theorem \ref{thm:upper-linear-o2o}. The first term scales as
$
\sqrt{{(d-s_0)T}{(1+H)^{-1}}},
$
showing that only the shared coordinates benefit from additional source markets. The second term scales as
$
s_0 \sqrt{T},
$
reflecting the unavoidable cost of learning target-specific deviations. Crucially, this term does not improve with $H$, since source observations carry no information about shifted coordinates.

Comparing Theorems~\ref{thm:upper-linear-o2o} and~\ref{thm:lower-linear-o2o}, we see that TJAP achieves the optimal dependence on $H$, $s_0$, and $T$ up to logarithmic factors.

\subsection{Structural Implications of Transfer in Joint Assortment-Pricing}
Theorems~\ref{thm:upper-linear-o2o} and~\ref{thm:lower-linear-o2o} isolate two distinct forces that
govern the value of transfer: \emph{statistical uncertainty}, which can be reduced by pooling data from
multiple markets, and \emph{structural mismatch}, which cannot. The upper bound~\eqref{eqn:regret-upper}
makes this separation explicit, and the lower bound~\eqref{eq:LB-final-s0} shows that the same
qualitative trade-off is unavoidable over the problem class.

\paragraph{Transfer accelerates learning along shared directions.}
The leading regret term scales as
$
\sqrt{T/(1+H)},
$
demonstrating that additional source markets reduce statistical uncertainty in proportion to pooled curvature. When covariates are homogeneous, source observations contribute information in the same geometric directions as target data. As a result, estimation along shared preference coordinates benefits from an effective sample size proportional to $1+H$. The diminishing marginal improvement in $H$ reflects classical concentration behavior.

\paragraph{Heterogeneity imposes an intrinsic ceiling.}
The second term, proportional to
$
s_0 \sqrt{T}
$
is independent of $H$. This term corresponds to coordinates where the target preference differs from all sources. Because source data contain no information about these shifted components, their estimation must rely entirely on target observations. The lower bound confirms that this cost is unavoidable: no algorithm can eliminate the $s_0$-driven contribution without additional structural assumptions.
Thus, transfer is fundamentally asymmetric. It improves learning only along directions where markets are behaviorally aligned and provides no benefit where structural mismatch persists.

\paragraph{Sparse heterogeneity determines the value of transfer.} The sparsity level $s_0$ emerges as the key measure of cross-market similarity. When heterogeneity is localized (small $s_0$), transfer yields substantial gains. When heterogeneity is diffuse (large $s_0$), the bias term dominates and the marginal benefit of additional source markets diminishes. In the extreme case $s_0 \approx d$, transfer offers little improvement over target-only learning.

\paragraph{Exploration overhead is secondary.}
The logarithmic term in the upper bound arises from episodic identifiability control. Its contribution is lower order relative to the $\sqrt{T}$ terms and does not alter the fundamental variance–bias structure.

\section{Numerical Experiments}\label{sec:exp}

We evaluate TJAP on synthetic instances designed to reflect the structured preference-shift model in Section \ref{sec:problem-formulation}. Our experiments aim to examine three operational questions:
\begin{itemize}
\item {\it How much do additional markets accelerate joint learning?} 
Regret should decrease with the number of source markets $H$, at a rate consistent with the $1/\sqrt{1+H}$ scaling in Theorem 2.
\item {\it When does transfer meaningfully improve joint assortment-pricing?} 
When cross-market differences are localized (small $s_0$), transfer should substantially improve decision quality; when heterogeneity is widespread, the gains should taper off.
\item {\it Is naive data pooling operationally safe?} 
Aggregating data without correcting for cross-market shifts should degrade performance when markets differ structurally.
\end{itemize}

\subsection{Synthetic Data Generator}

All markets follow the contextual MNL model with a common covariate distribution, so cross-market heterogeneity arises exclusively through preference shifts (Assumption \ref{assump:task-simi}). We vary:
\begin{itemize}
\item Feature dimension $d \in \{10,20,50\}$,
\item Sparsity level $s_0 \in \{0.2d, 0.3d\}$,
\item Catalog size $N \in \{30,100\}$,
\item Capacity $K = 5$,
\item Number of source markets $H \in \{0,1,3,5\}$,
\item Horizon $T = 2000$.
\end{itemize}
Contexts are generated i.i.d.~across periods and shared across markets to isolate preference heterogeneity from covariate shift. Target parameters are drawn randomly, and each source market is generated by introducing an $s_0$-sparse utility shift relative to the target.
Details of the generator are summarized as follows.
\begin{itemize}[leftmargin=1.4em,itemsep=2pt,topsep=3pt]
\item \textbf{Catalog and contexts.}
At each period $t=1,\dots,T$, we draw a context matrix $\bX_t\in\mathbb{R}^{N\times d}$ with rows $\bx_{it}\in\mathbb{R}^d$ sampled i.i.d.\ from a common distribution. Concretely, we generate $\bx_{it}$ by drawing $z_{it}\sim\mathcal{N}(0,I_d)$ and setting $\bx_{it}=|z_{it}|$, followed by entrywise clipping to $[0,1]$ to ensure $\|\bx_{it}\|_\infty\le 1$ (\ Assumption~\ref{assump:param-space}). We use the same $\bX_t$ across all markets at time $t$ to isolate cross-market differences to preference parameters rather than realized contexts.

\item \textbf{Target parameters.}
We sample $\btheta^{(0)}\sim\mathcal{N}(0,I_d)$ and draw $\bgamma^{(0)}$ with strictly positive entries. Since $\bx_{it}\ge 0$ componentwise by construction, we can rescale $\bgamma^{(0)}$ by a positive scalar to satisfy Assumption~\ref{assump:price-sensitivity}. 

\item \textbf{Source parameters.}
For each source market $h\in[H]$, we generate a sparse shift $\bdelta^{(h)}\in\mathbb{R}^{2d}$ with support size $s_0$ drawn uniformly at random: $\bdelta^{(h)}_j=\pm\Delta$ on the support and $0$ otherwise. We set
\[
\btheta^{(h)}=\btheta^{(0)}+\bdelta^{(h)}_{1:d},
\qquad
\bgamma^{(h)}=\max\{10^{-3},\,\bgamma^{(0)}+\bdelta^{(h)}_{d+1:2d}\}
\]
(componentwise maximum) to maintain positive price sensitivity. We choose $\Delta$ small enough so that the lower bound in Assumption~\ref{assump:price-sensitivity} continues to hold across markets. This construction matches the task-similarity model with an $s_0$-sparse discrepancy (\ Assumption~\ref{assump:task-simi}).
\end{itemize}
This construction allows us to directly test the variance-bias tradeoff characterized in Section \ref{sec:theory}.

\subsection{Baselines}

We compare \textsc{TJAP} with four baselines:
\textsc{CAP}~\citep{erginbas2025online},
\textsc{M3P}~\citep{javanmard2020multi},
\textsc{ONS-MPP}~\citep{perivier2022dynamic}, and a naive pooled estimator \textsc{Pool}$(H)$. 

\textsc{CAP} is a contextual joint assortment-pricing algorithm and serves as a natural benchmark in the single-market setting; we run it on the target market. \textsc{M3P} and \textsc{ONS-MPP} are pricing-centric methods; to respect the capacity constraint, at each period we rank items by the current estimated systematic utilities and offer the top $K$ items, while using the method’s posted prices for this subset. The pooled estimator \textsc{Pool}$(H)$ aggregates all observations from the target and $H$ sources and fits a single parameter $\bnu$ via the same likelihood update as \textsc{TJAP}, ignoring cross-market heterogeneity and performing no debiasing. 
Each configuration is repeated 10 times with independent seeds, and we report mean cumulative regret.

\begin{figure}[htb!]
  \centering

  \begin{minipage}[t]{0.32\linewidth}
    \centering
    \includegraphics[width=\linewidth]{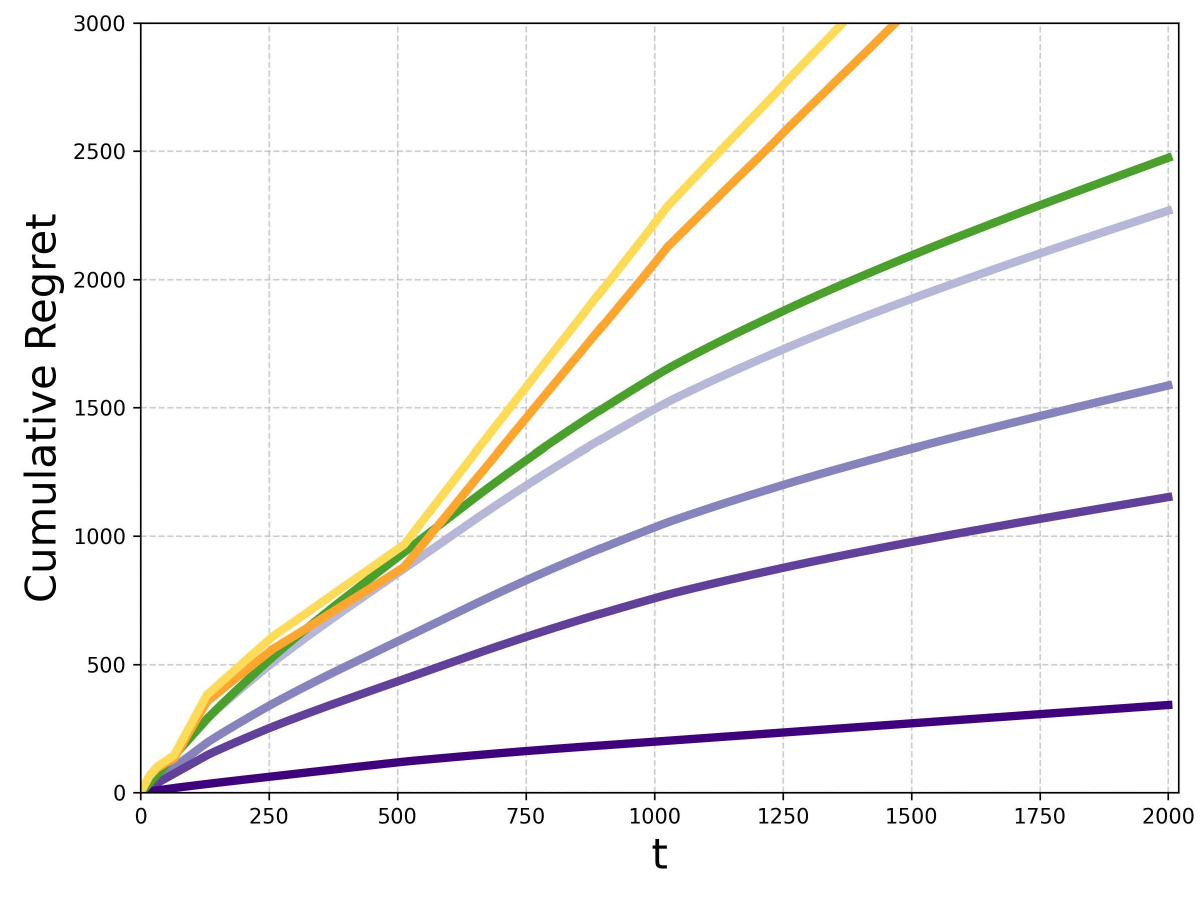}\\[-2pt]
    {\footnotesize (a) $d=10,\ s_0=2,\ K=5,\ N=30$}
    \label{fig:regret_2_10_5_30}
  \end{minipage}\hfill
  \begin{minipage}[t]{0.32\linewidth}
    \centering
    \includegraphics[width=\linewidth]{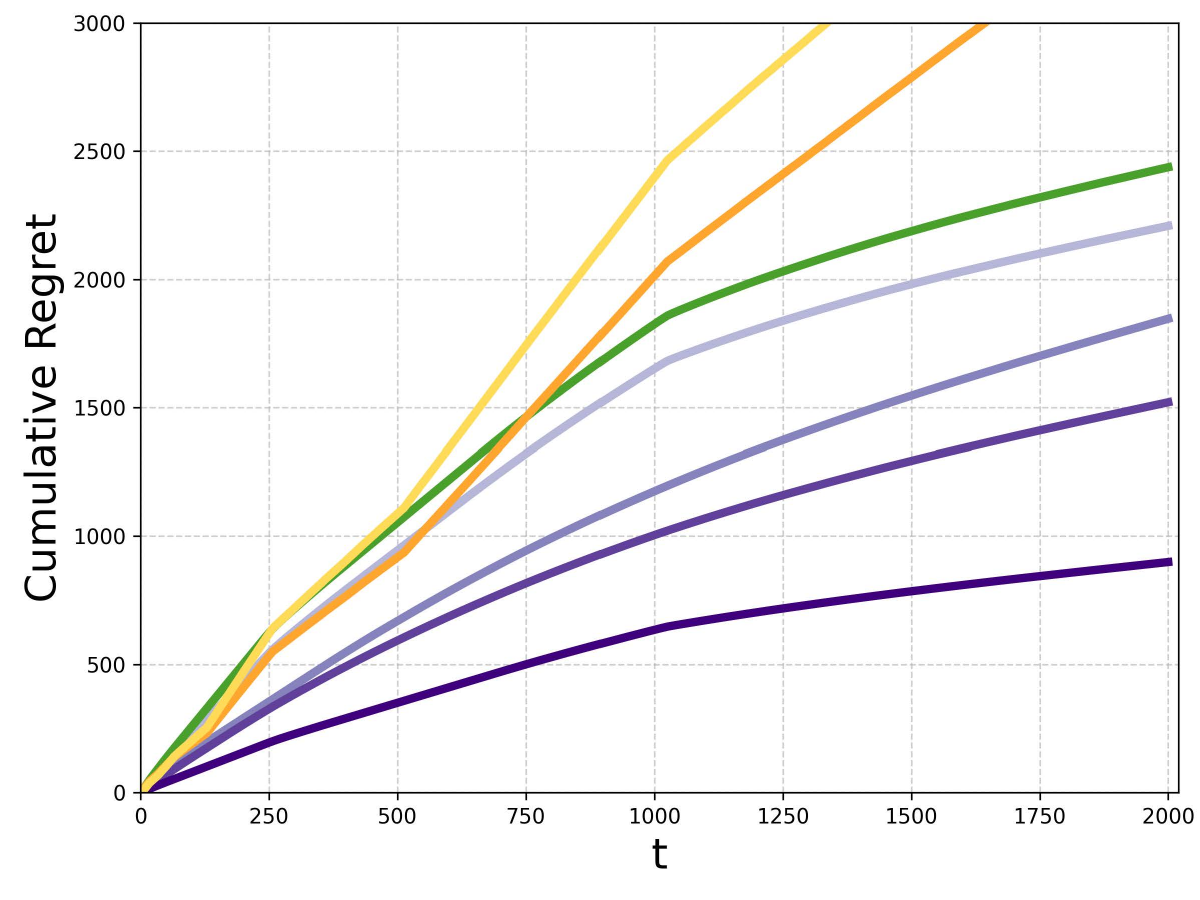}\\[-2pt]
    {\footnotesize (b) $d=20,\ s_0=6,\ K=5,\ N=100$}
    \label{fig:regret_6_20_5_100}
  \end{minipage}\hfill
  \begin{minipage}[t]{0.32\linewidth}
    \centering
    \includegraphics[width=\linewidth]{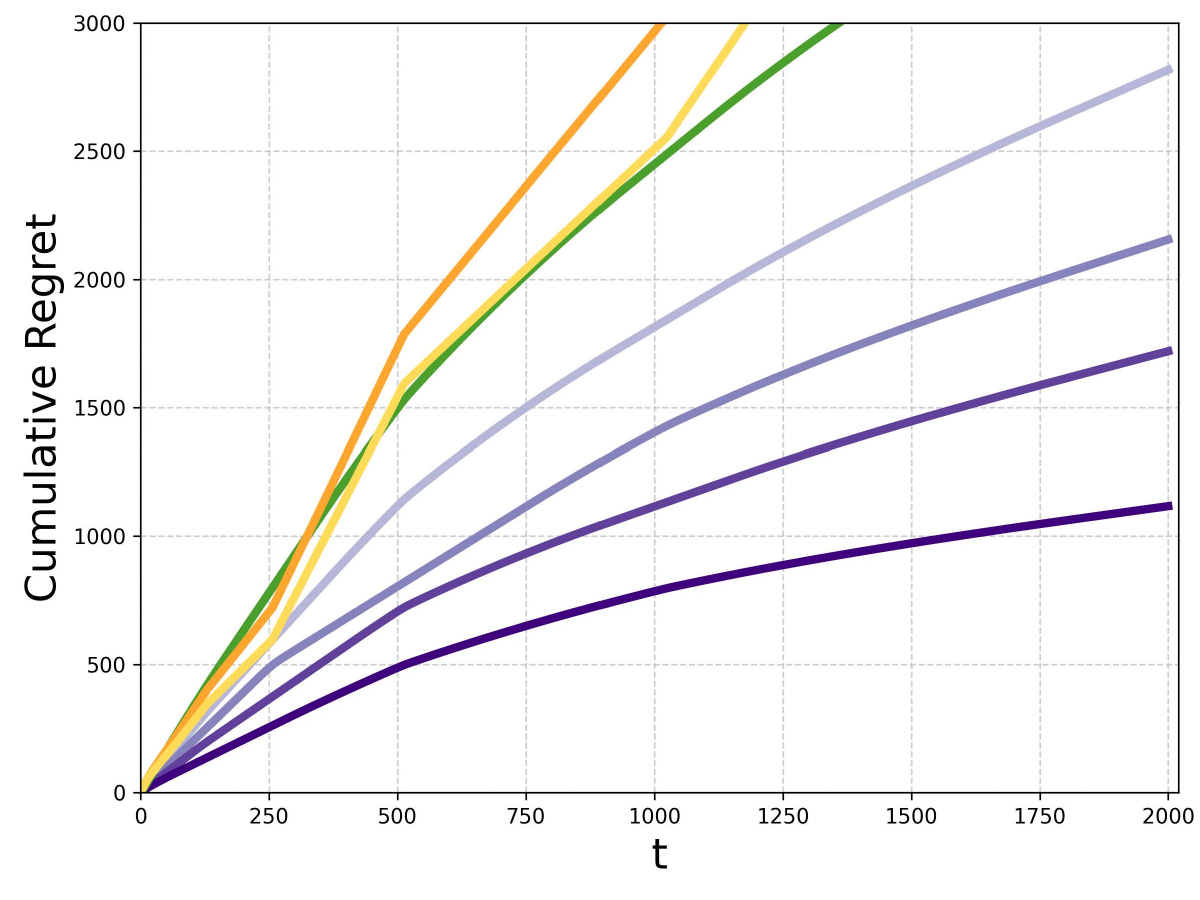}\\[-2pt]
    {\footnotesize (c) $d=50,\ s_0=15,\ K=5,\ N=100$}
    \label{fig:regret_15_50_5_100}
  \end{minipage}

  \vspace{-10pt}

  \begin{minipage}[t]{0.32\linewidth}
    \centering
    \includegraphics[width=\linewidth]{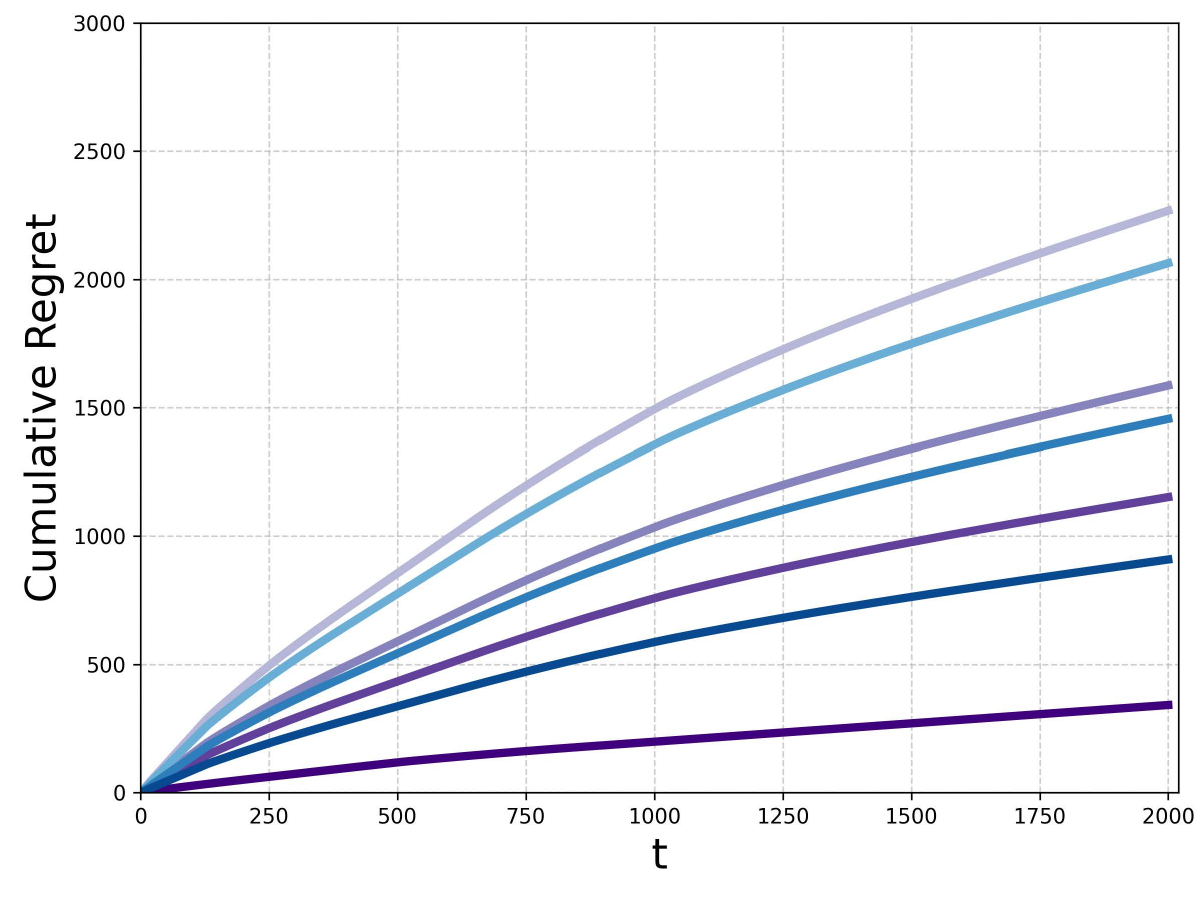}\\[-2pt]
    {\footnotesize (d) $d=10,\ s_0=2,\ K=5,\ N=30$}
    \label{fig:regret_2_10_5_30_pool}
  \end{minipage}\hfill
  \begin{minipage}[t]{0.32\linewidth}
    \centering
    \includegraphics[width=\linewidth]{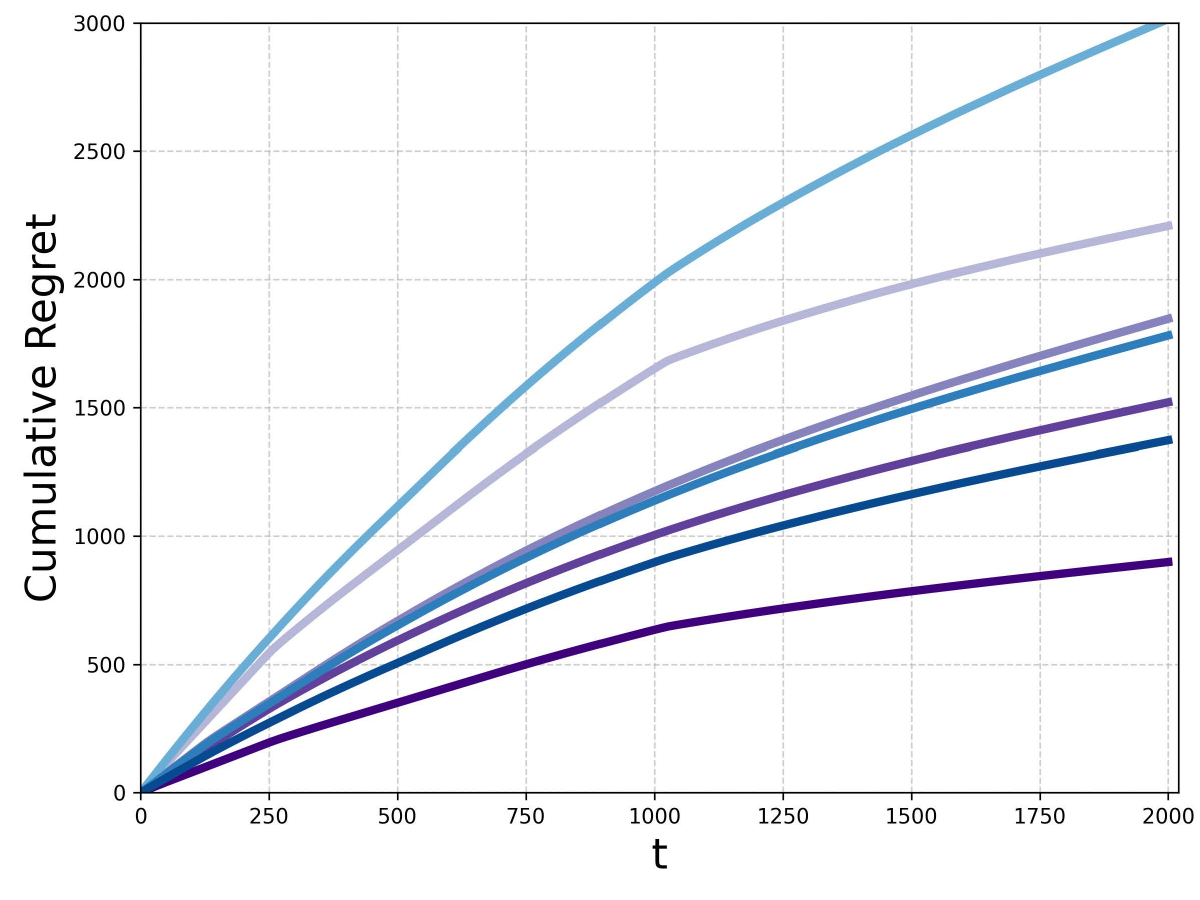}\\[-2pt]
    {\footnotesize (e) $d=20,\ s_0=6,\ K=5,\ N=100$}
    \label{fig:regret_6_20_5_100_pool}
  \end{minipage}\hfill
  \begin{minipage}[t]{0.32\linewidth}
    \centering
    \includegraphics[width=\linewidth]{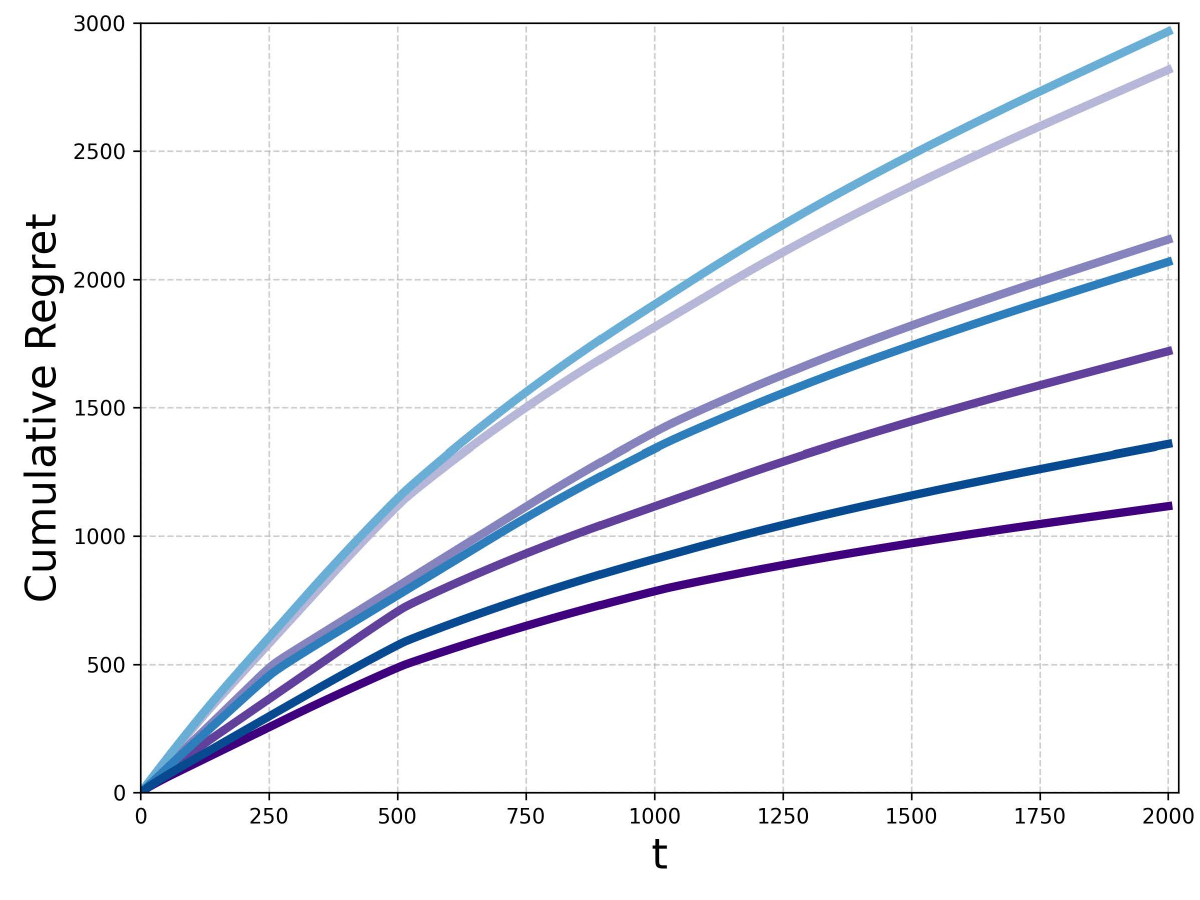}\\[-2pt]
    {\footnotesize (f) $d=50,\ s_0=15,\ K=5,\ N=100$}
    \label{fig:regret_15_50_5_100_pool}
  \end{minipage}

  \vspace{6pt}

  \begin{minipage}[t]{\linewidth}
    \centering
    \includegraphics[width=0.9\linewidth]{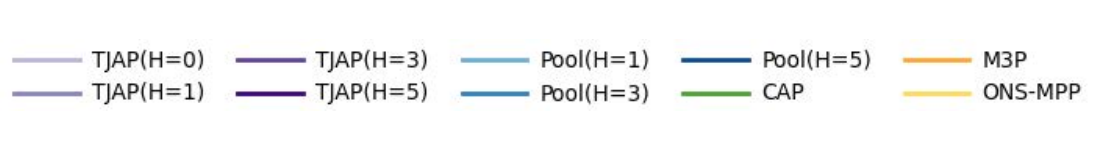}
  \end{minipage}
  \caption{Cumulative regret on synthetic instances under varying feature dimension $d$, sparsity level $s_0$, and catalog size $N$. \textbf{Top row:} TJAP with $H\in\{0,1,3,5\}$ compared against CAP, M3P, and ONS--MPP. \textbf{Bottom row:} TJAP with $H\in\{0,1,3,5\}$ compared against the pooled estimator \textsc{Pool}$(H)$ for $H\in\{1,3,5\}$. Each curve is averaged over $10$ independent runs; all methods share the same price range $[0,\overline P]$ and observe identical contexts.}
  \label{fig:regret-1}
\end{figure}

\subsection{Main findings}

Figure~\ref{fig:regret-1} reports cumulative regret for 3 configurations (with the remaining 9 configurations deferred to Figure~\ref{fig:regret-2}).
The experimental results closely mirror the structural insights developed in Section \ref{sec:theory} and provide clear evidence on when and how transfer improves joint assortment-pricing.

\subsubsection*{(i) Additional markets accelerate joint learning.}

Across all configurations in Figure \ref{fig:regret-1}, cumulative regret under TJAP decreases systematically as the number of source markets $H$ increases. The improvement is monotone in $H$, and the gap between $H=0$ and $H=5$ is substantial when preference shifts are sparse.

Operationally, this indicates that incorporating related markets materially accelerates convergence of both pricing and assortment decisions in the target market. The observed improvement is consistent with the theoretical scaling in Theorem~\ref{thm:upper-linear-o2o}, where pooled curvature effectively increases the information available for shared preference directions.

\subsubsection*{(ii) Transfer-aware learning dominates single-market learning.}
With transfer enabled, \textsc{TJAP} uniformly outperforms \textsc{CAP}, confirming that auxiliary markets can materially accelerate learning in the target market. In addition, even at $H=0$, \textsc{TJAP} improves upon \textsc{CAP} in our implementation, consistent with the benefits of episodic updates and information-matrix control. We also observe lower runtime for \textsc{TJAP} due to parameter updates being performed only at episode boundaries.

\subsubsection*{(iii) Naive pooling is operationally unsafe under heterogeneity.}
The pooled estimator POOL(H), which aggregates data across markets without debiasing, reduces variance but ignores cross-market shifts. As shown in the bottom panels of Figure \ref{fig:regret-1}, POOL(H) is uniformly dominated by TJAP with the same $H$, and the performance gap widens as $s_0$ increases.

This confirms that simply aggregating historical markets can lead to systematically distorted pricing and assortment decisions when markets differ. The aggregate-then-debias design of TJAP successfully captures the informational benefit of pooling while correcting for structural differences.

\subsubsection*{(iv) Joint modeling of assortment and pricing is essential.}
Even without transfer ($H=0$), \textsc{TJAP} outperforms \textsc{M3P} and \textsc{ONS-MPP}. This highlights the value of joint decision making: learning the preference parameters through the MNL choice structure and optimizing assortments and prices jointly yields performance gains that pricing-only or pricing-centric methods coupled with a heuristic assortment rule do not capture.

\subsubsection*{(v) Exploration overhead is limited.} Forced exploration is invoked only when target-market curvature is insufficient. Empirically, the number of forced rounds is small relative to the horizon, and their contribution to regret is negligible compared with the dominant learning terms. This supports the theoretical claim that identifiability control introduces only lower-order regret.

Overall, the experiments validate the central message of the paper: cross-market information can significantly accelerate joint assortment-pricing decisions when heterogeneity is structured, but indiscriminate pooling can degrade performance when structural differences are not accounted for.
\section{Proof of Main Results}
\label{sec:proof-sketch}
In this section, we outline the main steps in the proofs of
Theorems~\ref{thm:upper-linear-o2o} and~\ref{thm:lower-linear-o2o}.
We emphasize the proof map and intuition; all technical proofs and complete lemma statements
are deferred to the appendix.

\subsection{Important Lemmas}
\label{subsec:important-lemmas}

We restate three core lemmas that drive the upper-bound proof sketch. For readability,
we provide streamlined statements here; the appendix contains the full statements and proofs.

\begin{restatelemma}[Self-normalized concentration]{lem:SN-MNL-main}
On the event \(\mathcal E_{m-1}^{(\alpha)}\), the aggregated estimator satisfies the
self-normalized bound
\begin{equation*}
\bigl\|\hat\bnu^{(\mathrm{ag})}_{m-1} - \bnu^{(\mathrm{ag})}_{m-1}\bigr\|_{\bar W_{m-1}}
~\le~
\alpha_{m-1},
\end{equation*}
where \(\alpha_{m-1}\) scales as
\[
\alpha_{m-1}
~=~
\Theta\!\left(\sqrt{d\log\!\Bigl(1+\tfrac{\mathrm{tr}(\bar W_{m-1})}{d}\Bigr)
+\log\tfrac{1}{\eta_{m-1}^{(\alpha)}}}\right),
\]
with constants depending only on \((L_0,\overline P,C_{\min},C_{\max})\).
\end{restatelemma}

Lemma~\ref{lem:SN-MNL-main} provides the \emph{variance radius} in the geometry induced by
\(\bar W_{m-1}\). In the regret analysis, \(\alpha_{m-1}\) always appears multiplied by a design norm
\(\|\cdot\|_{\bar W_{m-1}^{-1}}\), so larger pooled information directly shrinks the effective statistical error.

\begin{restatelemma}[Target-only debiasing]{lem:debias-main}
Choose the \(\ell_1\)-penalty as \(\lambda_{m-1}=\Theta\!\bigl(\sqrt{\log(d/\eta_{m-1}^{(\beta)})/|\calT_{m-1}|}\bigr)\),
with constants depending only on \((L_0,\overline P)\).
On the event \(\mathcal E_{m-1}^{(\beta)}\), the debiasing estimator obeys
\begin{equation*}
\|\hat\bdelta_{m-1}-\bdelta_{m-1}^*\|_1
~\le~
\beta_{m-1}
~:=~
\frac{c_\beta\,s_0\,\lambda_{m-1}}{\phi_{m-1}^2},
\end{equation*}
for a constant \(c_\beta>0\) depending only on \((L_0,\overline P,C_{\min},C_{\max})\).
\end{restatelemma}

Lemma~\ref{lem:debias-main} controls the \emph{transfer bias} due to an \(s_0\)-sparse target--source shift.
This radius is driven by target data and the target restricted eigenvalue \(\phi_{m-1}\), and therefore does
not improve with more source markets.

\begin{restatelemma}[Revenue optimism]{lem:price-optimism}
Fix a finite set \(S\) and \(P=[0,\overline P]\). Let \(v_i,\tilde v_i:P\to\mathbb R\) be decreasing functions,
and define the price-optimized revenue functional \(\mathcal V_S(\cdot)\) as in the appendix.
If \(v_i(p)\le \tilde v_i(p)\) for all \(i\in S\) and all \(p\in P\), then
\[
\mathcal V_S(v)\le \mathcal V_S(\tilde v).
\]
\end{restatelemma}

Lemma~\ref{lem:price-optimism} ensures that optimism is preserved \emph{after} optimizing over prices (and assortments):
upper confidence envelopes at the utility level imply an upper bound on the optimized revenue, despite the
\(\arg\max\) depending on the utilities.

\subsection{Proof Sketch of Upper Bound (Theorem~\ref{thm:upper-linear-o2o})}

\paragraph{Episodic geometry and confidence radii.}
At the start of episode \(m\), we freeze the episodic information matrix.
Lemma~\ref{lem:SN-MNL-main} yields a self-normalized concentration event on which the aggregated estimator
lies in an ellipsoid of radius \(\alpha_{m-1}\) under the metric induced by \(\bar W_{m-1}\).
Under homogeneous covariates (Assumption~\ref{assump:Homo-Cov}), the pooled Fisher information grows
proportionally to \((1+H)\tau_{m-1}\), implying the geometric shrinkage
\(\|\cdot\|_{\bar W_{m-1}^{-1}}\asymp 1/\sqrt{(1+H)\tau_{m-1}}\) (up to constants).

\paragraph{Target-only debiasing and transfer bias.}
We first compute the aggregated center \(\hat{\bnu}^{(\mathrm{ag})}_{m-1}\) using all markets in episode \(m-1\),
and then correct it using only target data via an \(\ell_1\)-regularized debiasing step.
With \(\lambda_{m-1}=\Theta\!\bigl(\sqrt{\log(d/\eta_{m-1})/|\calT_{m-1}|}\bigr)\),
Lemma~\ref{lem:debias-main} yields an \(\ell_1\) error bound
\(\|\hat\bdelta_{m-1}-\bdelta_{m-1}^*\|_1\le \beta_{m-1}\),
where \(\beta_{m-1}=\Theta(s_0\lambda_{m-1})\) up to the target restricted eigenvalue factor \(\phi_{m-1}^{-2}\).
Consequently, the bias contribution scales linearly in \(s_0\) and does not benefit from additional sources.

\paragraph{Optimism and instantaneous regret.}
The algorithm selects \((S_t^{(0)},\bp_t^{(0)})\) by maximizing an optimistic revenue objective constructed from
upper confidence envelopes \(\tilde v_{it}\).
Lemma~\ref{lem:price-optimism} ensures that such utility-level optimism is preserved after optimizing over prices,
so the optimized revenue under the true utilities is upper bounded by the algorithm's optimistic value.
Combining this with Lipschitz continuity of MNL revenue in utilities and the variance--bias decomposition
induced by the radii \((\alpha_{m-1},\beta_{m-1})\), we obtain a per-round regret bound of the form
\[
\text{Regret}_t
\;\lesssim\;
K\overline P\Biggl[
\alpha_{m-1}\sqrt{\frac{d(1+\overline P^2)}{\lambda_{\min}(\bar W_{m-1})}}
+
\beta_{m-1}(1+\overline P)
\Biggr],
\qquad t\in\mathcal T_m,
\]
where the first term is the variance contribution and the second term is the bias contribution.

\paragraph{Summation over episodes and top-up rounds.}
Summing the variance part over \(t\in\mathcal T_m\) and using the episodic geometry above gives
\[
\sum_{t\in \mathcal{T}_m}\text{Regret}_t^{\mathrm{(var)}}
\;\lesssim\;
K\overline P\, d\,
\sqrt{\frac{|\mathcal T_m|}{1+H}}\,
\sqrt{\log\bigl((1+H)T^3\bigr)},
\]
and summing the bias part yields
\[
\sum_{t\in \mathcal{T}_m}\text{Regret}_t^{\mathrm{(bias)}}
\;\lesssim\;
K\overline P(1+\overline P)\, s_0\,
\sqrt{|\mathcal T_m|\log(4dT^2)}.
\]
Using \(\sum_m \sqrt{|\mathcal T_m|}\lesssim \sqrt{T}\) delivers the leading-order terms
\(K\overline P\,d\sqrt{T/(1+H)}\) and \(K\overline P\,s_0\sqrt{T}\) in Theorem~\ref{thm:upper-linear-o2o}.
Finally, the forced-exploration gate contributes only a lower-order ``top-up'' term because the total number
of forced rounds across episodes is \(\tilde O(d\log T)\), and each such round incurs at most \(K\overline P\) regret.
Combining these bounds and integrating over the failure probabilities yields the expected regret guarantee.

\subsection{Proof Sketch of Lower Bound (Theorem~\ref{thm:lower-linear-o2o})}

\paragraph{Hard instance and task similarity.}
We consider \(N=d\) items with deterministic covariates \(\bx_{jt}=e_j\), and set
\(\bgamma^{(h)}=L_0\mathbf{1}_d\) for all \(h\in\{0\}\cup[H]\), so Assumption~\ref{assump:price-sensitivity} holds.
Partition coordinates into a shared block \(J_{\mathrm{var}}\) and a shifted block \(J_{\mathrm{sh}}\):
\[
[d]=J_{\mathrm{var}}\cup J_{\mathrm{sh}},\qquad |J_{\mathrm{var}}|=d-s_0,\ \ |J_{\mathrm{sh}}|=s_0.
\]
Draw independent Rademacher signs \(u=(u_j)_{j\in J_{\mathrm{var}}}\) and \(w=(w_j)_{j\in J_{\mathrm{sh}}}\), and set
\[
\beta_j^{(h)} =
\begin{cases}
\Delta_{\mathrm{var}}\,u_j, & j\in J_{\mathrm{var}},\ h\in\{0\}\cup[H],\\
0, & j\in J_{\mathrm{sh}},\ h\in[H],
\end{cases}
\qquad
\beta_j^{(0)} =
\begin{cases}
\Delta_{\mathrm{var}}\,u_j, & j\in J_{\mathrm{var}},\\
\Delta_{\mathrm{shf}}\,w_j, & j\in J_{\mathrm{sh}}.
\end{cases}
\]
The target--source shift is supported on \(J_{\mathrm{sh}}\) with \(\ell_0\)-size at most \(s_0\), so
Assumption~\ref{assump:task-simi} holds. We analyze a genie-aided model in which each target action is also played
in all sources and all \(H+1\) outcomes are revealed; this can only make learning easier, so any lower bound remains valid.

\paragraph{Revenue structure.}
Lemma~\ref{lem:1item-lb} shows that the single-item price-optimized revenue \(r^*(\beta)\) is differentiable and
strictly increasing on a compact cube, with \(r^*(+\Delta)-r^*(-\Delta)\asymp \Delta\) for small \(\Delta>0\).
Appendix Lemma~\ref{lem:dilution-lb} shows that, when each item is priced at \(p_i=p^*(\beta_i)\),
the multi-item revenue is bounded below by a diluted sum of single-item revenues and that the clairvoyant top-\(K\)
assortment \(S_K^*(\beta)\) consists of the \(K\) largest coordinates of \(\beta\). Hence flipping the sign of
\(\beta_j\) changes the clairvoyant revenue by order \(\Delta\) per round whenever \(j\in S_K^*(\beta)\), and these gaps add across coordinates.

\paragraph{KL control and exposure.}
Let \(\mathbb P_{u,w}\) be the joint law of the (target+source) trajectory under \((u,w)\) and a fixed policy.
For \(j\in J_{\mathrm{var}}\) (shared block), Lemma~\ref{lem:KL-lb} yields
\[
\mathrm{KL}\!\left(\mathbb P_{u,w}\ \big\|\ \mathbb P_{u^{(j)},w}\right)
\;\lesssim\;
(1+H)\,\Delta_{\mathrm{var}}^2\,\mathbb E[N_j(T)].
\]
For \(j\in J_{\mathrm{sh}}\) (shift block), source observations are identical under both signs, so
\[
\mathrm{KL}\!\left(\mathbb P_{u,w}\ \big\|\ \mathbb P_{u,w^{(j)}}\right)
\;\lesssim\;
\Delta_{\mathrm{shf}}^2\,\mathbb E[N_j(T)],
\]
with no \((1+H)\) factor. Randomizing item labels and using exchangeability,
Lemma~\ref{lem:exposure-lb} gives \(\mathbb E[N_j(T)]=KT/d\) for all \(j\in[d]\).
Pinsker's inequality then converts these KL bounds into total-variation bounds for the corresponding binary tests.

\paragraph{From testing to regret.}
For \(j\in J_{\mathrm{var}}\), Le Cam's lemma lower bounds the Bayes error in terms of the total variation bound above.
Whenever \(j\) belongs to the clairvoyant top-\(K\) set, misclassifying \(u_j\) reverses its contribution and causes a
per-round loss of order \(\Delta_{\mathrm{var}}\).
Aggregating over the \(d-s_0\) shared coordinates via an Assouad-type argument and calibrating
\(\Delta_{\mathrm{var}}^2\asymp \frac{d}{(1+H)\,KT}\) yields a cumulative contribution
\(\Omega\!\big(\sqrt{\tfrac{K(d-s_0)T}{1+H}}\big)\).
For \(j\in J_{\mathrm{sh}}\), source observations carry no information, so the same testing argument has no \((1+H)\)
benefit. Choosing \(\Delta_{\mathrm{shf}}^2\asymp \frac{d}{KT}\) (and taking \(\Delta_{\mathrm{shf}}\) large enough so that
shifted coordinates with \(w_j=+1\) enter the clairvoyant top-\(K\) set when \(s_0\le K\)) yields a cumulative
contribution \(\Omega(s_0\sqrt{KT})\). Adding the two contributions and applying Yao's minimax principle gives
Theorem~\ref{thm:lower-linear-o2o}.
\section{Conclusion and Future Directions}
\label{sec:conc}

This paper develops a transfer learning framework for contextual joint assortment-pricing under an MNL choice model with bandit feedback. We leverage data from multiple source markets to accelerate learning in a designated target market while explicitly accounting for cross-market preference shifts. Our approach builds on an aggregate-then-debias estimation pipeline: it first pools information across source markets to estimate the shared contextual preference structure, and then applies a target-driven debiasing step to adapt to sparse market-specific deviations. Combined with an optimistic learning rule and information-matrix control to ensure identifiability under adaptively chosen assortments and prices, this design enables effective transfer while guarding against negative transfer when markets differ.

On the theoretical side, we establish matching minimax upper and lower regret bounds that quantify both the benefit and the limitation of transfer. The results show that additional source markets improve performance precisely along shared preference directions, yielding substantial gains when heterogeneity between source and target markets is sufficiently sparse, while target-specific directions incur an unavoidable cost. On the empirical side, our numerical experiments corroborate these insights: incorporating source-market data consistently reduces regret relative to baselines that learn solely from the target market. Together, these findings demonstrate that cross-market information can materially reduce exploration in joint assortment-pricing, and they provide a principled variance-bias perspective on when transfer is beneficial.

Several directions remain open. 
First, we assume that a set of informative source markets is available, whereas in practice the learner may need to discover which markets are reliably informative and to downweight or discard harmful sources online. 
Second, our analysis focuses on sparse preference shifts; extending the framework to richer and more flexible notions of relatedness, such as $\ell_q$-type approximate sparsity, low-rank or latent-factor structure across markets, and distributional notions of similarity that couple contexts with preferences, would broaden applicability and sharpen guidance for heterogeneous multi-market deployments.
Progress on these questions would further clarify how transfer learning can be deployed safely and effectively in operational decision-making systems.

\spacingset{1.18}
\bibliographystyle{agsm}
\bibliography{main}

\newpage
\begin{appendices}

\begin{center}
{\large\bf SUPPLEMENTARY MATERIAL of \\
``\TITLE''}
\end{center}


\begin{center}
{\large Supplemental Material of ``Transfer Learning for Contextual Joint Assortment-Pricing: Multi-Source Utility Shift under Multinomial Logit Model''}
\end{center}

\bigskip

\section{Notations}
We use the standard Landau notation, where \( f(t) = \mathcal{O}(g(t)) \) (equivalently, \( g(t) = \Omega(f(t))) \) if \( f(t) \leq Cg(t) \) holds for “large” \( t \) (not necessarily asymptotic) and for some positive constant \( C \) independent of \( t \). Similarly, for \( t \leq T \), we write \( f(t) = \tilde{\mathcal{O}}(g(t)) \) if \( f(t) \leq Cg(t)(\log T)^c \) holds for “large” \( t \), where \( C > 0 \) and \( c \in \mathbb{R} \) are constants independent of \( t \) and \( T \). 

Let lowercase letter $x$, boldface letter $\bx$, boldface capital letter $\bX$, and blackboard-bold letter $\XX$ represent scalar, vector, matrix, and tensor, respectively. The calligraphy letter $\calX$ represents operator.
We use the notation $[N]$ to refer to the positive integer set $\braces{1, \ldots, N}$ for $N \in \ZZ_+$.
Let $C, c, C_0, c_0, \ldots$ denote generic constants, where the uppercase and lowercase letters represent large and small constants, respectively. The actual values of these generic constants may vary from time to time.
For any matrix $\bX$,  we use $\bx_{i \cdot}$, $\bx_{j}$, and $x_{ij}$ to refer to its $i$-th row, $j$-th column, and $ij$-th entry, respectively. 
All vectors are column vectors and row vectors are written as $\bx^\top$ for any vector $\bx$. 


For any vector $\bx=(x_1,\ldots,x_p)^\top$, let $\|\bx\|:= \|\bx\|_2=(\sum_{i=1}^p x_i^2)^{1/2}$ be the $\ell_2$-norm, and let $\|\bx\|_1=\sum_{i=1}^p |x_i|$ be the $\ell_1$-norm. 
When $\bX$ is a square matrix, we denote by $\Tr \paren{\bX}$, $\lambda_{max} \paren{\bX}$, and $\lambda_{min} \paren{\bX}$ the trace, maximum and minimum singular value of $\bX$, respectively.
For two matrices of the same dimension, define the inner product $\angles{\bX_1,\bX_2} = \Tr(\bX_1^\top \bX_2)$.

\section{Proof of Theorem~\ref{thm:stats_error_bound}}
\label{app:proof-thm-stats}

Using \(\bnu^{(0)} = \bnu^{\mathrm{ag}}_{m-1} + \bdelta_{m-1}^*\), we have the exact decomposition
\begin{equation}
\label{eq:thm1:error-decomp}
\hat\bnu_{m-1} - \bnu^{(0)}
=
\bigl(\hat\bnu^{\mathrm{ag}}_{m-1} - \bnu^{\mathrm{ag}}_{m-1}\bigr)
+
\bigl(\hat\bdelta_{m-1} - \bdelta_{m-1}^*\bigr).
\end{equation}
The proof proceeds by bounding:
(i) the \emph{variance term} \(\hat\bnu^{\mathrm{ag}}_{m-1} - \bnu^{\mathrm{ag}}_{m-1}\) in the pooled
geometry and converting it to an \(\ell_2\) bound; and
(ii) the \emph{bias-correction term} \(\hat\bdelta_{m-1} - \bdelta_{m-1}^*\) in \(\ell_1\) (hence in \(\ell_2\)).
We then combine them via \eqref{eq:thm1:error-decomp}.

\subsection{Tuning parameters and global good event}
\label{app:thm1:tuning}

Fix a global failure probability \(\eta\in(0,1)\). We define four families of high-probability events:

\begin{itemize}
\item \(\mathcal E^{(\alpha)}\): variance event controlling the self-normalized radius \(\alpha_{m-1}\);
\item \(\mathcal E^{(\beta)}\): debiasing event controlling the transfer-bias radius \(\beta_{m-1}\);
\item \(\mathcal E^{(\mathrm{gate})}\): forced-exploration gate event ensuring adequate target curvature;
\item \(\mathcal E^{(W)}\): pooled Fisher-growth event ensuring \(\lambda_{\min}(W_{m-1})\) grows linearly.
\end{itemize}

\paragraph{Episode-wise budgets (for \(\alpha,\beta,\mathrm{gate}\)).}
To avoid ambiguity, we allocate separate per-episode budgets using the summable schedule:
\[
\bar\eta_m := \frac{6(\eta/2)}{\pi^2 m^2}=\frac{3\eta}{\pi^2 m^2},\qquad
\eta_m^{(\alpha)}=\eta_m^{(\beta)}=\eta_m^{(\mathrm{gate})}
:=\frac{\bar\eta_m}{3}=\frac{\eta}{\pi^2 m^2}.
\]
Then \(\sum_{m\ge 1}(\eta_m^{(\alpha)}+\eta_m^{(\beta)}+\eta_m^{(\mathrm{gate})})\le \eta/2\).

\paragraph{Single budget (for pooled growth).}
Set \(\eta^{(W)}:=\eta/2\) and construct \(\mathcal E^{(W)}\) with
\(\mathbb P(\mathcal E^{(W)})\ge 1-\eta^{(W)}\).

\paragraph{Global good event.}
Let \(M:=\lceil \log_2 T\rceil\) be the number of episodes. Define
\begin{equation}
\label{eq:global-good-event}
\mathcal E
:=
\mathcal E^{(W)}
\cap
\bigcap_{m=1}^M
\Bigl(
\mathcal E_m^{(\alpha)}\cap \mathcal E_m^{(\beta)}\cap \mathcal E_m^{(\mathrm{gate})}
\Bigr),
\end{equation}
where \(\mathcal E_m^{(\alpha)}\) is the event asserted by Lemma~\ref{lem:SN-MNL-main},
\(\mathcal E_m^{(\beta)}\) is the event asserted by Lemma~\ref{lem:debias-main}, and
\(\mathcal E_m^{(\mathrm{gate})}\) is the event asserted by Lemma~\ref{lem:topup-chernoff}.
By a union bound and the budget construction, \(\mathbb P(\mathcal E)\ge 1-\eta\).

In the rest of the proof we work conditionally on \(\mathcal E\).

\subsection{Variance control via self-normalized concentration}
\label{app:thm1:variance}

Recall the episodic pooled information matrix
\begin{equation}
\label{eq:W-episodic}
W_{m-1}
~:=~
V^{(0)}_{\tau_{m-1}}
+ \sum_{h=1}^H V^{(h)}_{\tau_{m-1}}.
\end{equation}
For analysis introduce the ridge-regularized matrix \(\bar W_{m-1}:=W_{m-1}+\lambda_0 I_{2d}\),
with fixed \(\lambda_0>0\).

\begin{lemma}[Variance radius \(\alpha_{m-1}\)]
\label{lem:SN-MNL-main}
Consider episode \(m\). There exists a constant \(c_\alpha>0\), depending only on
\((L_0,\overline P,C_{\min},C_{\max})\), such that on \(\mathcal E_{m-1}^{(\alpha)}\),
\begin{equation}
\label{eq:alpha-event}
\|\hat\bnu^{\mathrm{ag}}_{m-1} - \bnu^{\mathrm{ag}}_{m-1}\|_{\bar W_{m-1}}
~\le~
\alpha_{m-1}
~:=~
c_\alpha \sqrt{\,
d\log\!\Bigl(1+\frac{\mathrm{tr}(\bar W_{m-1})}{d\lambda_0}\Bigr)
+ \log\!\frac{2}{\eta_{m-1}^{(\alpha)}}\, }.
\end{equation}
\end{lemma}

A detailed proof is given in Appendix~\ref{app:proof-SN-MNL}.

\subsection{Converting pooled-geometry control to an \texorpdfstring{$\ell_2$}{l2} bound}
\label{app:thm1:pooled-growth}

Lemma~\ref{lem:SN-MNL-main} controls the aggregation error in the \(\bar W_{m-1}\)-geometry.
To obtain an \(\ell_2\) bound, we need a lower bound on \(\lambda_{\min}(\bar W_{m-1})\).
This is supplied by the pooled Fisher-growth event \(\mathcal E^{(W)}\).

\begin{lemma}[Pooled Fisher growth]\label{lem:pooled-fisher-growth}
There exist constants \(c_W>0\) and \(m_0\in\mathbb N\), depending only on
\((L_0,\overline P,C_{\min},C_{\max},\kappa)\), such that for any \(\eta^{(W)}\in(0,1)\)
we can construct an event \(\mathcal E^{(W)}\) with
\(\mathbb P(\mathcal E^{(W)})\ge 1-\eta^{(W)}\) on which
\[
\lambda_{\min}\!\bigl(W_{m-1}\bigr)
\;\ge\;
c_W (1+H)\,C_{\min}\,\tau_{m-1}
\qquad\text{for all episodes } m\ge m_0.
\]
Consequently, for all such \(m\),
\[
\lambda_{\min}\!\bigl(\bar W_{m-1}\bigr)
=
\lambda_{\min}\!\bigl(W_{m-1}+\lambda_0 I_{2d}\bigr)
\;\ge\;
\lambda_0 + c_W(1+H)\,C_{\min}\,\tau_{m-1}
\;\asymp\;
(1+H)\,C_{\min}\,\tau_{m-1}.
\]
\end{lemma}

The proof is given in Appendix~\ref{app:proof-pooled-fisher-growth}.

\paragraph{Immediate \(\ell_2\) consequence.}
On \(\mathcal E\), for all \(m\ge m_0\), combine Lemma~\ref{lem:SN-MNL-main} with the eigenvalue lower bound:
\begin{align}
\label{eq:l2-agg-bound}
\|\hat\bnu^{\mathrm{ag}}_{m-1}-\bnu^{\mathrm{ag}}_{m-1}\|_2
&\le
\frac{\|\hat\bnu^{\mathrm{ag}}_{m-1}-\bnu^{\mathrm{ag}}_{m-1}\|_{\bar W_{m-1}}}{\sqrt{\lambda_{\min}(\bar W_{m-1})}}
\le
\frac{\alpha_{m-1}}{\sqrt{\lambda_{\min}(\bar W_{m-1})}}
\;\lesssim\;
\alpha_{m-1}\sqrt{\frac{1}{(1+H)\tau_{m-1}}}.
\end{align}
For the finitely many initial episodes \(m<m_0\), \(\tau_{m-1}\) is \(O(1)\), so the same display holds after enlarging constants.

\subsection{Debiasing and transfer-bias radius \texorpdfstring{$\beta_{m-1}$}{beta}}
\label{app:thm1:debias}

We next control the error of the \(\ell_1\)-penalized target correction.
Let \(\bdelta_{m-1}^*\) be defined by \eqref{eq:def-delta-star-pop}, and let
\[
S^* := \mathrm{supp}(\bdelta_{m-1}^*),
\qquad |S^*|\le s_0
\quad\text{(Assumption~\ref{assump:task-simi}).}
\]

TJAP estimates \(\bdelta_{m-1}^*\) from target data by solving
\begin{equation}
\label{eq:l1-debias}
\hat\bdelta_{m-1}
\in
\arg\min_{\bdelta\in\mathbb R^{2d}}
\Biggl\{
\frac{1}{|\calT_{m-1}|}\sum_{t\in\calT_{m-1}}
\ell_t^{(0)}\bigl(\hat\bnu^{\mathrm{ag}}_{m-1}+\bdelta\bigr)
+ \lambda_{m-1}\|\bdelta\|_1
\Biggr\}.
\end{equation}

Let \(\calL_{m-1}(\bdelta) := |\calT_{m-1}|^{-1}\sum_{t\in \calT_{m-1}}
\ell_t^{(0)}(\hat\bnu^{\mathrm{ag}}_{m-1}+\bdelta)\) and denote its Hessian by
\(H(\bdelta):=\nabla^2\calL_{m-1}(\bdelta)\).

\begin{definition}[Target-only restricted eigenvalue]
\label{def:RE-target}
In episode \(m\), the target restricted eigenvalue (RE) constant on the \(2s_0\)-sparse cone is
\begin{equation}
\label{eq:RE-target}
\phi_{m-1}^2
~:=~
\min_{\|u\|_0\le 2s_0}
\frac{u^\top H(\bdelta_{m-1}^*)u}{\|u\|_2^2}.
\end{equation}
\end{definition}

\begin{lemma}[Uniform target restricted eigenvalue]\label{lem:target-re}
Suppose the forced-exploration gate in Lemma~\ref{lem:topup-chernoff} is used with a schedule
\(\{\Lambda_m\}_{m\ge 1}\) such that, on the global good event \(\mathcal E\),
\[
\lambda_{\min}\bigl(V^{(0)}_{\tau_m-1}\bigr) \;\ge\; \Lambda_{m-1}
\quad \text{for all episodes } m.
\]
Then there exists a constant \(\phi_*>0\), depending only on
\((L_0,\overline P,C_{\min},C_{\max},\kappa,r)\), such that on \(\mathcal E\),
\[
\phi_{m-1}^2 \;\ge\; \phi_* \qquad \text{for all episodes } m.
\]
\end{lemma}

We defer the proof of Lemma~\ref{lem:target-re} to Appendix~\ref{app:proof-target-re}.

\begin{lemma}[Transfer-bias radius \(\beta_{m-1}\)]
\label{lem:debias-main}
For the failure budget \(\eta_{m-1}^{(\beta)}\in(0,1)\), choose
\begin{equation}
\label{eq:lambda-choice-new}
\lambda_{m-1}
=
c_\lambda
\sqrt{
\frac{\log\bigl(2d/\eta_{m-1}^{(\beta)}\bigr)}{|\calT_{m-1}|}
},
\end{equation}
for a constant \(c_\lambda>0\) depending only on \((L_0,\overline P)\).
There exists \(c_\beta>0\), depending only on \((L_0,\overline P,C_{\min},C_{\max})\), such that
on the event \(\mathcal E_{m-1}^{(\beta)}\),
\begin{equation}
\label{eq:l1-lb-main}
\|\hat\bdelta_{m-1}-\bdelta_{m-1}^*\|_1
~\le~
\frac{c_\beta\,s_0\,\lambda_{m-1}}{\phi_{m-1}^2}.
\end{equation}
Consequently, define the debiasing radius
\[
\beta_{m-1} := \frac{c_\beta\,s_0\,\lambda_{m-1}}{\phi_{m-1}^2}.
\]
\end{lemma}

A detailed proof is given in Appendix~\ref{app:proof-debias}.

\subsection{Total error bound and completion of Theorem~\ref{thm:stats_error_bound}}
\label{app:thm1:total-error}

Work on the global good event \(\mathcal E\).
Fix any episode \(m\). Using the decomposition \eqref{eq:thm1:error-decomp} and the triangle inequality,
\begin{equation}
\label{eq:thm1:triangle}
\|\hat\bnu_{m-1}-\bnu^{(0)}\|_2
\le
\|\hat\bnu^{\mathrm{ag}}_{m-1}-\bnu^{\mathrm{ag}}_{m-1}\|_2
+
\|\hat\bdelta_{m-1}-\bdelta_{m-1}^*\|_2.
\end{equation}

\paragraph{Variance term.}
By \eqref{eq:l2-agg-bound} and the definition of \(\alpha_{m-1}\), and using that
\(\mathrm{tr}(\bar W_{m-1})\lesssim d + (1+H)\tau_{m-1}\) on \(\mathcal E\),
we obtain, for a constant \(C_v>0\),
\begin{equation}
\label{eq:thm1:variance-final}
\|\hat\bnu^{\mathrm{ag}}_{m-1}-\bnu^{\mathrm{ag}}_{m-1}\|_2
\le
C_v
\sqrt{\frac{d\log\bigl((1+H)T/\eta\bigr)}{(1+H)\tau_{m-1}}}.
\end{equation}

\paragraph{Bias term.}
By Lemma~\ref{lem:debias-main} and \(\|x\|_2\le \|x\|_1\),
\[
\|\hat\bdelta_{m-1}-\bdelta_{m-1}^*\|_2
\le
\|\hat\bdelta_{m-1}-\bdelta_{m-1}^*\|_1
\le
\frac{c_\beta s_0}{\phi_{m-1}^2}\lambda_{m-1}.
\]
Using Lemma~\ref{lem:target-re} to substitute \(\phi_{m-1}^2\ge \phi_*\) and
\(|\calT_{m-1}|\asymp \tau_{m-1}\) under the doubling schedule yields, for a constant \(C_b>0\),
\begin{equation}
\label{eq:thm1:bias-final}
\|\hat\bdelta_{m-1}-\bdelta_{m-1}^*\|_2
\le
C_b\,s_0\sqrt{\frac{\log(dT/\eta)}{\tau_{m-1}}}.
\end{equation}

\paragraph{Combine.}
Substituting \eqref{eq:thm1:variance-final} and \eqref{eq:thm1:bias-final} into
\eqref{eq:thm1:triangle} proves that on \(\mathcal E\), for every episode \(m\),
\[
\|\hat\bnu_{m-1}-\bnu^{(0)}\|_2
\le
C_v
\sqrt{\frac{d\log\bigl((1+H)T/\eta\bigr)}{(1+H)\tau_{m-1}}}
+
C_b\,s_0
\sqrt{\frac{\log(dT/\eta)}{\tau_{m-1}}}.
\]
Since \(\mathbb P(\mathcal E)\ge 1-\eta\), this completes the proof of
Theorem~\ref{thm:stats_error_bound}.

\hfill\(\square\)

\subsection{Technical lemmas for Theorem~\ref{thm:stats_error_bound}}
\subsubsection{Proof of Lemma~\ref{lem:SN-MNL-main}}
\label{app:proof-SN-MNL}

We write the negative log-likelihood of all observations used to form the aggregation center
\(\hat\bnu^{\mathrm{ag}}_{m-1}\) (and the corresponding pooled geometry \(W_{m-1}\)) as
\[
\ell_{m-1}(\bnu):=-\sum_{t\in\mathcal I_{m-1}}\log\Pr_{\bnu}(Y_t\mid X_t),
\]
where \(\mathcal I_{m-1}\) is the union of all rounds contributing to \(W_{m-1}\)
under the episode-freezing rule. Denote the gradient by
\[
g_{m-1}(\bnu) := \nabla\ell_{m-1}(\bnu) = \sum_{t\in\mathcal I_{m-1}} s_t(\bnu),
\]
where \(s_t(\bnu)\) is the single-round score. For an assortment \((S_t,\bp_t)\) and parameter \(\bnu\),
\[
s_t(\bnu)
=
\sum_{i\in S_t} \tilde{\bx}_{it}
\bigl(\mathbf 1\{Y_t=i\} - q_{it}(\bnu)\bigr),
\]
with \(q_{it}(\bnu)\) the MNL choice probability and \(\tilde{\bx}_{it}=(\bx_{it},-p_{it}\bx_{it})\).
The per-round Fisher increment is
\[
I_t(\bnu)
=
\tilde{\bX}_{t}^\top
(\mathrm{Diag}(q_t(\bnu)) - q_t(\bnu)q_t(\bnu)^\top)\tilde{\bX}_{t}.
\]

The proof proceeds in two steps:
(i) a self-normalized concentration for the score at \(\bnu^{\mathrm{ag}}_{m-1}\) in the geometry \(\bar W_{m-1}\);
(ii) conversion to a parameter error bound using local strong convexity.

\medskip
\paragraph{Self-normalized score bound.}

Fix any \(u\in\mathbb R^{2d}\). Define the \emph{centered} single-round score at the
population pooled center \(\bnu^{\mathrm{ag}}_{m-1}\) by
\[
\tilde s_t
:=
s_t\bigl(\bnu^{\mathrm{ag}}_{m-1}\bigr)
-
\mathbb E\!\left[s_t\bigl(\bnu^{\mathrm{ag}}_{m-1}\bigr)\,\middle|\,\mathcal F_{t-1}\right].
\]
Define the scalar martingale
\[
M_t(u)
:=
\sum_{s\in\mathcal I_{m-1}: s\le t} \langle u,\tilde s_s\rangle,
\qquad
V_t(u)
:=
\sum_{s\in\mathcal I_{m-1}: s\le t}
\mathrm{Var}\bigl(\langle u,\tilde s_s\rangle\mid\mathcal F_{s-1}\bigr).
\]
Then \(M_t(u)\) is a martingale with bounded increments.

Using
\(\|\bx_{it}\|_\infty\le 1\) and \(p_{it}\in[0,\overline P]\), we have
\(\|\tilde{\bx}_{it}\|_2^2 \le d(1+\overline P^2)\) and \(|S_t|\le K\), hence
\[
\bigl|\langle u,\tilde s_t\rangle\bigr|
\le 2\sum_{i\in S_t} |\langle u,\tilde{\bx}_{it}\rangle|
\le 2\sqrt{K\,d(1+\overline P^2)}\,\|u\|_2.
\]
Therefore, Freedman's inequality for martingales with bounded increments
implies that for any \(\delta\in(0,1)\),
\begin{equation}
\label{eq:freedman-scalar-sn}
\mathbb P\!\left\{
M_{|\mathcal I_{m-1}|}(u)
\ge
\sqrt{2V_{|\mathcal I_{m-1}|}(u)\log\frac{1}{\delta}}
+
\frac{2}{3}\sqrt{K\,d(1+\overline P^2)}\,\|u\|_2\,\log\frac{1}{\delta}
\right\}
\le \delta.
\end{equation}

Moreover, since \(\mathrm{Var}(Z\mid\mathcal F_{t-1})\le \mathbb E[Z^2\mid\mathcal F_{t-1}]\), we have
\[
V_{|\mathcal I_{m-1}|}(u)
\le
\sum_{t\in\mathcal I_{m-1}}
\mathbb E\!\left[\langle u,\tilde s_t\rangle^2 \,\middle|\, \mathcal F_{t-1}\right]
\le
u^\top\!\left(\sum_{t\in\mathcal I_{m-1}} I_t\bigl(\bnu^{\mathrm{ag}}_{m-1}\bigr)\right)u,
\]
where the last inequality uses the standard quadratic-form upper bound induced by the
model-based Fisher increment \(I_t(\cdot)\).
By Assumption~\ref{assump:fisher-invertible} and boundedness of covariates, the right-hand side
is dominated by \(u^\top \bar W_{m-1}u\) up to constants; in particular,
\(V_{|\mathcal I_{m-1}|}(u)\le u^\top\bar W_{m-1}u\).

Let \(\mathcal E := \{u\in\mathbb R^{2d}: u^\top\bar W_{m-1}u\le 1\}\), and
let \(\mathcal N\) be a \(1/2\)-net of \(\mathcal E\) under the norm
\(\|\cdot\|_{\bar W_{m-1}}\). Standard volumetric estimates give
\[
|\mathcal N|
\le
\Bigl(1+\frac{\mathrm{tr}(\bar W_{m-1})}{d\lambda_0}\Bigr)^d.
\]
Applying \eqref{eq:freedman-scalar-sn} to each \(u\in\mathcal N\) with
\(\delta := \eta_{m-1}^{(\alpha)}/(2|\mathcal N|)\), and using
\(V_{|\mathcal I_{m-1}|}(u)\le 1\) and \(\|u\|_2\le \lambda_0^{-1/2}\) for
\(u\in\mathcal E\), we obtain that with probability at least
\(1-\eta_{m-1}^{(\alpha)}\),
\[
\max_{u\in\mathcal N}
\frac{|M_{|\mathcal I_{m-1}|}(u)|}{\sqrt{u^\top\bar W_{m-1}u}}
\le
C'
\sqrt{
d\log\!\Bigl(1+\frac{\mathrm{tr}(\bar W_{m-1})}{d\lambda_0}\Bigr)
+\log\!\frac{2}{\eta_{m-1}^{(\alpha)}}
}
\]
for some constant \(C'>0\). A standard net-to-uniform argument (lifting from
\(\mathcal N\) to \(\mathcal E\)) then yields
\[
\sup_{u\ne 0}\frac{|\langle u,\sum_{t\in\mathcal I_{m-1}}\tilde s_t\rangle|}
{\sqrt{u^\top\bar W_{m-1}u}}
\le
2\max_{u\in\mathcal N}
\frac{|M_{|\mathcal I_{m-1}|}(u)|}{\sqrt{u^\top\bar W_{m-1}u}}
\]
with the same probability.

\medskip
\noindent\emph{Identification of the centered score sum.}
Under Option~I, the ground-truth aggregate center \(\bnu^{\mathrm{ag}}_{m-1}\) is defined as a
population minimizer of the pooled objective corresponding to \(\ell_{m-1}(\cdot)\).
To make this definition compatible with adaptive designs, we define the episode-\(m\!-\!1\)
\emph{conditional pooled risk}
\[
\bar \ell_{m-1}(\bnu)
:=
\mathbb E\!\left[\ell_{m-1}(\bnu)\,\middle|\,\mathcal F^{\mathrm{fr}}_{m-1}\right]
=
\sum_{t\in\mathcal I_{m-1}}
\mathbb E\!\left[-\log \Pr_{\bnu}(Y_t\mid X_t)\,\middle|\,\mathcal F_{t-1}\right],
\]
where \(\mathcal F^{\mathrm{fr}}_{m-1}\) is the $\sigma$-field generated by the episode-freezing rule,
i.e., the predictable design \((S_t,\bp_t,\tilde\bX_t)_{t\in\mathcal I_{m-1}}\).
Let
\[
\bnu^{\mathrm{ag}}_{m-1}\in\arg\min_{\bnu\in\mathbb R^{2d}} \bar \ell_{m-1}(\bnu).
\]
Then the first-order optimality condition yields
\[
\sum_{t\in\mathcal I_{m-1}}
\mathbb E\!\left[s_t\bigl(\bnu^{\mathrm{ag}}_{m-1}\bigr)\,\middle|\,\mathcal F_{t-1}\right]
=
\nabla \bar \ell_{m-1}\!\left(\bnu^{\mathrm{ag}}_{m-1}\right)
=0,
\]
and hence
\[
g_{m-1}\!\left(\bnu^{\mathrm{ag}}_{m-1}\right)
=
\sum_{t\in\mathcal I_{m-1}} s_t\!\left(\bnu^{\mathrm{ag}}_{m-1}\right)
=
\sum_{t\in\mathcal I_{m-1}} \tilde s_t .
\]
Therefore,
\[
\bigl\|g_{m-1}(\bnu^{\mathrm{ag}}_{m-1})\bigr\|_{\bar W_{m-1}^{-1}}
=
\sup_{u\ne 0}\frac{|\langle u,\sum_{t\in\mathcal I_{m-1}}\tilde s_t\rangle|}
{\sqrt{u^\top\bar W_{m-1}u}}
\le
2\max_{u\in\mathcal N}
\frac{|M_{|\mathcal I_{m-1}|}(u)|}{\sqrt{u^\top\bar W_{m-1}u}},
\]
and absorbing constants into \(C_1:=2C'\) gives
\[
\bigl\|g_{m-1}(\bnu^{\mathrm{ag}}_{m-1})\bigr\|_{\bar W_{m-1}^{-1}}
\le
C_1
\sqrt{
d\log\!\Bigl(1+\frac{\mathrm{tr}(\bar W_{m-1})}{d\lambda_0}\Bigr)
+\log\!\frac{2}{\eta_{m-1}^{(\alpha)}}
},
\]
as claimed.

\medskip
\paragraph{From score to parameter.}
By optimality of the MLE, \(g_{m-1}(\hat\bnu^{\mathrm{ag}}_{m-1})=0\), and by convexity,
\(\ell_{m-1}(\hat\bnu^{\mathrm{ag}}_{m-1})-\ell_{m-1}(\bnu^{\mathrm{ag}}_{m-1})\ge 0\).
A second-order Taylor expansion around \(\bnu^{\mathrm{ag}}_{m-1}\), together with local strong convexity
and the self-normalized bound, yields \eqref{eq:alpha-event}. \hfill\(\square\)

\subsubsection{Proof of Lemma~\ref{lem:target-re}}
\label{app:proof-target-re}
Recall that the normalized target loss around $\bnu^{(ag)}_{m-1}$ is
\[
L_{m-1}(\bdelta) \;=\; \frac{1}{|\calT_{m-1}|} \sum_{t\in \mathcal{T}^{(0)}_{m-1}} \ell_t(\bdelta),
\]
with Hessian
\[
H(\bdelta)
\;=\; \nabla^2 L_{m-1}(\bdelta)
\;=\; \frac{1}{|\calT_{m-1}|} \sum_{t\in \mathcal{T}^{(0)}_{m-1}}
\tilde{\bX}_{t}^\top \Bigl(\mathrm{Diag}(q_t(\bdelta)) - q_t(\bdelta) q_t(\bdelta)^\top\Bigr) \tilde{\bX}_{t}.
\]

By Assumption~\ref{assump:Homo-Cov}, all choice probabilities are uniformly bounded away from zero on a ball of radius $r$ around the true target parameter $\bnu^{(0)}$, for all feasible assortments and prices. Since the aggregate-and-debias estimator stays in this ball on the global good event $\mathcal{E}$ (cf. Lemmas~\ref{lem:SN-MNL-main} and~\ref{lem:debias-main}), there exist constants $0<c_{\mathrm{curv}}\le C_{\mathrm{curv}}<\infty$, depending only on $(L_0,\overline P,\kappa,r)$, such that for every episode $m$ and every $\bdelta$ on the line segment joining $0$ and $\bdelta_{m-1}^*$,
\begin{equation}\label{eq:H-vs-Fisher}
c_{\mathrm{curv}} \,\frac{V^{(0)}_{\tau_m-1}}{|\calT_{m-1}|}
\;\preceq\;
H(\bdelta)
\;\preceq\;
C_{\mathrm{curv}} \,\frac{V^{(0)}_{\tau_m-1}}{|\calT_{m-1}|}.
\end{equation}
The lower bound in~\eqref{eq:H-vs-Fisher} is standard in high-dimensional GLM analysis: it follows from the bounded covariates in Assumption~\ref{assump:Homo-Cov}, the non-degeneracy of the MNL link in Assumption~\ref{assump:fisher-invertible}, and the fact that the Fisher information is the conditional expectation of the Hessian~\citep{tian2023transfer}.

On the other hand, Assumption~\ref{assump:task-simi} states that the augmented covariates $\tilde{\bx}_{it}^{(0)}$ have covariance matrix
$\tilde{\Sigma} := \mathbb{E}[\tilde{\bx}\tilde{\bx}^\top]$ with $\lambda_{\min}(\tilde{\Sigma})\ge C_{\min}>0$. Combining this with~\eqref{eq:H-vs-Fisher} and the gate condition in Lemma~\ref{lem:topup-chernoff}, we obtain that on $\mathcal{E}$
\[
u^\top H(\bdelta_{m-1}^*) u
\;\ge\; c_{\mathrm{curv}} \,\frac{\Lambda_{m-1}}{|\calT_{m-1}|} \,\|u\|_2^2
\;\ge\; c_{\mathrm{curv}} C_{\min} \,\|u\|_2^2,
\qquad \forall u\in\mathbb{R}^{2d}
\]
for all $m$ large enough.
Restricting $u$ to the $2s_0$-sparse cone in Definition~\ref{def:RE-target}, this implies
\[
\phi_{m-1}^2
\;=\;
\min_{\|u\|_0\le 2s_0} \frac{u^\top H(\bdelta_{m-1}^*) u}{\|u\|_2^2}
\;\ge\;
\phi_*
\;:=\; c_{\mathrm{curv}} C_{\min} \;>\; 0
\]
for all episodes $m$ on the global good event~$\mathcal{E}$. The constant $\phi_*$ depends only on
$(L_0,\overline P,C_{\min},C_{\max},\kappa,r)$ and is independent of $H$ and $m$, which proves Lemma~\ref{lem:target-re}.
\hfill\(\square\)

\subsubsection{Proof of Lemma~\ref{lem:debias-main}}
\label{app:proof-debias}

The proof follows the standard Lasso template:
(i) control of \(\|g(\bdelta_{m-1}^*)\|_\infty\);
(ii) a cone condition from the basic inequality using the sparsity of \(\bdelta_{m-1}^*\);
(iii) restricted curvature via the RE constant \(\phi_{m-1}^2\).

\paragraph{Gradient \(\ell_\infty\)-bound at \(\bdelta_{m-1}^*\).}
Define \(g(\bdelta):=\nabla\calL_{m-1}(\bdelta)\). For any coordinate \(j\in[2d]\),
\[
g_j(\bdelta)
=
\frac{1}{|\calT_{m-1}|}\sum_{t\in\calT_{m-1}} Z_{t,j}(\bdelta),
\]
where \(Z_{t,j}(\bdelta)\) is the \(j\)-th coordinate of the target score at parameter
\(\hat\bnu^{\mathrm{ag}}_{m-1}+\bdelta\).
Using bounded covariates and prices, each \(Z_{t,j}(\bdelta)\) is bounded by a constant \(B\) depending only on \((\overline P)\).

At \(\bdelta=\bdelta_{m-1}^*\), the parameter equals
\[
\hat\bnu^{\mathrm{ag}}_{m-1}+\bdelta_{m-1}^*
=
\bnu^{(0)} + \bigl(\hat\bnu^{\mathrm{ag}}_{m-1}-\bnu^{\mathrm{ag}}_{m-1}\bigr).
\]
On \(\mathcal E_{m-1}^{(\alpha)}\), the aggregation error is controlled in a neighborhood where the score is Lipschitz.
Thus, writing \(e_{m-1}:=\hat\bnu^{\mathrm{ag}}_{m-1}-\bnu^{\mathrm{ag}}_{m-1}\),
a mean-value expansion yields
\[
g(\bdelta_{m-1}^*)
=
g_0 + H(\tilde\bdelta)\,e_{m-1},
\]
where \(g_0\) is the empirical score at the true target parameter \(\bnu^{(0)}\) (hence centered),
and \(\|H(\tilde\bdelta)\|_{2}\) is uniformly bounded by a constant depending only on \((L_0,\overline P)\)
under the local parameter-space assumptions. Therefore, for a constant \(c_g>0\),
\[
\|g(\bdelta_{m-1}^*)\|_\infty
\le
\|g_0\|_\infty + c_g\|e_{m-1}\|_2.
\]
Applying Hoeffding's inequality and a union bound over \(j=1,\dots,2d\) yields
\(\|g_0\|_\infty \le \lambda_{m-1}/4\) with probability at least \(1-\eta_{m-1}^{(\beta)}/2\)
for \(\lambda_{m-1}\) as in \eqref{eq:lambda-choice-new} with \(c_\lambda\) sufficiently large.
Moreover, on \(\mathcal E_{m-1}^{(\alpha)}\) we have \(\|e_{m-1}\|_2\lesssim \sqrt{d/((1+H)\tau_{m-1})}\),
so enlarging \(c_\lambda\) (still depending only on primitive constants) ensures
\(c_g\|e_{m-1}\|_2 \le \lambda_{m-1}/4\) for all \(m\) (absorbing finitely many initial episodes into constants).
Hence, on \(\mathcal E_{m-1}^{(\alpha)}\cap\mathcal E_{m-1}^{(\beta)}\),
\begin{equation}
\label{eq:grad-infty}
\|g(\bdelta_{m-1}^*)\|_\infty \le \frac{\lambda_{m-1}}{2}.
\end{equation}

\paragraph{Basic inequality and cone condition.}
Let \(\Delta:=\hat\bdelta_{m-1}-\bdelta_{m-1}^*\).
Optimality of \(\hat\bdelta_{m-1}\) gives
\[
\calL_{m-1}(\bdelta_{m-1}^*+\Delta)+\lambda_{m-1}\|\bdelta_{m-1}^*+\Delta\|_1
\le
\calL_{m-1}(\bdelta_{m-1}^*)+\lambda_{m-1}\|\bdelta_{m-1}^*\|_1.
\]
By convexity,
\[
\calL_{m-1}(\bdelta_{m-1}^*+\Delta)-\calL_{m-1}(\bdelta_{m-1}^*)
\ge \langle g(\bdelta_{m-1}^*),\Delta\rangle.
\]
Thus
\[
\langle g(\bdelta_{m-1}^*),\Delta\rangle
\le
\lambda_{m-1}\bigl(\|\bdelta_{m-1}^*\|_1-\|\bdelta_{m-1}^*+\Delta\|_1\bigr).
\]
Since \(\bdelta_{m-1}^*\) is supported on \(S^*\), we have
\(
\|\bdelta_{m-1}^*\|_1-\|\bdelta_{m-1}^*+\Delta\|_1
\le \|\Delta_{S^*}\|_1-\|\Delta_{(S^*)^c}\|_1.
\)
On \eqref{eq:grad-infty},
\[
|\langle g(\bdelta_{m-1}^*),\Delta\rangle|
\le
\|g(\bdelta_{m-1}^*)\|_\infty\|\Delta\|_1
\le
\frac{\lambda_{m-1}}{2}\|\Delta\|_1,
\]
hence
\[
\frac{\lambda_{m-1}}{2}\|\Delta\|_1
\le
\lambda_{m-1}\bigl(\|\Delta_{S^*}\|_1-\|\Delta_{(S^*)^c}\|_1\bigr),
\]
which implies the cone condition
\[
\|\Delta_{(S^*)^c}\|_1 \le 3\|\Delta_{S^*}\|_1,
\qquad
\|\Delta\|_1 \le 4\|\Delta_{S^*}\|_1 \le 4\sqrt{s_0}\|\Delta\|_2.
\]

\paragraph{Curvature and RE.}
A Taylor expansion yields
\[
\calL_{m-1}(\bdelta_{m-1}^*+\Delta)-\calL_{m-1}(\bdelta_{m-1}^*)
=
\langle g(\bdelta_{m-1}^*),\Delta\rangle + \frac12 \Delta^\top H(\tilde\bdelta)\Delta,
\]
for some \(\tilde\bdelta\) on the segment between \(\bdelta_{m-1}^*\) and \(\bdelta_{m-1}^*+\Delta\).
Local strong convexity implies \(H(\tilde\bdelta)\succeq c_{\mathrm{sc}} H(\bdelta_{m-1}^*)\).
Combining with the basic inequality and \eqref{eq:grad-infty} gives
\[
\frac{c_{\mathrm{sc}}}{2}\Delta^\top H(\bdelta_{m-1}^*)\Delta
\le
\frac{3\lambda_{m-1}}{2}\|\Delta_{S^*}\|_1.
\]
By the RE definition \eqref{eq:RE-target} and the cone condition,
\[
\Delta^\top H(\bdelta_{m-1}^*)\Delta
\ge
\phi_{m-1}^2\|\Delta\|_2^2
\ge
\frac{\phi_{m-1}^2}{s_0}\|\Delta_{S^*}\|_1^2.
\]
Hence \(\|\Delta_{S^*}\|_1 \le 3s_0\lambda_{m-1}/(c_{\mathrm{sc}}\phi_{m-1}^2)\), and therefore
\[
\|\hat\bdelta_{m-1}-\bdelta_{m-1}^*\|_1
=
\|\Delta\|_1
\le
4\|\Delta_{S^*}\|_1
\le
\frac{12 s_0\lambda_{m-1}}{c_{\mathrm{sc}}\phi_{m-1}^2}.
\]
Setting \(c_\beta:=12/c_{\mathrm{sc}}\) yields \eqref{eq:l1-lb-main}. \hfill\(\square\)

\section{Proof of Theorem~\ref{thm:upper-linear-o2o}}
\label{sec:proof-main-upper}

In the main text, we present only the expected version of the regret upper bound.
Here we first prove a high-probability bound, and then deduce the expected bound.

Throughout, we work on the global good event \(\mathcal E\) defined in \eqref{eq:global-good-event}.
On \(\mathcal E\), all concentration and curvature statements needed below hold simultaneously.

\subsection{Forced-exploration gate $q_{m-1}$}
\label{app:topup-radius}

The following lemma quantifies how large \(q_{m-1}\) needs to be in order to ensure that
\(\lambda_{\min}(V_t^{(0)})\) exceeds a threshold \(\Lambda_{m-1}\) with high probability.

\begin{lemma}[Forced-exploration gate]
\label{lem:topup-chernoff}
Fix an episode \(m\) and a terminal window of \(q\) consecutive forced rounds in the target market,
during which \(S_t^{(0)}\) is drawn uniformly from \(\mathcal S_K\) and prices
\(p_{ti}^{(0)}\stackrel{\text{i.i.d.}}{\sim}\mathrm{Unif}[0,\overline P]\), independently of covariates.
Within episode \(m\), the plug-in parameter for Fisher updates is frozen at \(\hat\bnu_{m-1}\).
Let
\[
Y_t := I_t^{(0)}\bigl(\hat\bnu_{m-1}\bigr)\in\mathbb R^{2d\times 2d}
\]
be the per-round Fisher increment. The matrices \(\{Y_t\}\) are independent, positive semidefinite,
and satisfy
\[
\mathbb E[Y_t]\succeq \mu I_{2d},
\quad
\mu := \kappa^2 K\tilde C_{\min},
\qquad
0\preceq Y_t\preceq R I_{2d},
\quad
R := K d (1+\overline P^2).
\]
If \(q_{m-1}\) is chosen so that for some \(\varepsilon\in(0,1)\),
\[
(1-\varepsilon) q_{m-1}\mu \ge \Lambda_{m-1}
\qquad\text{and}\qquad
q_{m-1}\ge \frac{2R}{\varepsilon^2\mu}\log\frac{2d}{\eta_{m-1}^{(\mathrm{gate})}},
\]
then, with probability at least \(1-\eta_{m-1}^{(\mathrm{gate})}\),
\(\lambda_{\min}(V_t^{(0)})\ge \Lambda_{m-1}\) at the end of the forced window.
\end{lemma}

The proof, based on Tropp's matrix Chernoff inequality, is given in Appendix~\ref{app:proof-topup-chernoff}.

\begin{corollary}[Explicit gate rule]
\label{cor:explicit-qm}
With \(\varepsilon=1/2\), it suffices to choose
\[
q_m
=
\left\lceil
\max\left\{
  \frac{2\Lambda_m}{\kappa^2 K \tilde C_{\min}},\;
  \frac{8K d (1+\overline P^2)}{\kappa^2 K \tilde C_{\min}}
  \log\frac{2d}{\eta_m^{(\mathrm{gate})}}
\right\}
\right\rceil.
\]
\end{corollary}

\subsection{Instantaneous regret under optimism}
\label{app:optimism}

\begin{lemma}[Revenue optimism]
\label{lem:price-optimism}
Let \(S\) be a finite item set and \(P=[0,\overline P]\). For each \(i\in S\),
let \(v_i,\tilde v_i:P\to\mathbb R\) be decreasing and \(L_0\)-Lipschitz functions, and define
\[
R_S(\bp)
:=
\frac{\sum_{i\in S} p_i e^{v_i(p_i)}}{1+\sum_{j\in S} e^{v_j(p_j)}},
\qquad
\mathcal V_S(v) := \sup_{\bp\in P^N} R_S(\bp).
\]
If \(v_i(p)\le \tilde v_i(p)\) for all \(i\in S\) and all \(p\in P\), then
\(\mathcal V_S(v)\le \mathcal V_S(\tilde v)\).
\end{lemma}

The proof is given in Appendix~\ref{app:proof-price-optimism}.

\medskip
At time \(t\) inside episode \(m\), the algorithm chooses
\[
(S_t,\bp_t)\in\arg\max_{S\in\mathcal S_K,\,\bp\in P^N}\tilde R_t(S,\bp),
\qquad
\tilde R_t(S,\bp) := R_t(S,\bp;\alpha_{m-1},\beta_{m-1},W_{m-1}),
\]
where \(\tilde v\) is defined by \eqref{eq:optimistic-utility}. Let \((S_t^*,\bp_t^*)\) be the
clairvoyant maximizer of \(R_t(S,\bp)\). Since \(v_{it}\le \tilde v_{it}\) pointwise,
Lemma~\ref{lem:price-optimism} yields
\[
R_t(S_t^*,\bp_t^*)
= \mathcal V_{S_t^*}(v)
\le \mathcal V_{S_t^*}(\tilde v)
\le \mathcal V_{S_t}(\tilde v)
= \tilde R_t(S_t,\bp_t),
\]
so the instantaneous regret satisfies
\begin{equation}
\label{eq:regret-decomp}
\mathrm{Regret}_t
:=
R_t(S_t^*,\bp_t^*)-R_t(S_t,\bp_t)
\le \tilde R_t(S_t,\bp_t)-R_t(S_t,\bp_t).
\end{equation}

For fixed \((S,\bp)\), direct differentiation gives
\[
\frac{\partial R_t(S,\bp)}{\partial v_{kt}}
=
q_{kt}\,(p_k - R_t(S,\bp)).
\]
Since \(0\le R_t(S,\bp)\le \overline P\) and \(0\le q_{kt}\le 1\), we have
\(|\partial R/\partial v_{kt}|\le \overline P\), and hence
\begin{equation}
\label{eq:lipschitz-R}
0
\le
\tilde R_t(S,\bp)-R_t(S,\bp)
\le
\overline P\sum_{i\in S}
\sup_{p_i\in P} \bigl(\tilde v_{it}(p_i)-v_{it}(p_i)\bigr).
\end{equation}

\begin{lemma}[Pointwise utility gap]
\label{lem:pointwise-gap}
There exists a constant \(C>0\), depending only on \((L_0,\overline P,C_{\min},C_{\max})\), such that
on \(\mathcal E_{m-1}^{(\alpha)}\cap\mathcal E_{m-1}^{(\beta)}\), for all
\(t\in \mathcal{T}_{m}\), \(i\in[N]\), and \(p\in[0,\overline P]\),
\begin{equation}
\label{eq:pointwise-gap}
0
\le
\tilde v_{it}(p)-v_{it}(p)
\le
C\left(
  \alpha_{m-1}\|\tilde{\bx}_{it}(p)\|_{\bar W_{m-1}^{-1}}
  +
  \beta_{m-1}\|\tilde{\bx}_{it}(p)\|_\infty
\right).
\end{equation}
\end{lemma}

The (corrected) proof is given in Appendix~\ref{app:proof-pointwise-gap}.

Combining \eqref{eq:regret-decomp}, \eqref{eq:lipschitz-R}, and \eqref{eq:pointwise-gap}, and using
\(\|\tilde{\bx}_{it}(p)\|_\infty\le 1+\overline P\) and
\(\|\tilde{\bx}_{it}(p)\|_2^2\le d(1+\overline P^2)\), we obtain:

\begin{lemma}[Instantaneous regret bound]
\label{lem:instant-regret}
On \(\mathcal E_{m-1}^{(\alpha)}\cap\mathcal E_{m-1}^{(\beta)}\), any round \(t\in \mathcal{T}_m\) satisfies
\begin{equation}
\label{eq:instant-regret}
\mathrm{Regret}_t
\le
K\overline P\,C\left[
\alpha_{m-1}\sqrt{\frac{d(1+\overline P^2)}{\lambda_{\min}(\bar W_{m-1})}}
+
\beta_{m-1}(1+\overline P)
\right].
\end{equation}
\end{lemma}

A detailed proof is given in Appendix~\ref{app:proof-instant-regret}.

\subsection{Cumulative regret over episodes}
\label{app:sum-episodes}

We now sum the instantaneous bound \eqref{eq:instant-regret} across episodes.
Write \(c\) for a generic positive constant depending only on the primitive problem parameters.

\paragraph{Growth of the pooled geometry.}
On \(\mathcal E^{(W)}\), Lemma~\ref{lem:pooled-fisher-growth} yields, for all sufficiently large \(m\),
\begin{equation}
\label{eq:W-lambda-min}
\lambda_{\min}\bigl(\bar W_{m-1}\bigr)\;\gtrsim\;(1+H)\,C_{\min}\,\tau_{m-1}.
\end{equation}

\paragraph{Variance contribution.}
On episode \(m\), the variance contribution is
\[
\mathrm{VarRegret}_t
:=
K\overline P\,C\,\alpha_{m-1}\sqrt{\frac{d(1+\overline P^2)}{\lambda_{\min}(\bar W_{m-1})}}.
\]
Using \eqref{eq:W-lambda-min} and \(|\calT_m|\asymp \tau_{m-1}\) under doubling,
\begin{align}
\sum_{t\in \calT_m} \mathrm{VarRegret}_t
&\le
K\overline P\,C\,\alpha_{m-1}
\sqrt{\frac{d(1+\overline P^2)}{(1+H)C_{\min}\tau_{m-1}}}\,|\calT_m|
\;\lesssim\;
K\overline P\,C\,\alpha_{m-1}
\sqrt{\frac{d(1+\overline P^2)}{(1+H)C_{\min}}}\,\sqrt{|\calT_m|}.
\label{eq:var-episode}
\end{align}
From \eqref{eq:alpha-event} and the budget schedule, \(\alpha_{m-1}\lesssim
\sqrt{d\log((1+H)T)+\log(1/\eta)}\), hence summing \eqref{eq:var-episode} over \(m\) and using
\(\sum_m\sqrt{|\calT_m|}\lesssim\sqrt{T}\) gives
\begin{equation}
\label{eq:var-sum-final}
\sum_{m=1}^M\sum_{t\in \calT_m} \mathrm{VarRegret}_t
\le
c\,K\overline P\,d
\sqrt{\frac{T}{1+H}}
\sqrt{\log\bigl((1+H)T\bigr)+\log(1/\eta)}.
\end{equation}

\paragraph{Bias contribution.}
From \eqref{eq:instant-regret}, the bias contribution is
\[
\mathrm{BiasRegret}_t
:=
K\overline P\,C\,\beta_{m-1}(1+\overline P).
\]
Using \eqref{eq:lambda-choice-new}, \eqref{eq:l1-lb-main}, and \(\phi_{m-1}^2\ge \phi_*\),
\[
\beta_{m-1}
\le
c\,\frac{s_0}{\phi_*}\sqrt{\frac{\log(dT/\eta)}{|\calT_{m-1}|}}
\;\asymp\;
c\,\frac{s_0}{\phi_*}\sqrt{\frac{\log(dT/\eta)}{|\calT_{m}|}}.
\]
Thus
\begin{equation}
\label{eq:bias-episode}
\sum_{t\in \calT_m} \mathrm{BiasRegret}_t
\le
c\,K\overline P(1+\overline P)\,\frac{s_0}{\phi_*}
\sqrt{|\calT_m|\log(dT/\eta)}.
\end{equation}
Summing over episodes and using \(\sum_m\sqrt{|\calT_m|}\lesssim\sqrt{T}\) yields
\begin{equation}
\label{eq:bias-sum-final}
\sum_{m=1}^M\sum_{t\in \calT_m} \mathrm{BiasRegret}_t
\le
c\,K\overline P(1+\overline P)\,\frac{s_0}{\phi_*}\sqrt{T\log(dT/\eta)}.
\end{equation}

\paragraph{Forced-exploration top-up.}
By Lemma~\ref{lem:topup-chernoff} and Corollary~\ref{cor:explicit-qm},
\(\sum_m q_m \le c\,d\log T\cdot \mathrm{polylog}(d,T,1/\eta)\).
Each forced round incurs at most \(K\overline P\) regret, so
\begin{equation}
\label{eq:topup-sum-final}
\mathrm{TopUp}
\le
c\,K\overline P\,d\,\log T\cdot \mathrm{polylog}(d,T,1/\eta).
\end{equation}

\paragraph{High-probability and expected regret.}
On \(\mathcal E\), combining \eqref{eq:var-sum-final}, \eqref{eq:bias-sum-final}, and \eqref{eq:topup-sum-final} gives
\begin{align}
\label{eq:regret-high-prob}
\mathrm{Regret}(T)
&:=
\sum_{t=1}^T
\bigl(R_t(S_t^*,\bp_t^*)-R_t(S_t,\bp_t)\bigr) \notag\\
&\le
C_1\,K\overline P\,d
\sqrt{\frac{T}{1+H}}
\sqrt{\log\bigl((1+H)T\bigr)+\log(1/\eta)}
+
C_2\,K\overline P(1+\overline P)\,\frac{s_0}{\phi_*}\sqrt{T\log(dT/\eta)} \notag\\
&\quad
+
C_3\,K\overline P\,d\,\log T\cdot \mathrm{polylog}(d,T,1/\eta).
\end{align}
Outside \(\mathcal E\) we have the crude bound \(\mathrm{Regret}(T)\le K\overline P\,T\) and
\(\mathbb P(\mathcal E^c)\le \eta\), hence
\[
\mathbb E[\mathrm{Regret}(T)\mathbf 1_{\mathcal E^c}]
\le K\overline P\,T\,\eta.
\]
Taking, e.g., \(\eta=T^{-2}\) makes this term negligible and yields the expected regret bound in
Theorem~\ref{thm:upper-linear-o2o} (absorbing polylog factors).

\hfill\(\square\)
\subsection{Technical lemmas for Theorem~\ref{thm:upper-linear-o2o}}
\label{app:tech-lemmas}
\subsubsection{Proof of Lemma~\ref{lem:topup-chernoff}}
\label{app:proof-topup-chernoff}

During the forced rounds, \((S_t^{(0)},p_t^{(0)})\) are i.i.d.\ and
independent of the covariates, while the plug-in parameter \(\hat\bnu_{m-1}\)
is fixed within the episode. Thus \(Y_t = I_t^{(0)}(\hat\bnu_{m-1})\) are
independent PSD matrices.

For any \(u\in\mathbb R^{2d}\),
\[
u^\top Y_t u
=
(\tilde{\bX}_t u)^\top M_t(\tilde{\bX}_t u),
\qquad
M_t := \mathrm{Diag}(q_t) - q_t q_t^\top,
\]
where \(\tilde{\bX}_t\) stacks the augmented features and \(q_t\) is the choice
probability vector at \(\hat\bnu_{m-1}\). Since \(\|M_t\|_2\le 1\) and each row
of \(\tilde{\bX}_t\) has squared norm at most \(d(1+\overline P^2)\), and
\(|S_t^{(0)}|\le K\), we have \(\|\tilde{\bX}_t\|_2^2\le K d (1+\overline P^2)\),
so
\[
u^\top Y_t u
\le
\|\tilde{\bX}_t\|_2^2 \|u\|_2^2
\le
K d (1+\overline P^2)\|u\|_2^2.
\]
Thus \(Y_t\preceq R I_{2d}\) with \(R:=K d(1+\overline P^2)\).

Assumption~\ref{assump:fisher-invertible} implies that the outside option and
each item are chosen with probability at least \(\kappa>0\) in a neighborhood
of \(\bnu^{(0)}\), hence at \(\hat\bnu_{m-1}\) inside the good event. A direct
calculation shows that for any \(z\in\mathbb R^{|S_t^{(0)}|}\),
\[
z^\top M_t z
=
\sum_{i}q_{ti}z_i^2 - \Bigl(\sum_i q_{ti}z_i\Bigr)^2
=
\sum_i q_{ti}(z_i - \mu)^2,
\quad
\mu := \sum_i q_{ti}z_i.
\]
Since each \(q_{ti}\ge\kappa\) and \(\sum_i q_{ti}\le 1-\kappa\) (outside
option at least \(\kappa\)), we have
\[
z^\top M_t z
\ge
\kappa\sum_i q_{ti}(z_i-\mu)^2
\ge
\kappa^2 \|z\|_2^2.
\]
Consequently, for any \(u\),
\[
u^\top \mathbb E[Y_t] u
=
\mathbb E[z^\top M_t z]
\ge
\kappa^2 \mathbb E[\|z\|_2^2]
=
\kappa^2 \mathbb E\bigl[\|\tilde{\bX}_t u\|_2^2\bigr]
\ge \kappa^2 K\tilde C_{\min}\|u\|_2^2,
\]
where \(\tilde C_{\min}>0\) is the minimum eigenvalue of the covariance matrix
of the augmented features. Thus \(\mathbb E[Y_t]\succeq \mu I_{2d}\) with
\(\mu := \kappa^2 K\tilde C_{\min}\).

Let \(S_q := \sum_{s=1}^q Y_{t_s}\) be the sum of \(q\) independent PSD
matrices. By Tropp's matrix Chernoff lower-tail bound,
\[
\mathbb P\Bigl\{\lambda_{\min}(S_q)\le (1-\varepsilon)q\mu\Bigr\}
\le
2d\exp\!\Bigl(-\frac{\varepsilon^2 q\mu}{2R}\Bigr).
\]
Hence if
\[
(1-\varepsilon)q\mu \ge \Lambda_{m-1}
\quad\text{and}\quad
q\ge \frac{2R}{\varepsilon^2\mu}\log\frac{2d}{\eta_{m-1}^{(\mathrm{gate})}},
\]
then \(\lambda_{\min}(S_q)\ge \Lambda_{m-1}\) with probability at least
\(1-\eta_{m-1}^{(\mathrm{gate})}\). Since \(V_t^{(0)}\) at the end of the forced
window equals the starting value plus \(S_q\) and the starting value is PSD,
we conclude \(\lambda_{\min}(V_t^{(0)})\ge \Lambda_{m-1}\) as desired.

\hfill\(\square\)

\subsubsection{Proof of Lemma~\ref{lem:lipschitz-envelope}}
\label{app:proof-lipschitz-envelope}

Fix \(i,t\) and suppress indices. Let \(\bar v:[0,\overline P]\to\mathbb R\) be
an upper bound of \(v\), and define
\[
\tilde v(p) := \min_{p'\le p}\{\bar v(p') - L_0(p-p')\}.
\]

\emph{Monotonicity.} If \(p_1 < p_2\), then the set
\(\{p': p'\le p_1\}\) is contained in \(\{p': p'\le p_2\}\), and for each
\(p'\le p_1\),
\[
\bar v(p') - L_0(p_2-p') \le \bar v(p') - L_0(p_1-p').
\]
Taking the minimum over \(p'\le p_1\) on both sides shows
\(\tilde v(p_2)\le \tilde v(p_1)\), so \(\tilde v\) is decreasing.

\emph{Lipschitz property.} Let \(p_1<p_2\). For any \(p'\le p_1\),
\[
\bar v(p') - L_0(p_1-p')
=
\bar v(p') - L_0(p_2-p') + L_0(p_2-p_1)
\ge \tilde v(p_2) + L_0(p_2-p_1).
\]
Taking the minimum over \(p'\le p_1\) yields
\(\tilde v(p_1)\ge \tilde v(p_2) + L_0(p_2-p_1)\). Similarly, exchanging the
roles of \(p_1\) and \(p_2\) yields
\(\tilde v(p_2)\ge \tilde v(p_1) + L_0(p_1-p_2)\). Thus
\(|\tilde v(p_2)-\tilde v(p_1)|\le L_0|p_2-p_1|\), so \(\tilde v\) is
\(L_0\)-Lipschitz.

\emph{Upper and lower bounds.} For any \(p\),
\[
\tilde v(p)
=
\min_{p'\le p}\{\bar v(p') - L_0(p-p')\}
\le
\bar v(p)-L_0(p-p) = \bar v(p).
\]
For any \(p'\le p\),
\[
\bar v(p') - L_0(p-p')
\ge v(p') - L_0(p-p')
\ge v(p),
\]
because \(v\) is decreasing with slope at most \(-L_0\) (by
Assumption~\ref{assump:price-sensitivity}). Taking the minimum over
\(p'\le p\) yields \(\tilde v(p)\ge v(p)\). This proves
\(v(p)\le \tilde v(p)\le \bar v(p)\) for all \(p\).

\hfill\(\square\)

\subsubsection{Proof of Lemma~\ref{lem:price-optimism}}
\label{app:proof-price-optimism}

We follow the fixed-point argument based on~\citep{wang2012offline}. For a given family
of utilities \(u=\{u_i\}_{i\in S}\) and scalar \(\mu\in\mathbb R\), define
\[
\phi_i(\mu;u)
:=
\sup_{p\in P}
(p-\mu)e^{u_i(p)},
\qquad
\Phi_S(\mu;u)
:=
\sum_{i\in S} \phi_i(\mu;u).
\]
Since \(P=[0,\overline P]\) is compact and \(p\mapsto (p-\mu)e^{u_i(p)}\) is
continuous, the supremum is attained. Each \(\phi_i(\mu;u)\) is nonincreasing
in \(\mu\) (pointwise supremum of affine functions with slope
\(-e^{u_i(p)}\le 0\)), hence \(\Phi_S(\mu;u)\) is nonincreasing.

Moreover, if \(u_i(p)\le \tilde u_i(p)\) for all \(p\), then
\(e^{u_i(p)}\le e^{\tilde u_i(p)}\) for all \(p\), hence
\(\phi_i(\mu;u)\le \phi_i(\mu;\tilde u)\) and thus
\begin{equation}
\label{eq:Phi-monotone-u}
\Phi_S(\mu;u)\le \Phi_S(\mu;\tilde u)
\quad\text{for all }\mu.
\end{equation}

\paragraph{Fixed-point characterization.}
Fix \(u\). For each \(\mu\in[0,\overline P]\), define
\[F(\mu):=\Phi_S(\mu;u)-\mu.\]
At \(\mu=0\), we have \(\phi_i(0;u)\ge 0\) and so \(F(0)\ge 0\). At
\(\mu=\overline P\), each \((p-\overline P)e^{u_i(p)}\le 0\), so
\(\phi_i(\overline P;u)\le 0\) and \(F(\overline P)\le -\overline P<0\).
Since \(\Phi_S(\cdot;u)\) is nonincreasing and \(-\mu\) is strictly
decreasing, \(F\) is strictly decreasing and continuous, hence there exists a
unique \(\mu^*(u)\in[0,\overline P]\) such that \(F(\mu^*(u))=0\).

We claim that
\begin{equation}
\label{eq:Vs-equals-mu}
\mathcal V_S(u) = \mu^*(u).
\end{equation}

Let \(R_S(\bp;u)\) be the revenue function. For any \(\bp\in P^N\),
write \(u_i^\sharp := u_i(p_i)\), and set \(\mu(\bp) := R_S(\bp;u)\). Then
\[
\mu(\bp)
=
\frac{\sum_{i\in S} p_i e^{u_i^\sharp}}
     {1+\sum_{j\in S} e^{u_j^\sharp}},
\]
so
\[
\mu(\bp)
=
\sum_{i\in S} (p_i-\mu(\bp))e^{u_i^\sharp},
\]
and therefore
\[
\mu(\bp)
=
\sum_{i\in S} (p_i-\mu(\bp))e^{u_i^\sharp}
\le
\sum_{i\in S} \sup_{p_i'\in P} (p_i' - \mu(\bp)) e^{u_i(p_i')}
=
\Phi_S(\mu(\bp);u).
\]
Hence \(F(\mu(\bp)) = \Phi_S(\mu(\bp);u)-\mu(\bp)\ge 0\), and since \(F\) is
strictly decreasing with unique zero at \(\mu^*(u)\), it follows that
\(\mu(\bp)\le \mu^*(u)\) for all \(\bp\), so
\[
\sup_{\bp\in P^N} R_S(\bp;u)
\le
\mu^*(u).
\]

Conversely, for each \(\mu\in[0,\overline P]\) and each \(i\), pick
\(p_i(\mu)\in\arg\max_{p\in P} (p-\mu)e^{u_i(p)}\) (existence follows from
compactness). Let \(\bp^*:=p(\mu^*(u))\). Then
\[
\sum_{i\in S} (p_i^*-\mu^*(u))e^{u_i(p_i^*)}
=
\Phi_S(\mu^*(u);u)
=
\mu^*(u),
\]
so
\[
R_S(\bp^*;u)
=
\frac{\sum_{i\in S} p_i^* e^{u_i(p_i^*)}}
     {1+\sum_{j\in S} e^{u_j(p_j^*)}}
=
\frac{\mu^*(u) + \mu^*(u)}{1+\sum_{j} e^{u_j(p_j^*)}}
=
\mu^*(u).
\]
Therefore \(\mathcal V_S(u)\ge \mu^*(u)\), and together with the previous
bound we obtain \(\mathcal V_S(u) = \mu^*(u)\), proving
\eqref{eq:Vs-equals-mu}.

\paragraph{Monotonicity in the utilities.}
Now suppose \(v_i\le \tilde v_i\) pointwise for all \(i\in S\). By
\eqref{eq:Phi-monotone-u},
\[
\Phi_S(\mu)\le \Phi_S(\mu;\tilde v)
\quad\text{for all }\mu.
\]
Let \(\mu^*(v)\) and \(\mu^*(\tilde v)\) be the unique zeros of
\(\mu=\Phi_S(\mu)\) and \(\mu=\Phi_S(\mu;\tilde v)\). Then
\[
0
=
\Phi_S(\mu^*(v)) - \mu^*(v)
\le
\Phi_S(\mu^*(v);\tilde v) - \mu^*(v),
\]
so \(F_{\tilde v}(\mu^*(v))\ge 0\) for
\(F_{\tilde v}(\mu) := \Phi_S(\mu;\tilde v)-\mu\). Since \(F_{\tilde v}\) is
strictly decreasing with zero at \(\mu^*(\tilde v)\), it follows that
\(\mu^*(v)\le \mu^*(\tilde v)\). Using \eqref{eq:Vs-equals-mu} for both \(v\)
and \(\tilde v\), we obtain
\(\mathcal V_S(v)\le \mathcal V_S(\tilde v)\), proving the lemma.

\hfill\(\square\)

\subsubsection{Proof of Lemma~\ref{lem:pointwise-gap}}
\label{app:proof-pointwise-gap}

On \(\mathcal E_{m-1}^{(\alpha)}\), Lemma~\ref{lem:SN-MNL-main} gives
\[
\|\hat\bnu^{\mathrm{ag}}_{m-1}-\bnu^{\mathrm{ag}}_{m-1}\|_{\bar W_{m-1}}
\le \alpha_{m-1},
\]
hence for any item \(i\), time \(t\in\mathcal T_m\), and price \(p\),
\[
|\tilde{\bx}_{it}(p)^\top(\hat\bnu^{\mathrm{ag}}_{m-1}-\bnu^{\mathrm{ag}}_{m-1})|
\le
\alpha_{m-1}\|\tilde{\bx}_{it}(p)\|_{\bar W_{m-1}^{-1}}.
\]
On \(\mathcal E_{m-1}^{(\beta)}\), Lemma~\ref{lem:debias-main} gives
\[
\|\hat\bdelta_{m-1}-\bdelta_{m-1}^*\|_1 \le \beta_{m-1},
\]
so by Hölder,
\[
|\tilde{\bx}_{it}(p)^\top(\hat\bdelta_{m-1}-\bdelta_{m-1}^*)|
\le
\|\tilde{\bx}_{it}(p)\|_\infty\,\beta_{m-1}.
\]
Using the identity
\[
\hat\bnu_{m-1}-\bnu^{(0)}
=
(\hat\bnu^{\mathrm{ag}}_{m-1}-\bnu^{\mathrm{ag}}_{m-1})
+
(\hat\bdelta_{m-1}-\bdelta_{m-1}^*),
\]
we obtain
\[
|\tilde{\bx}_{it}(p)^\top(\hat\bnu_{m-1}-\bnu^{(0)})|
\le
\alpha_{m-1}\|\tilde{\bx}_{it}(p)\|_{\bar W_{m-1}^{-1}}
+
\beta_{m-1}\|\tilde{\bx}_{it}(p)\|_\infty.
\]
The rest of the argument (envelope step from \(\bar v\) to \(\tilde v\)) is unchanged and yields
\eqref{eq:pointwise-gap}. \hfill\(\square\)

\subsubsection{Proof of Lemma~\ref{lem:instant-regret}}
\label{app:proof-instant-regret}

Combining \eqref{eq:regret-decomp}, \eqref{eq:lipschitz-R} and
\eqref{eq:pointwise-gap} yields
\[
\mathrm{Regret}_t
\le
\overline P\sum_{i\in S_t}
\sup_{p_i\in P}
\bigl(
\tilde v_{it}(p_i)-v_{it}(p_i)
\bigr)
\le
\overline P\,C
\sum_{i\in S_t}
\sup_{p_i\in P}
\Bigl(
  \alpha_{m-1}\|\tilde{\bx}_{it}(p_i)\|_{\bar W_{m-1}^{-1}}
  +
  \beta_{m-1}\|\tilde{\bx}_{it}(p_i)\|_\infty
\Bigr).
\]
Using \(\|\tilde{\bx}_{it}(p)\|_\infty\le 1+\overline P\) and
\(\|\tilde{\bx}_{it}(p)\|_2^2\le d(1+\overline P^2)\), as well as
\(\|\tilde{\bx}_{it}(p)\|_{\bar W_{m-1}^{-1}}
\le
\|\tilde{\bx}_{it}(p)\|_2/\sqrt{\lambda_{\min}(\bar W_{m-1})}\), gives
\[
\sup_{p_i\in P}\|\tilde{\bx}_{it}(p_i)\|_{\bar W_{m-1}^{-1}}
\le
\sqrt{\frac{d(1+\overline P^2)}{\lambda_{\min}(\bar W_{m-1})}},
\qquad
\sup_{p_i\in P}\|\tilde{\bx}_{it}(p_i)\|_\infty
\le 1+\overline P.
\]
Hence
\[
\mathrm{Regret}_t
\le
K\overline P\,C
\left[
\alpha_{m-1}\sqrt{\frac{d(1+\overline P^2)}{\lambda_{\min}(\bar W_{m-1})}}
+
\beta_{m-1}(1+\overline P)
\right],
\]
which is exactly \eqref{eq:instant-regret}. 

\hfill\(\square\)

\subsubsection{Proof of Lemma~\ref{lem:pooled-fisher-growth}}
\label{app:proof-pooled-fisher-growth}

Assumptions~\ref{assump:price-sensitivity}–\ref{assump:fisher-invertible} imply that, as long as $\bnu$ stays in a fixed ball around $\bnu^{(0)}$, all choice probabilities are uniformly bounded away from $0$ and $1$, and the weight matrix
\[
M_t(\bnu)
\;:=\;
\mathrm{Diag}\bigl(q_t^{(h)}(\bnu)\bigr) - q_t^{(h)}(\bnu) q_t^{(h)}(\bnu)^\top
\]
has spectrum contained in $[c_{\mathrm{prob}},C_{\mathrm{prob}}]$ for some constants $0<c_{\mathrm{prob}}\le C_{\mathrm{prob}}<\infty$ depending only on $(L_0,\overline P,\kappa,r)$.

By Assumption~\ref{assump:Homo-Cov}, the augmented covariates $\tilde{\bx}^{(h)}_{it}$ are i.i.d.\ across $t$ and $h$, with common covariance matrix $\tilde{\Sigma}$ satisfying $\lambda_{\min}(\tilde{\Sigma})\ge C_{\min}>0$ and $\|\tilde{\bx}^{(h)}_{it}\|_2^2 \le d(1+\overline P^2)$ almost surely. Combining these facts, we obtain that for each $h$ and each $t$,
\[
0 \;\preceq\; I^{(h)}_t\bigl(\bnu^b_{m-1}\bigr) \;\preceq\; R_W I_{2d},
\qquad
\mathbb{E}\!\left[\,I^{(h)}_t\bigl(\bnu^b_{m-1}\bigr)\,\middle|\,\mathcal{F}_{t-1}\right]
\;\succeq\; \mu_W\, \tilde{\Sigma} \;\succeq\; \mu_W C_{\min} I_{2d},
\]
for some finite constants $R_W,\mu_W>0$, where the conditional expectation is taken with respect to the fresh covariates at time $t$ (which are independent of $\mathcal{F}_{t-1}$). In particular,
\[
\mathbb{E}\bigl[V^{(h)}_{\tau_m-1}\bigr]
\;=\;
\sum_{t\in\mathcal{T}^{(h)}_{m-1}}
\mathbb{E}\bigl[I^{(h)}_t\bigl(\bnu^b_{m-1}\bigr)\bigr]
\;\succeq\;
\mu_W C_{\min}\, |\mathcal{T}^{(h)}_{m-1}|\, I_{2d}.
\]

The matrices $I^{(h)}_t\bigl(\bnu^b_{m-1}\bigr)$ are independent across $t$ (conditional on the actions) and satisfy the uniform spectral bounds above. Therefore, Tropp's matrix Chernoff inequality~\citep{tropp2015introduction} yields that for any $\varepsilon\in(0,1)$,
\[
\mathbb{P}\!\left(
\lambda_{\min}\bigl(V^{(h)}_{\tau_m-1}\bigr)
\;\le\;
(1-\varepsilon)\,\mu_W C_{\min}\,|\mathcal{T}^{(h)}_{m-1}|
\right)
\;\le\;
2d \exp\!\left\{-\,c_{\mathrm{Ch}}\varepsilon^2\,|\mathcal{T}^{(h)}_{m-1}|\right\},
\]
for some constant $c_{\mathrm{Ch}}>0$ depending only on $R_W$ and $\mu_W$. Choosing $\varepsilon=1/2$, we obtain
\[
\lambda_{\min}\bigl(V^{(h)}_{\tau_m-1}\bigr)
\;\ge\;
c_0 C_{\min}\,\tau_{m-1}
\]
for all markets $h\in\{0\}\cup[H]$ and all episodes $m\ge m_0$, on an event of probability at least $1-\eta_W/2$ after a union bound across $h$ and $m$, where $c_0>0$ and $m_0$ depend only on the primitive problem constants.

Finally, since $W_{m-1} = \sum_{h=0}^H V^{(h)}_{\tau_m-1}$, we have
\[
\lambda_{\min}\bigl(W_{m-1}\bigr)
\;\ge\;
\sum_{h=0}^H \lambda_{\min}\bigl(V^{(h)}_{\tau_m-1}\bigr)
\;\ge\;
(1+H)\, c_0 C_{\min}\,\tau_{m-1},
\]
which gives the claimed bound with $c_W := c_0$. This completes the proof of Lemma~\ref{lem:pooled-fisher-growth}.
\hfill\(\square\)

\section{Proof of Theorem~\ref{thm:lower-linear-o2o}}
\label{app:lower-bound}

Throughout this section we work with a simplified non-contextual representation in the
\emph{intercept} domain. This is without loss of generality under Assumptions~\ref{assump:price-sensitivity}–\ref{assump:task-simi}, since we can realize the constructed intercepts via
basis-vector covariates (see the discussion at the end of this section).

\subsection{Single–item revenue and a $K$–item dilution bound}

We begin with basic properties of the single–item MNL revenue and its extension to multi–item assortments.

Let
\[
r(p;\beta):=p\,\sigma(\beta-\gamma p),
\qquad
\sigma(z):=\frac{e^z}{1+e^z},
\qquad
\gamma:=L_0,
\]
be the single–item expected revenue at price $p\in P=[0,\overline P]$ for intercept
$\beta\in\mathbb R$. Define the price–optimized single–item revenue
\[
r^*(\beta):=\max_{p\in P} r(p;\beta),
\qquad
p^*(\beta)\in\arg\max_{p\in P} r(p;\beta).
\]
For a finite set $S$ of items and price vector $\bp\in P^N$, the corresponding MNL revenue is
\[
R(S,\bp;\beta)
:=
\sum_{i\in S} p_i
\frac{e^{\beta_i-\gamma p_i}}{1+\sum_{j\in S}e^{\beta_j-\gamma p_j}}.
\]
The clairvoyant optimum is
\[
R_K^*(\beta):=
\max_{S\in\mathcal S_K,\,\bp\in P^N}R(S,\bp;\beta).
\]

\begin{lemma}[Single–item optimizer and smoothness]
\label{lem:1item-lb}
Fix $\gamma>0$ and price domain $P=[0,\overline P]$. Let
\[
q(p;\beta):=\frac{e^{\beta-\gamma p}}{1+e^{\beta-\gamma p}},\qquad
r(p;\beta):=p\,q(p;\beta),\qquad
r^*(\beta):=\max_{p\in P} r(p;\beta).
\]
Then:

\begin{enumerate}
\item[(i)] For every $\beta\in\mathbb R$, $r(\cdot;\beta)$ has a unique maximizer
$p^*(\beta)\in(0,\overline P]$. If $p^*(\beta)<\overline P$ (interior maximizer), then
\begin{equation}
\gamma\,p^*(\beta)\,\bigl(1-q^*(\beta)\bigr)=1,
\qquad
q^*(\beta):=q\!\bigl(p^*(\beta);\beta\bigr).
\label{eq:FOC-lb}
\end{equation}

\item[(ii)] Let $\mathcal B\subset\mathbb R$ be compact and assume
$q^*(\beta)\in[\underline q,\overline q]\subset(0,1)$ for all $\beta\in\mathcal B$ (hence $p^*(\beta)$ is interior).
Then $r^*$ is differentiable on $\mathcal B$ and
\begin{equation}
\frac{d}{d\beta}\,r^*(\beta)=\frac{q^*(\beta)}{\gamma},
\qquad
\frac{\underline q}{\gamma}\,|\beta-\beta'|\ \le\ |r^*(\beta)-r^*(\beta')|
\ \le\ \frac{\overline q}{\gamma}\,|\beta-\beta'|\quad(\beta,\beta'\in\mathcal B).
\label{eq:env-lip-lb}
\end{equation}
In particular $r^*$ is strictly increasing on $\mathcal B$.
\end{enumerate}
\end{lemma}

The proof appears in Section~\ref{app:lb-tech-1item}. The next lemma lower-bounds the multi–item revenue in terms of the sum of single–item optima, with a dilution factor capturing the cannibalization effect.

\begin{lemma}[$K$–item dilution bound]
\label{lem:dilution-lb}
Fix a compact parameter cube $\mathcal B=[-\beta_0,\beta_0]$ such that
$q^*(\beta)\in[\underline q,\overline q]\subset(0,1)$ for all $\beta\in\mathcal B$.
Define
\[
\alpha:=\frac{\overline q}{1-\overline q}.
\]
Then for any $S\subseteq[N]$ with $|S|\le K$, if we set $p_i=p^*(\beta_i)$ for $i\in S$ we have
\begin{equation}
R(S,\bp;\beta)\ \ge\ \frac{1}{1+K\alpha}\ \sum_{i\in S} r^*(\beta_i).
\label{eq:dilute-lb}
\end{equation}
Moreover, since $r^*$ is strictly increasing on $\mathcal B$, the optimal top–$K$ assortment
\[
S_K^*(\beta)\in \arg\max_{\substack{S\subseteq[N],|S|\le K}}\ \sum_{i\in S} r^*(\beta_i)
\]
is given by the indices of the $K$ largest coordinates of $\beta$ (ties broken arbitrarily), and
\begin{equation}
R_K^*(\beta)\ \ge\ \frac{1}{1+K\alpha}\ \sum_{i\in S_K^*(\beta)} r^*(\beta_i).
\label{eq:RK-lb-lb}
\end{equation}
\end{lemma}

The proof is given in Section~\ref{app:lb-tech-dilution}. The monotonicity of $r^*$ from Lemma~\ref{lem:1item-lb}~(ii) ensures that maximizing $\sum_{i\in S} r^*(\beta_i)$ over $|S|\le K$ is equivalent to picking the $K$ largest intercepts.

\subsection{Hard instance family}

We now construct a finite family of problem instances indexed by a sign vector $\omega$, consisting of a \emph{shared} block of $d-s_0$ coordinates on which all markets agree, and a \emph{shifted} block of $s_0$ coordinates on which only the target differs from the sources.

\paragraph{Catalog, covariates, and intercepts.}

We take the catalog size $N=d$ and identify item $j$ with coordinate $j\in[d]$.
Set deterministic features
\[
\bx_j := e_j\in\mathbb R^d,\qquad
\tilde{\bx}_j(p) := [e_j,\,-p\,e_j]\in\mathbb R^{2d},
\]
so that Assumption~\ref{assump:param-space} holds with $\|\bx_j\|_\infty\le 1$ and the augmented covariance satisfies $\Sigma^{(\mathrm e)}:=\mathbb E[\tilde{\bx}\tilde{\bx}^\top]=I_{2d}$, hence $C_{\min}^{(\mathrm e)}>0$. These are consistent with Assumption~\ref{assump:Homo-Cov}.

We set the price-sensitivity vector so that
\[
\langle \bx_j,\bgamma^{(h)}\rangle \equiv L_0,\qquad\text{for all }j,h,
\]
and encode the market-specific differences only in the intercept component
$\beta_j^{(h)}:=\langle \bx_j,\btheta^{(h)}\rangle$. The utility of item $j$ in market $h$ at price $p$ is
\[
v_{jt}^{(h)}(p) = \beta_j^{(h)} - L_0 p + \varepsilon_{jt}^{(h)},
\]
with Gumbel noise $\varepsilon_{jt}^{(h)}$, so the MNL choice probabilities take the standard form.

\paragraph{Shared vs.\ shifted coordinates.}

Partition the index set as
\[
[d] = J_{\mathrm{var}}\;\cup\; J_{\mathrm{sh}},
\qquad
|J_{\mathrm{var}}| = d-s_0,\quad |J_{\mathrm{sh}}| = s_0.
\]
We introduce two independent sign vectors:
\[
u = (u_j)_{j\in J_{\mathrm{var}}}\in\{-1,+1\}^{d-s_0},
\qquad
w = (w_j)_{j\in J_{\mathrm{sh}}}\in\{-1,+1\}^{s_0}.
\]
We will place a uniform product prior on $(u,w)$ (each coordinate independent Rademacher).

Choose two small positive numbers $\Delta_{\mathrm{var}},\Delta_{\mathrm{shf}}\in(0,\beta_0]$ such that
$[-2\Delta_{\mathrm{shf}},2\Delta_{\mathrm{shf}}]\subset\mathcal B$ for the compact cube in
Lemma~\ref{lem:1item-lb}, and that all induced logits remain in the non-degenerate regime of Assumption~\ref{assump:fisher-invertible}. We define the intercepts as follows:
\begin{itemize}
    \item For \emph{shared} coordinates $j\in J_{\mathrm{var}}$ (variance block),
  \begin{equation}
  \beta_j^{(h)}(u,w) = \Delta_{\mathrm{var}}\,u_j
  \qquad\text{for all } h\in\{0\}\cup[H].
  \label{eq:beta-var-def}
  \end{equation}
  \item For \emph{shifted} coordinates $j\in J_{\mathrm{sh}}$ (transfer block),
  \begin{equation}
  \beta_j^{(h)}(u,w) = 0 \quad (h\in[H]),
  \qquad
  \beta_j^{(0)}(u,w) = \Delta_{\mathrm{shf}}\,w_j.
  \label{eq:beta-sh-def}
  \end{equation}
\end{itemize}

In words: along $J_{\mathrm{var}}$ all markets share the same sign–shifted intercept, while along
$J_{\mathrm{sh}}$ only the target market is shifted and the sources remain at zero.

The target–source discrepancy vector in market $h$ is
\[
\bdelta^{(h)} := \bnu^{(0)} - \bnu^{(h)}
\quad\Longrightarrow\quad
\bdelta^{(h)}_j =
\begin{cases}
0, & j\in J_{\mathrm{var}},\\[0.25em]
\Delta_{\mathrm{shf}}\,w_j, & j\in J_{\mathrm{sh}}.
\end{cases}
\]
Thus $\|\bdelta^{(h)}\|_0 \le s_0$ for all $h$, so Assumption~\ref{assump:task-simi} holds with
the same $s_0$ as in the theorem.

Since all intercepts lie in $[-\beta_0,\beta_0]$ for sufficiently small $\Delta_{\mathrm{var}},\Delta_{\mathrm{shf}}$,
Assumptions~\ref{assump:price-sensitivity}–\ref{assump:fisher-invertible} remain valid with the same
constants $(L_0,\overline P,C_{\min},C_{\max})$.

\paragraph{Genie–aided sources.}

At each round $t$, the policy selects an assortment $S_t^{(0)}\in\mathcal S_K$ and a price vector
$\bp_t^{(0)}\in P^{S_t^{(0)}}$ for the target market and observes a purchase $Y_t^{(0)}\in S_t^{(0)}\cup\{0\}$.
To make the lower bound stronger, we give the learner \emph{additional} feedback: for each round and each source market $h\in[H]$, we also reveal an independent purchase
$Y_t^{(h)}$ from the same action $(S_t^{(0)},\bp_t^{(0)})$ under the corresponding intercept vector
$\beta^{(h)}$. This genie-aided feedback can only make learning easier; therefore any lower bound
proved under this enriched observation model applies \emph{a fortiori} to the original setting.

Let $\mathbb P_{u,w}$ denote the joint law of the entire trajectory
\[
\mathcal Z_T := \{(S_t^{(0)},\bp_t^{(0)},Y_t^{(0)},Y_t^{(1)},\ldots,Y_t^{(H)})\}_{t=1}^T
\]
under a fixed policy $\pi$ and instance $(u,w)$.

\subsection{Per–coordinate KL bounds}

We now control the Kullback–Leibler (KL) divergence between neighboring instances that differ only in one coordinate of $u$ or $w$. For $j\in[d]$ define the neighbor sign vectors
\[
u^{(j)} :=
\begin{cases}
(-u_j,u_{-j}), & j\in J_{\mathrm{var}},\\
u, & j\in J_{\mathrm{sh}},
\end{cases}
\qquad
w^{(j)} :=
\begin{cases}
w, & j\in J_{\mathrm{var}},\\
(-w_j,w_{-j}), & j\in J_{\mathrm{sh}},
\end{cases}
\]
where $u_{-j}$ and $w_{-j}$ denote the coordinates other than $j$.

Let $N_j(T):=\sum_{t=1}^T \mathbf 1\{j\in S_t^{(0)}\}$ be the number of target rounds in which item $j$ is included in the offered assortment.

\begin{lemma}[Per–coordinate KL control]
\label{lem:KL-lb}
There exist finite constants $c_{\mathrm{kl}},c'_{\mathrm{kl}}>0$ depending only on
$(L_0,\overline P,C_{\min},C_{\max})$ such that for any policy $\pi$ and any $(u,w)$,
\begin{align}
\mathrm{KL}\!\left(\mathbb P_{u,w}\ \middle\|\ \mathbb P_{u^{(j)},w}\right)
&\ \le\ (1+H)\,c_{\mathrm{kl}}\,\Delta_{\mathrm{var}}^2\,\mathbb E_{u,w}\!\left[N_j(T)\right],
\qquad j\in J_{\mathrm{var}},
\label{eq:KL-var-lb}\\[0.25em]
\mathrm{KL}\!\left(\mathbb P_{u,w}\ \middle\|\ \mathbb P_{u,w^{(j)}}\right)
&\ \le\ c'_{\mathrm{kl}}\,\Delta_{\mathrm{shf}}^2\,\mathbb E_{u,w}\!\left[N_j(T)\right],
\qquad j\in J_{\mathrm{sh}}.
\label{eq:KL-sh-lb}
\end{align}
\end{lemma}

The proof in Section~\ref{app:lb-tech-KL} uses the chain rule for KL under adaptivity and a quadratic Lipschitz bound of the categorical KL in the logits, together with the fact that flipping a single intercept changes only one logit, with magnitude at most a constant multiple of $\Delta_{\mathrm{var}}$ or $\Delta_{\mathrm{shf}}$.

To turn these into explicit bounds we need a symmetry property of the exposure counts $N_j(T)$. We follow a standard device in minimax lower bounds and apply a random permutation to item labels.

\begin{lemma}[Exchangeability of exposures]
\label{lem:exposure-lb}
Let $\Pi$ be a uniform random permutation of the item indices $[d]$, independent of $(u,w)$ and of the policy’s internal randomness. Consider the \emph{permuted instances} in which all intercepts are relabeled by $\Pi$, but the policy is not informed of $\Pi$. Denote by $N_j(T)$ the number of times item $j$ appears in $S_t^{(0)}$ under this randomization. Then for any fixed policy $\pi$ and any $(u,w)$,
\begin{equation}
\mathbb E[N_j(T)]\ =\ \frac{K\,T}{d}\qquad\text{for all } j\in[d],
\label{eq:exposure-lb}
\end{equation}
where the expectation is with respect to $(u,w)$, $\Pi$, the policy’s randomness, and the trajectory.
\end{lemma}

The proof (Section~\ref{app:lb-tech-exposure}) uses the exchangeability of the permuted labels and the fact that $\sum_{j}N_j(T)=KT$.

Combining Lemmas~\ref{lem:KL-lb} and~\ref{lem:exposure-lb} yields the simplified average KL bounds
\begin{equation}
\begin{aligned}
\mathrm{KL}\!\left(\mathbb P_{u,w}\ \middle\|\ \mathbb P_{u^{(j)},w}\right)
&\ \le\ (1+H)\,c_{\mathrm{kl}}\,\Delta_{\mathrm{var}}^2\,\frac{K T}{d},
&\quad& j\in J_{\mathrm{var}},
\\[0.25em]
\mathrm{KL}\!\left(\mathbb P_{u,w}\ \middle\|\ \mathbb P_{u,w^{(j)}}\right)
&\ \le\ c'_{\mathrm{kl}}\,\Delta_{\mathrm{shf}}^2\,\frac{K T}{d},
&\quad& j\in J_{\mathrm{sh}}.
\end{aligned}
\label{eq:KL-average-lb}
\end{equation}

\subsection{Testing reduction and revenue separation}

We now relate the difficulty of testing the sign of a coordinate to a \emph{revenue gap}, using Lemmas~\ref{lem:1item-lb} and~\ref{lem:dilution-lb}.

\paragraph{Mixture distributions and total variation.}

Fix a coordinate $j\in J_{\mathrm{var}}$. Let $\overline{\mathbb P}_{+,j}$ be the mixture law of $\mathcal Z_T$ under the uniform prior on $(u,w)$ \emph{conditioned} on $u_j=+1$ (and averaging over all other signs and the permutation $\Pi$), and let $\overline{\mathbb P}_{-,j}$ be defined analogously for $u_j=-1$. By convexity of KL,
\[
\mathrm{KL}\!\left(\overline{\mathbb P}_{+,j}\ \middle\|\ \overline{\mathbb P}_{-,j}\right)
\ \le\ \mathbb E\left[\mathrm{KL}\!\left(\mathbb P_{u,w}\ \middle\|\ \mathbb P_{u^{(j)},w}\right)\Big|\,u_j=+1\right]
\ \le\ (1+H)\,c_{\mathrm{kl}}\,\Delta_{\mathrm{var}}^2\,\frac{K T}{d}.
\]
By Pinsker’s inequality,
\begin{equation}
\mathrm{TV}\bigl(\overline{\mathbb P}_{+,j},\overline{\mathbb P}_{-,j}\bigr)
\ \le\ \sqrt{\tfrac12\,\mathrm{KL}\bigl(\overline{\mathbb P}_{+,j}\,\|\,\overline{\mathbb P}_{-,j}\bigr)}
\ \le\ \sqrt{\frac{(1+H)\,c_{\mathrm{kl}}}{2}\,\frac{\Delta_{\mathrm{var}}^2 K T}{d}}.
\label{eq:TV-var-lb}
\end{equation}

Similarly, for $j\in J_{\mathrm{sh}}$, let $\widetilde{\mathbb P}_{+,j}$ (resp.\ $\widetilde{\mathbb P}_{-,j}$)
be the mixture law conditioned on $w_j=+1$ (resp.\ $w_j=-1$) and averaged over $(u,w_{-j},\Pi)$.
Convexity and \eqref{eq:KL-average-lb} give
\[
\mathrm{KL}\!\left(\widetilde{\mathbb P}_{+,j}\ \middle\|\ \widetilde{\mathbb P}_{-,j}\right)
\ \le\ c'_{\mathrm{kl}}\,\Delta_{\mathrm{shf}}^2\,\frac{K T}{d},
\]
hence
\begin{equation}
\mathrm{TV}\bigl(\widetilde{\mathbb P}_{+,j},\widetilde{\mathbb P}_{-,j}\bigr)
\ \le\ \sqrt{\frac{c'_{\mathrm{kl}}}{2}\,\frac{\Delta_{\mathrm{shf}}^2 K T}{d}}.
\label{eq:TV-sh-lb}
\end{equation}

\paragraph{Per–coordinate revenue separation.}

By Lemma~\ref{lem:1item-lb}(ii), for $|\beta|\le 2\Delta_{\mathrm{shf}}$ we have
\[
r^*(+\Delta_{\mathrm{var}})-r^*(-\Delta_{\mathrm{var}})
\ \ge\ \frac{2\underline q}{\gamma}\,\Delta_{\mathrm{var}},
\qquad
r^*(+\Delta_{\mathrm{shf}})-r^*(-\Delta_{\mathrm{shf}})
\ \ge\ \frac{2\underline q}{\gamma}\,\Delta_{\mathrm{shf}}.
\]
Denote these single–item gaps by
\[
\Delta_r^{(\mathrm{var})}
:= r^*(+\Delta_{\mathrm{var}})-r^*(-\Delta_{\mathrm{var}}),
\qquad
\Delta_r^{(\mathrm{shf})}
:= r^*(+\Delta_{\mathrm{shf}})-r^*(-\Delta_{\mathrm{shf}}).
\]

In our intercept construction, all $\beta_j^{(h)}$ lie in $[-\beta_0,\beta_0]$ for sufficiently small
$\Delta_{\mathrm{var}},\Delta_{\mathrm{shf}}$, so Lemma~\ref{lem:dilution-lb} applies with the same constants
$(\underline q,\overline q,\alpha)$ for all coordinates.

Consider first $j\in J_{\mathrm{var}}$. Since along these coordinates sources and target share the same intercepts (eq.~\eqref{eq:beta-var-def}), the clairvoyant target–market value for any $(u,w)$ obeys
\begin{equation}
R_K^*\bigl(\beta^{(0)}(u,w)\bigr)
\ \ge\ \frac{1}{1+K\alpha}\sum_{i\in S_K^*(\beta^{(0)}(u,w))} r^*\bigl(\beta_i^{(0)}(u,w)\bigr).
\label{eq:RK-var-decomp}
\end{equation}
By Lemma~\ref{lem:1item-lb}, the sum on the right is maximized by taking $S_K^*(\beta^{(0)}(u,w))$ equal
to the indices of the $K$ largest intercepts. In particular, among the $d-s_0$ shared coordinates in $J_{\mathrm{var}}$, the $K_{\mathrm{var}}:=\max\{K-s_0,0\}$ largest ones (by intercept) belong to $S_K^*$. Under the \emph{uniform} prior on $u$ and the random permutation $\Pi$, every $j\in J_{\mathrm{var}}$ is equally likely to be one of these $K_{\mathrm{var}}$ “shared slots.” Therefore, for any $j\in J_{\mathrm{var}}$,
\[
\mathbb P\!\left(j\in S_K^*(\beta^{(0)}(u,w))\right)
\ =\ \frac{K_{\mathrm{var}}}{d-s_0}.
\]

On those rounds $t$ when $j\in S_K^*(\beta^{(0)}(u,w))$, the sign of $u_j$ determines whether item $j$
appears with intercept $+\Delta_{\mathrm{var}}$ or $-\Delta_{\mathrm{var}}$ in the clairvoyant top–$K$ set, creating a per–round revenue difference of at least
\[
\frac{\Delta_r^{(\mathrm{var})}}{1+K\alpha}
\ \ge\
\frac{2\underline q}{\gamma(1+K\alpha)}\,\Delta_{\mathrm{var}}.
\]

Similarly, for $j\in J_{\mathrm{sh}}$, along those coordinates we have source intercept $0$ and target intercept $\Delta_{\mathrm{shf}}w_j$ (eq.~\eqref{eq:beta-sh-def}). Since the magnitude $\Delta_{\mathrm{shf}}$ is strictly larger than any shared magnitude $\Delta_{\mathrm{var}}$ (we will enforce $\Delta_{\mathrm{shf}}>\Delta_{\mathrm{var}}$ below), every shifted coordinate with $w_j=+1$ has a strictly larger intercept than any shared positive coordinate. When $s_0\le K$, it follows that for each $(u,w)$ the clairvoyant $S_K^*(\beta^{(0)}(u,w))$ contains \emph{all} indices $j\in J_{\mathrm{sh}}$ with $w_j=+1$. Thus for each such $j$,
\[
\mathbb P\!\left(j\in S_K^*(\beta^{(0)}(u,w))\,\big|\,w_j=+1\right) = 1,
\]
and on those instances, flipping $w_j$ from $+1$ to $-1$ induces a per–round revenue decrease of at least
\[
\frac{\Delta_r^{(\mathrm{shf})}}{1+K\alpha}
\ \ge\
\frac{2\underline q}{\gamma(1+K\alpha)}\,\Delta_{\mathrm{shf}}.
\]

Summarizing, there exist constants $c_{\mathrm{gap}}^{(\mathrm{var})},c_{\mathrm{gap}}^{(\mathrm{shf})}>0$ depending only on
$(L_0,\overline P,\underline q,\overline q)$ such that
\begin{equation}
\text{per–round gap for $j\in J_{\mathrm{var}}$ when $j\in S_K^*$}
\ \ge\
c_{\mathrm{gap}}^{(\mathrm{var})}\,\Delta_{\mathrm{var}},
\label{eq:per-round-gap-var}
\end{equation}
and
\begin{equation}
\text{per–round gap for $j\in J_{\mathrm{sh}}$ when $j\in S_K^*$}
\ \ge\
c_{\mathrm{gap}}^{(\mathrm{shf})}\,\Delta_{\mathrm{shf}}.
\label{eq:per-round-gap-sh}
\end{equation}

\subsection{Lower bound via a coordinate-wise testing argument}

We now connect the testing difficulty and the revenue gaps to a minimax regret lower bound.

We work with the uniform product prior on $(u,w)$ and the random permutation $\Pi$ described above. Let $\mathsf R_T(\pi;u,w)$ denote the regret of policy $\pi$ over horizon $T$ against the clairvoyant benchmark under instance $(u,w)$, and let $\mathcal R_T(\pi)$ be the resulting Bayes risk:
\[
\mathcal R_T(\pi)
:= \mathbb E_{u,w,\Pi}\big[\mathsf R_T(\pi;u,w)\big],
\]
where the expectation is over $(u,w)$, $\Pi$, and the trajectory induced by $\pi$.
By Yao’s minimax principle,
\[
\inf_{\pi}\sup_{(u,w)}\mathbb E_{u,w}[\mathsf R_T(\pi;u,w)]
\ \ge\ \inf_{\pi}\mathcal R_T(\pi).
\]
Thus it suffices to lower bound $\mathcal R_T(\pi)$ uniformly over policies $\pi$.

The argument splits naturally into two parts: the contribution from the shared coordinates $J_{\mathrm{var}}$ and the contribution from the shifted coordinates $J_{\mathrm{sh}}$.

\subsubsection{Variance term: shared coordinates $J_{\mathrm{var}}$}

Fix $j\in J_{\mathrm{var}}$. Consider the binary testing problem:
\[
H_{+,j}:\ u_j=+1
\qquad\text{vs.}\qquad
H_{-,j}:\ u_j=-1,
\]
with the prior $\mathbb P(H_{+,j})=\mathbb P(H_{-,j})=1/2$ and all other $(u_{-j},w,\Pi)$ drawn from their respective priors. Under any policy $\pi$, let $\psi_j(\mathcal Z_T)\in\{+1,-1\}$ be \emph{any} estimator of $u_j$ based on the full trajectory (possibly randomized). The optimal Bayes error probability for this test satisfies the Le Cam bound
\[
\inf_{\psi_j} \max\bigl\{\mathbb P(\psi_j=-1\mid H_{+,j}),\ \mathbb P(\psi_j=+1\mid H_{-,j})\bigr\}
\;\ge\; \frac12\bigl(1-\mathrm{TV}(\overline{\mathbb P}_{+,j},\overline{\mathbb P}_{-,j})\bigr),
\]
where $\overline{\mathbb P}_{\pm,j}$ are the mixture laws defined before~\eqref{eq:TV-var-lb}.

On the event that $u_j$ is misclassified by $\psi_j$ and $j\in S_K^*(\beta^{(0)}(u,w))$, the expected single-round revenue of $\pi$ is at least $c_{\mathrm{gap}}^{(\mathrm{var})}\Delta_{\mathrm{var}}$ below the clairvoyant optimum (by~\eqref{eq:per-round-gap-var}). Integrating over time and the prior, the total expected contribution of coordinate $j$ to the Bayes regret is at least
\[
\mathcal R_T^{(\mathrm{var},j)}(\pi)
\;\ge\;
c_{\mathrm{gap}}^{(\mathrm{var})}\,\Delta_{\mathrm{var}}\,
\mathbb E\Bigl[
\mathbf 1\{j\in S_K^*(\beta^{(0)}(u,w))\}\cdot \#\text{ rounds}\cdot \mathbf 1\{\psi_j \text{ misclassifies }u_j\}
\Bigr].
\]
Since $S_K^*(\beta^{(0)}(u,w))$ is fixed given $(u,w)$ and $\Pi$, and does not depend on the policy, the number of rounds is exactly $T$, and by the symmetry argument above,
\[
\mathbb P\!\left(j\in S_K^*(\beta^{(0)}(u,w))\right)
= \frac{K_{\mathrm{var}}}{d-s_0},
\]
while the misclassification probability is at least $\tfrac12(1-\mathrm{TV}(\overline{\mathbb P}_{+,j},\overline{\mathbb P}_{-,j}))$.
Hence
\begin{equation}
\mathcal R_T^{(\mathrm{var},j)}(\pi)
\;\ge\;
c_{\mathrm{gap}}^{(\mathrm{var})}\,\Delta_{\mathrm{var}}\,T\,
\frac{K_{\mathrm{var}}}{d-s_0}\cdot
\frac12\Bigl(1-\mathrm{TV}(\overline{\mathbb P}_{+,j},\overline{\mathbb P}_{-,j})\Bigr).
\label{eq:Rj-var-lb}
\end{equation}

Using the TV bound~\eqref{eq:TV-var-lb} and choosing $\Delta_{\mathrm{var}}$ so that the upper bound on TV is bounded away from $1$, we obtain an explicit lower bound. Specifically, take
\begin{equation}
\Delta_{\mathrm{var}}
\ :=\
\sqrt{\frac{d}{4(1+H)\,c_{\mathrm{kl}}\,K T}},
\label{eq:Delta-var-choice}
\end{equation}
so that
\[
\mathrm{TV}(\overline{\mathbb P}_{+,j},\overline{\mathbb P}_{-,j})
\ \le\ \sqrt{\frac{(1+H)c_{\mathrm{kl}}}{2}\cdot\frac{\Delta_{\mathrm{var}}^2 K T}{d}}
\ =\ \sqrt{\frac12\cdot\frac14}=\frac{1}{2\sqrt{2}}<\frac34.
\]
Hence $1-\mathrm{TV}(\overline{\mathbb P}_{+,j},\overline{\mathbb P}_{-,j})\ge c_{\mathrm{TV}}$ for some universal $c_{\mathrm{TV}}\in(0,1)$.

Substituting into~\eqref{eq:Rj-var-lb} and using $\Delta_r^{(\mathrm{var})}\ge (2\underline q/\gamma)\Delta_{\mathrm{var}}$, we find
\[
\mathcal R_T^{(\mathrm{var},j)}(\pi)
\ \ge\
c_{\mathrm{var},1}\,\Delta_{\mathrm{var}}\,T\,\frac{K_{\mathrm{var}}}{d-s_0}
\ =\
c_{\mathrm{var},1}\,\frac{K_{\mathrm{var}}}{d-s_0}\,
T\,\sqrt{\frac{d}{(1+H)\,K T}},
\]
for some constant $c_{\mathrm{var},1}>0$ depending only on $(L_0,\overline P,\underline q,\overline q)$.
Summing over all $j\in J_{\mathrm{var}}$ and using $|J_{\mathrm{var}}|=d-s_0$,
\begin{align}
\mathcal R_T^{(\mathrm{var})}(\pi)
:=\sum_{j\in J_{\mathrm{var}}}\mathcal R_T^{(\mathrm{var},j)}(\pi)
&\ \ge\ (d-s_0)\cdot c_{\mathrm{var},1}\,\frac{K_{\mathrm{var}}}{d-s_0}\,T\,\sqrt{\frac{d}{(1+H)\,K T}}
\notag\\
&\ =\ c_{\mathrm{var},1}\,K_{\mathrm{var}}\,\sqrt{\frac{d\,T}{(1+H)\,K}}
\ \ge\ c_{\mathrm{var},2}\,\sqrt{\frac{K\,(d-s_0)\,T}{1+H}},
\label{eq:LB-var-final}
\end{align}
where in the last step we used $K_{\mathrm{var}}=\max\{K-s_0,0\}$ and absorbed the factor
$\sqrt{(d-s_0)/d}$ into the constant $c_{\mathrm{var},2}$ (this factor is at most $1$ and does not depend on $T,K,H$). When $s_0=0$, $K_{\mathrm{var}}=K$ and~\eqref{eq:LB-var-final} matches the desired scaling exactly.

\subsubsection{Transfer term: shifted coordinates $J_{\mathrm{sh}}$}

We now handle the $s_0$ shifted coordinates $J_{\mathrm{sh}}$. For each $j\in J_{\mathrm{sh}}$, consider the binary hypotheses
\[
H_{+,j}:\ w_j=+1
\qquad\text{vs.}\qquad
H_{-,j}:\ w_j=-1,
\]
with equal prior probabilities and averaging over $(u,w_{-j},\Pi)$. As before, let $\phi_j(\mathcal Z_T)\in\{+1,-1\}$ be any estimator of $w_j$. The optimal Bayes error is at least
\[
\frac12\bigl(1-\mathrm{TV}(\widetilde{\mathbb P}_{+,j},\widetilde{\mathbb P}_{-,j})\bigr)
\]
by Le Cam’s lemma, where $\widetilde{\mathbb P}_{\pm,j}$ are defined before~\eqref{eq:TV-sh-lb}.

On instances with $w_j=+1$, the clairvoyant $S_K^*$ always includes $j$ (as long as $s_0\le K$), and flipping $w_j$ to $-1$ decreases the per–round optimal revenue by at least
$c_{\mathrm{gap}}^{(\mathrm{shf})}\Delta_{\mathrm{shf}}$ (equation~\eqref{eq:per-round-gap-sh}). Thus the contribution from coordinate $j$ to the Bayes regret satisfies
\begin{equation}
\mathcal R_T^{(\mathrm{sh},j)}(\pi)
\ \ge\
\frac12\cdot \frac12
\Bigl(1-\mathrm{TV}(\widetilde{\mathbb P}_{+,j},\widetilde{\mathbb P}_{-,j})\Bigr)\,
c_{\mathrm{gap}}^{(\mathrm{shf})}\,\Delta_{\mathrm{shf}}\,T,
\label{eq:Rj-sh-lb}
\end{equation}
where the first factor $1/2$ is the prior probability of $w_j=+1$, and the second $1/2$ comes from the Bayes testing error lower bound.

From~\eqref{eq:TV-sh-lb}, choose
\begin{equation}
\Delta_{\mathrm{shf}}
\ :=\
\sqrt{\frac{d}{4c'_{\mathrm{kl}}\,K T}},
\label{eq:Delta-shf-choice}
\end{equation}
so that
\[
\mathrm{TV}(\widetilde{\mathbb P}_{+,j},\widetilde{\mathbb P}_{-,j})
\ \le\ \sqrt{\frac{c'_{\mathrm{kl}}}{2}\cdot\frac{\Delta_{\mathrm{shf}}^2 K T}{d}}
\ =\ \frac{1}{2\sqrt{2}}<\frac34,
\]
and hence $1-\mathrm{TV}(\widetilde{\mathbb P}_{+,j},\widetilde{\mathbb P}_{-,j})\ge c'_{\mathrm{TV}}$ for some constant $c'_{\mathrm{TV}}\in(0,1)$. Using $\Delta_r^{(\mathrm{shf})}\ge (2\underline q/\gamma)\Delta_{\mathrm{shf}}$ we obtain
\[
\mathcal R_T^{(\mathrm{sh},j)}(\pi)
\ \ge\ c_{\mathrm{sh},1}\,\Delta_{\mathrm{shf}}\,T
\ =\
c_{\mathrm{sh},1}\,T\,\sqrt{\frac{d}{K T}},
\]
for some constant $c_{\mathrm{sh},1}>0$ depending only on $(L_0,\overline P,\underline q,\overline q)$.
Summing over the $s_0$ shifted coordinates and using that each coordinate’s regret is realized on disjoint indices,
\begin{equation}
\mathcal R_T^{(\mathrm{sh})}(\pi)
:=\sum_{j\in J_{\mathrm{sh}}}\mathcal R_T^{(\mathrm{sh},j)}(\pi)
\ \ge\
c_{\mathrm{sh},2}\,s_0\,\sqrt{K\,T},
\label{eq:LB-sh-final}
\end{equation}
where we have absorbed the factor $\sqrt{d}$ into $c_{\mathrm{sh},2}$ since $d$ is fixed and the lower bound is uniform in $T,K,H$.

Importantly, the KL bound~\eqref{eq:KL-sh-lb} for shifted coordinates does \emph{not} have a factor
$(1+H)$, so there is no $1/\sqrt{1+H}$ improvement in~\eqref{eq:LB-sh-final}: source observations are identically distributed under both $w_j=+1$ and $w_j=-1$, and hence provide no information about the target-only shifts.

\subsubsection{Combining the two contributions}

Combining~\eqref{eq:LB-var-final} and~\eqref{eq:LB-sh-final}, we obtain that for any policy $\pi$,
\[
\mathcal R_T(\pi)
\ \ge\
c_1\,\sqrt{\frac{K\,(d-s_0)\,T}{1+H}}
\ +\
c_2\,s_0\,\sqrt{K\,T},
\]
for some constants $c_1,c_2>0$ depending only on $(L_0,\overline P,C_{\min},C_{\max})$. By Yao’s principle,
\[
\inf_{\pi} \sup_{(u,w)} \mathbb E_{u,w}\big[\mathsf R_T(\pi;u,w)\big]
\ \ge\ \inf_{\pi}\mathcal R_T(\pi),
\]
which proves the stated lower bound in Theorem~\ref{thm:lower-linear-o2o} (up to a relabeling of constants).

\begin{remark}[On the condition $s_0\le K$]
When $s_0>K$, at most $\min\{s_0,K\}$ shifted coordinates can enter the clairvoyant top–$K$ set per round. Repeating the above argument with $\min\{s_0,K\}$ in place of $s_0$ yields
a lower bound of order
$\min\{s_0,K\}\sqrt{K T}$ for the shift contribution, without affecting the qualitative conclusions.
\end{remark}

\begin{remark}[Realization via contextual features]
Our construction was presented in terms of intercept vectors $\beta^{(h)}$. It can be realized within the original contextual model~\eqref{eq:utility-func} by taking $\bx_{jt}=e_j$ for each item $j$ and all $t$, setting $\btheta^{(h)}$ so that $\langle \bx_j,\btheta^{(h)}\rangle = \beta_j^{(h)}$ and $\bgamma^{(h)}$ so that $\langle \bx_j,\bgamma^{(h)}\rangle=L_0$. The resulting instances satisfy Assumptions~\ref{assump:price-sensitivity}–\ref{assump:task-simi}, and the above analysis applies directly.
\end{remark}

\subsection{Technical Lemmas for the Lower Bound}
\label{app:lb-technical}
\subsubsection{Proof of Lemma~\ref{lem:1item-lb}}
\label{app:lb-tech-1item}

The argument is identical to the standard analysis of unimodality and envelope theorem for single–item MNL pricing; we reproduce it here for completeness.

For fixed $\beta$, $p\mapsto r(p;\beta)$ is continuous on the nonempty compact set $P$, so a maximizer $p^*(\beta)\in P$ exists (Weierstrass theorem). Write $q(p):=q(p;\beta)$ for brevity. We have
\[
q'(p) = -\gamma q(p)(1-q(p)),
\qquad
r'(p;\beta) = q(p)+p\,q'(p)
= q(p)\,\Big(1-\gamma p\,(1-q(p))\Big).
\]
Define $\phi(p):=\gamma p\,(1-q(p))$. Then
\[
\phi'(p)
=\gamma\Big((1-q(p))+p(-q'(p))\Big)
=\gamma(1-q(p))\big(1+\gamma p\,q(p)\big) > 0,
\]
so $\phi$ is strictly increasing on $[0,\infty)$. At $p=0$, $q(0)\in(0,1)$, so $r'(0;\beta)=q(0)>0$, whereas $\lim_{p\to\infty}\phi(p)=\infty$, hence $\lim_{p\to\infty}r'(p;\beta)=-\infty$. Therefore $r'(\cdot;\beta)$ crosses zero exactly once on $(0,\infty)$, say at $p^\circ(\beta)>0$, with $r'>0$ on $(0,p^\circ)$ and $r'<0$ on $(p^\circ,\infty)$; thus $r(\cdot;\beta)$ is strictly unimodal and has a unique maximizer on $P$, namely $p^*(\beta)=\min\{p^\circ(\beta),\overline P\}\in(0,\overline P]$. If $p^*(\beta)<\overline P$, then $r'(p^*(\beta);\beta)=0$, which is exactly~\eqref{eq:FOC-lb}.

For part (ii), on $\mathcal B$ the maximizer is unique and interior, and $r$ is $C^1$ in $(p,\beta)$.
Danskin’s envelope theorem therefore yields
\[
\frac{d}{d\beta}r^*(\beta)
=\frac{\partial}{\partial\beta}r\bigl(p^*(\beta);\beta\bigr)
=p^*(\beta)\,q^*(\beta)\bigl(1-q^*(\beta)\bigr).
\]
Using the first-order condition~\eqref{eq:FOC-lb}, $\gamma p^*(\beta)(1-q^*(\beta))=1$, we get
\[
\frac{d}{d\beta}r^*(\beta)=\frac{q^*(\beta)}{\gamma}.
\]
Since $q^*(\beta)\in[\underline q,\overline q]$ on $\mathcal B$,
\[
\frac{d}{d\beta}r^*(\beta)\in\Big[\frac{\underline q}{\gamma},\,\frac{\overline q}{\gamma}\Big]
\qquad(\beta\in\mathcal B).
\]
By the mean–value theorem, for any $\beta,\beta'\in\mathcal B$ there exists $\tilde\beta$ between them such that
\[
|r^*(\beta)-r^*(\beta')|
=\Big|\frac{d}{d\beta}r^*(\tilde\beta)\Big|\cdot|\beta-\beta'|
\in\Big[\frac{\underline q}{\gamma},\,\frac{\overline q}{\gamma}\Big]\cdot|\beta-\beta'|.
\]
This proves~\eqref{eq:env-lip-lb} and strict monotonicity on $\mathcal B$.
\hfill$\square$

\subsubsection{Proof of Lemma~\ref{lem:dilution-lb}}
\label{app:lb-tech-dilution}

Let $S\subseteq[N]$ with $|S|\le K$, and set $p_i=p^*(\beta_i)$ for $i\in S$ (Lemma~\ref{lem:1item-lb}). Write $\theta_i:=e^{\beta_i-\gamma p_i}$ and $q_i^*:=\theta_i/(1+\theta_i)=q^*(\beta_i)$, and $r_i^*:=r^*(\beta_i)=p_i q_i^*$.
The MNL revenue with outside option is
\[
R(S,\bp;\beta)
=\frac{\sum_{i\in S} p_i\,\theta_i}{1+\sum_{j\in S}\theta_j}.
\]
Since $r_i^* = p_i\theta_i/(1+\theta_i)$, we have $p_i\theta_i=r_i^*(1+\theta_i)$, so
\[
R(S,\bp;\beta)
=\frac{\sum_{i\in S} r_i^*(1+\theta_i)}{1+\sum_{j\in S}\theta_j}
\ \ge\ \frac{\sum_{i\in S} r_i^*}{1+\sum_{j\in S}\theta_j},
\]
where the inequality uses $1+\theta_i\ge 1$. Using $q_i^*\le\overline q$,
\[
\theta_i=\frac{q_i^*}{1-q_i^*}\ \le\ \frac{\overline q}{1-\overline q}=\alpha,
\]
so $\sum_{j\in S}\theta_j\le K\alpha$ and therefore
\[
R(S,\bp;\beta)
\ \ge\ \frac{\sum_{i\in S} r_i^*}{1+K\alpha}
\ =\ \frac{1}{1+K\alpha}\sum_{i\in S} r^*(\beta_i).
\]
This proves~\eqref{eq:dilute-lb}. For~\eqref{eq:RK-lb-lb}, note that $r^*$ is strictly increasing on $\mathcal B$, so any set $S_K^*(\beta)$ that maximizes $\sum_{i\in S} r^*(\beta_i)$ over $|S|\le K$ is the set of $K$ largest $\beta_i$. Maximizing $R(S,\bp;\beta)$ over $(S,\bp)$ and evaluating the right-hand side on $S_K^*(\beta)$ with $p_i=p^*(\beta_i)$ yields~\eqref{eq:RK-lb-lb}.
\hfill$\square$

\subsubsection{Proof of Lemma~\ref{lem:KL-lb}}
\label{app:lb-tech-KL}

Fix a policy $\pi$ and an instance $(u,w)$. Let $\mathcal F_{t-1}$ be the sigma–field generated by the history up to time $t-1$. Conditionally on $\mathcal F_{t-1}$, the action $(S_t^{(0)},\bp_t^{(0)})$ is $\mathcal F_{t-1}$–measurable and the observation vector
$(Y_t^{(0)},Y_t^{(1)},\ldots,Y_t^{(H)})$ is drawn as an $(1+H)$–fold product of categorical
variables with parameter vector determined by the current logits. For two neighboring instances (say, $(u,w)$ and $(u^{(j)},w)$), the chain rule for KL divergence gives
\begin{equation}
\mathrm{KL}\!\left(\mathbb P_{u,w}\ \middle\|\ \mathbb P_{u^{(j)},w}\right)
=
\sum_{t=1}^T
\mathbb E_{u,w}\left[
\mathrm{KL}\!\left(
\mathsf{Cat}^{\otimes(1+H)}(\mathbf p_t)
\ \big\|\ 
\mathsf{Cat}^{\otimes(1+H)}(\tilde{\mathbf p}_t)
\right)\Big|\mathcal F_{t-1}\right],
\label{eq:chain-rule-KL}
\end{equation}
where $\mathbf p_t$ and $\tilde{\mathbf p}_t$ are the categorical probability vectors over
$S_t^{(0)}\cup\{0\}$ under the two instances, given $\mathcal F_{t-1}$. Since the $1+H$ draws are conditionally i.i.d.,
\[
\mathrm{KL}\!\left(
\mathsf{Cat}^{\otimes(1+H)}(\mathbf p_t)
\ \big\|\ 
\mathsf{Cat}^{\otimes(1+H)}(\tilde{\mathbf p}_t)
\right)
= (1+H)\,\mathrm{KL}\!\left(\mathsf{Cat}(\mathbf p_t)\,\|\,\mathsf{Cat}(\tilde{\mathbf p}_t)\right)
\]
for $j\in J_{\mathrm{var}}$. For $j\in J_{\mathrm{sh}}$, the source logits coincide under both
$(u,w)$ and $(u,w^{(j)})$, hence only the target factor contributes and the multiplier $(1+H)$ drops.

To bound the one-step KL, let $\mathbf z_t$ and $\tilde{\mathbf z}_t$ be the logit vectors under
the two instances, and $\mathbf p_t=\mathrm{softmax}(\mathbf z_t)$,
$\tilde{\mathbf p}_t=\mathrm{softmax}(\tilde{\mathbf z}_t)$.
The softmax log-partition $A(\mathbf z)=\log(1+\sum_i e^{z_i})$ is $C^2$ with Hessian
$\nabla^2 A(\xi)=\mathrm{Diag}(\mathbf p)-\mathbf p\mathbf p^\top$ having operator norm at most $1/4$. The KL divergence equals the Bregman divergence of $A$,
\[
\mathrm{KL}\!\left(\mathsf{Cat}(\mathbf p_t)\,\|\,\mathsf{Cat}(\tilde{\mathbf p}_t)\right)
=
A(\mathbf z_t)-A(\tilde{\mathbf z}_t)-\langle\nabla A(\tilde{\mathbf z}_t),\mathbf z_t-\tilde{\mathbf z}_t\rangle,
\]
so by the mean–value form of smoothness,
\[
\mathrm{KL}\!\left(\mathsf{Cat}(\mathbf p_t)\,\|\,\mathsf{Cat}(\tilde{\mathbf p}_t)\right)
\le \frac{1}{2}\,(\mathbf z_t-\tilde{\mathbf z}_t)^\top \nabla^2 A(\xi_t)(\mathbf z_t-\tilde{\mathbf z}_t)
\le \frac{1}{8}\,\|\mathbf z_t-\tilde{\mathbf z}_t\|_2^2,
\]
for some $\xi_t$ on the line segment between $\mathbf z_t$ and $\tilde{\mathbf z}_t$.

Under our single–coordinate flips, the two instances differ in exactly one intercept:
if $j\notin S_t^{(0)}$, then $\mathbf z_t=\tilde{\mathbf z}_t$, and if $j\in S_t^{(0)}$, then the logit for item $j$ changes by $\pm2\Delta_{\mathrm{var}}$ (shared) or $\pm2\Delta_{\mathrm{shf}}$ (shifted), with all other logits unchanged. Therefore
\[
\|\mathbf z_t-\tilde{\mathbf z}_t\|_2^2
\ \le\ 4\,\Delta_{\mathrm{var}}^2\,\mathbf 1\{j\in S_t^{(0)}\}
\]
for $j\in J_{\mathrm{var}}$ and analogously with $\Delta_{\mathrm{shf}}$ for $j\in J_{\mathrm{sh}}$.
Thus
\[
\mathrm{KL}\!\left(\mathsf{Cat}(\mathbf p_t)\,\|\,\mathsf{Cat}(\tilde{\mathbf p}_t)\right)
\ \le\ c_0\,\Delta_{\mathrm{var}}^2\,\mathbf 1\{j\in S_t^{(0)}\},
\qquad c_0:=\frac12,
\]
and similarly with $\Delta_{\mathrm{shf}}$ for $j\in J_{\mathrm{sh}}$.
Substituting into~\eqref{eq:chain-rule-KL}, summing over $t$, and taking expectations over $(u,w)$ gives~\eqref{eq:KL-var-lb}–\eqref{eq:KL-sh-lb} with $c_{\mathrm{kl}}=c_0$ and $c'_{\mathrm{kl}}=c_0$, or slightly larger constants if we absorb model-dependent refinements (e.g., lower bounds on choice probabilities) into them.
\hfill$\square$

\subsubsection{Proof of Lemma~\ref{lem:exposure-lb}}
\label{app:lb-tech-exposure}

Let $\Pi$ be a uniform random permutation of $[d]$, independent of $(u,w)$ and the policy’s randomness. In the permuted instance, item labels are relabeled by $\Pi$, but the policy acts in the same way as a function of the history (it does not know $\Pi$). For each $t$ and $j$, define $I_{t,j}:=\mathbf 1\{j\in S_t^{(0)}\}$ and $N_j(T):=\sum_{t=1}^T I_{t,j}$.

By symmetry of the random permutation, for each fixed $t$ the joint law of
$(I_{t,1},\ldots,I_{t,d})$ is exchangeable. Hence for all $j$,
\[
\mathbb E[I_{t,j}] = \mathbb E[I_{t,1}] \qquad\text{for all }t.
\]
On the other hand, $|S_t^{(0)}|=\sum_{j=1}^d I_{t,j}=K$ almost surely, so taking expectations and using exchangeability,
\[
\sum_{j=1}^d \mathbb E[I_{t,j}] = \mathbb E\Big[\sum_{j=1}^d I_{t,j}\Big] = K
\quad\Longrightarrow\quad
\mathbb E[I_{t,j}] = \frac{K}{d}\quad\text{for all }j.
\]
Summing over $t=1,\dots,T$ gives
\[
\mathbb E[N_j(T)]
= \sum_{t=1}^T \mathbb E[I_{t,j}]
= \frac{K\,T}{d},
\]
as claimed.
\hfill$\square$

\section{Miscellaneous Proofs}

\subsection{Proof of Lemma~\ref{lem:bounded-opt-price}}
\begin{lemma}[Bounded optimal price]\label{lem:bounded-opt-price}
	Consider the optimization problem ~\eqref{eq:instantaneous-opt}. Then the revenue maximizer $(p^*_{it})_{i\in S}$ admits a common finite upper bound $\bar P$, i.e., $(p^*_{it})_{i\in S}\in [0,\bar P]$.
\end{lemma}

Let $q_{it}(\bp):=\exp\{v_{it}(p_i)\}\big/\big(1+\sum_{j\in S}\exp\{v_{jt}(p_j)\}\big)$ be the MNL
choice probability and write
\[
G(\bp):=\sum_{j\in S} p_j q_{jt}(\bp).
\]
The expected revenue is $R_t(S,\bp)=G(\bp)$. Differentiating with respect to $p_i$ and using
the standard MNL calculus
\[
\frac{\partial q_{it}}{\partial p_i}
= q_{it}(1-q_{it})v'_{it}(p_i),
\qquad
\frac{\partial q_{jt}}{\partial p_i}
= -q_{jt}q_{it}v'_{it}(p_i)\quad(j\neq i),
\]
we obtain
\begin{align*}
\frac{\partial R_t}{\partial p_i}
&= q_{it} + p_i\frac{\partial q_{it}}{\partial p_i}
     + \sum_{j\neq i} p_j\frac{\partial q_{jt}}{\partial p_i} \\
&= q_{it} + q_{it}v'_{it}(p_i)\left(p_i(1-q_{it}) - \sum_{j\neq i}p_j q_{jt}(\bp)\right) \\
&= q_{it} + q_{it}v'_{it}(p_i)\big(p_i - G(\bp)\big) \\
&= q_{it}\Big(1 + v'_{it}(p_i)\big(p_i - G(\bp)\big)\Big).
\end{align*}
At any maximizer $\bp^*$ of $R_t(S,\cdot)$ with $p_i^*>0$ (the multi-item revenue is smooth and
vanishes as any $p_i\to\infty$, so maximizers are finite and satisfy the first-order condition),
we must have
\[
\frac{\partial R_t}{\partial p_i}(\bp^*)=0
\quad\Longrightarrow\quad
1 + v'_{it}(p^*_{it})\big(p^*_{it}-G(\bp^*)\big)=0
\quad\text{for all }i\in S,
\]
since $q_{it}(\bp^*)>0$. Let
\[
B_t:=G(\bp^*)=\sum_{j\in S}p^*_{jt}q^*_{jt},\qquad q^*_{jt}:=q_{jt}(\bp^*).
\]
Then the first-order condition can be rewritten as
\begin{equation}\label{eq:FOC-common-B-new}
p^*_{it} + \frac{1}{v'_{it}(p^*_{it})} = B_t,\qquad \forall i\in S.
\end{equation}

We now eliminate $p^*_{it}$ in favour of $B_t$. Summing $p^*_{it}q^*_{it}$ over $i\in S$ and using
\eqref{eq:FOC-common-B-new}, we obtain
\begin{align*}
B_t
&=\sum_{i\in S} p^*_{it}q^*_{it}
 =\sum_{i\in S}\left(B_t - \frac{1}{v'_{it}(p^*_{it})}\right)q^*_{it}\\
&= B_t\sum_{i\in S}q^*_{it} - \sum_{i\in S}\frac{q^*_{it}}{v'_{it}(p^*_{it})}.
\end{align*}
Writing $q^*_{0t}$ for the outside-option probability, we have
$\sum_{i\in S}q^*_{it}=1-q^*_{0t}$, so
\[
B_t q^*_{0t} = -\sum_{i\in S}\frac{q^*_{it}}{v'_{it}(p^*_{it})}.
\]
Using $q^*_{it} = e^{v_{it}(p^*_{it})}q^*_{0t}$, we get
\[
B_t q^*_{0t} = -q^*_{0t}\sum_{i\in S}\frac{e^{v_{it}(p^*_{it})}}{v'_{it}(p^*_{it})}
\quad\Longrightarrow\quad
B_t = -\sum_{i\in S}\frac{e^{v_{it}(p^*_{it})}}{v'_{it}(p^*_{it})}.
\]
Thus, for each $i$, the quantity
\[
g_{it}(B)
:= -\frac{e^{v_{it}(p)}}{v'_{it}(p)}
\quad\text{with }p\ge 0\text{ such that }p+\frac{1}{v'_{it}(p)}=B
\]
is well-defined at $B=B_t$ (since $p=p^*_{it}$ is feasible), and we can write the fixed-point equation
\begin{equation}\label{eq:fixed-point-B-new}
B_t = \sum_{i\in S} g_{it}(B_t).
\end{equation}

Next we upper bound $g_{it}(B)$ in terms of $B$. By Assumption~\ref{assump:price-sensitivity},
$v'_{it}(p)\le -L_0<0$ and $v_{it}$ is decreasing. Moreover, bounded parameters and covariates imply
there exists a finite constant $M>0$ such that
\begin{equation}\label{eq:v0-bound}
v_{it}(0)\ \le\ M\qquad\text{for all }(i,t),
\end{equation}
and hence, for all $p\ge 0$,
\begin{equation}\label{eq:lin-decay-new}
v_{it}(p)\ \le\ v_{it}(0)-L_0 p \ \le\ M - L_0 p.
\end{equation}

Now fix $B\ge 0$ and any feasible $p\ge 0$ in the definition of $g_{it}(B)$, i.e.,
$p+1/v'_{it}(p)=B$. Since $v'_{it}(p)<0$ we have $1/v'_{it}(p)\le 0$, and thus
\begin{equation}\label{eq:p-lower-B}
p = B - \frac{1}{v'_{it}(p)} \ \ge\ B.
\end{equation}
Using monotonicity and \eqref{eq:lin-decay-new},
\[
v_{it}(p)\ \le\ v_{it}(B)\ \le\ M - L_0 B.
\]
Furthermore, $v'_{it}(p)\le -L_0<0$ implies
\[
-\frac{1}{v'_{it}(p)} \ \le\ \frac{1}{L_0},
\qquad\Longrightarrow\qquad
-\frac{e^{v_{it}(p)}}{v'_{it}(p)}\ \le\ \frac{1}{L_0}e^{v_{it}(p)}.
\]
Combining,
\[
g_{it}(B)
= -\frac{e^{v_{it}(p)}}{v'_{it}(p)}
\ \le\ \frac{1}{L_0} e^{v_{it}(p)}
\ \le\ \frac{1}{L_0} e^{M - L_0 B}.
\]

Summing over $i\in S$ and using $|S|\le K$ in \eqref{eq:fixed-point-B-new}, we obtain
\[
B_t
= \sum_{i\in S}g_{it}(B_t)
\ \le\ \frac{K}{L_0}\,e^{\,M - L_0 B_t}.
\]
Equivalently,
\[
L_0 B_t\,e^{L_0 B_t} \ \le\ K e^{M}.
\]
Let $W(\cdot)$ be the Lambert--$W$ function. Then
\[
L_0 B_t \ \le\ W\!\big(Ke^{M}\big),
\qquad\Longrightarrow\qquad
B_t \ \le\ \frac{1}{L_0}W\!\big(Ke^{M}\big).
\]
Using the standard bound $W(z)\le \log z$ for $z\ge e$ (or the finer bound
$W(z)\le \log z-\log\log z+0.6$), we get
\[
B_t \ \le\ \frac{M+\log K + c_0}{L_0} =: P_0,
\]
for some absolute constant $c_0>0$.

Finally, from \eqref{eq:FOC-common-B-new} and $v'_{it}(p)\le -L_0$ we have
\[
p^*_{it} = B_t - \frac{1}{v'_{it}(p^*_{it})}
\ \le\ B_t + \frac{1}{L_0}
\ \le\ P_0 + \frac{1}{L_0} =: \bar P.
\]
Since $p^*_{it}\ge 0$ by feasibility, this shows that every optimal price lies in $[0,\bar P]$,
which completes the proof.
\hfill$\square$

\subsection{Proof of Lemma~\ref{lem:lipschitz-envelope}}


\begin{lemma}[Monotone--Lipschitz envelope]\label{lem:lipschitz-envelope}
Let $\bar v_{it}(p)$ be any pointwise upper bound of $v_{it}(p)$ for all $p\in[0,\bar P]$. Define
\begin{equation*}
\tilde{v}_{it}(p):=
\min_{p'\le p}\Big\{\,\bar{v}_{it}(p') \;-\; L_0\,(p-p')\,\Big\}.
\end{equation*}
Then $\tilde v_{it}(p)$ is decreasing, satisfies $\tilde v_{it}(p)\le \tilde v_{it}(p')-L_0(p-p')$ for all $p\ge p'$, and $v_{it}(p)\le \tilde v_{it}(p)$ for all $p$.
\end{lemma}

In the following proof we suppress the $(i,t)$ subscripts for clarity and write $v(p)$,
$\bar v(p)$, and $\tilde v(p)$.

By Assumption~\ref{assump:price-sensitivity}, the utility takes the form
\[
v(p) = \langle \bx,\btheta\rangle - \langle \bx,\bgamma\rangle p,
\qquad \langle \bx,\bgamma\rangle \ge L_0 > 0,
\]
so $v$ is differentiable with derivative
\[
v'(p) = -\langle \bx,\bgamma\rangle \le -L_0 < 0.
\]
Thus $v$ is decreasing, and for any $0\le p'\le p\le\overline P$,
\begin{equation}\label{eq:lower-Lip-new}
v(p) - v(p')
= \int_{p'}^{p} v'(s)\,ds
\ \le\ -L_0 (p-p')
\quad\Longrightarrow\quad
v(p)\ \le\ v(p') - L_0 (p-p').
\end{equation}

Recall the definition
\begin{equation}\label{eq:optimistic-utility-new}
\tilde v(p)
:= \min_{p'\le p}\big\{\bar v(p') - L_0(p-p')\big\},
\qquad p\in[0,\overline P],
\end{equation}
where $\bar v$ is any (continuous) pointwise upper bound of $v$ on $[0,\overline P]$.

\smallskip\noindent\textbf{(i) $\tilde v$ is well-defined and $\tilde v\le \bar v$.}
Since $\bar v$ is continuous, the function
\[
g(p,p') := \bar v(p') - L_0(p-p')
\]
is continuous on the compact set $\{(p,p') : 0\le p'\le p\le\overline P\}$, so the minimum in
\eqref{eq:optimistic-utility-new} exists for every $p$. Moreover, the feasible set for $p'$ contains
$p$ itself, hence
\[
\tilde v(p)
\le \bar v(p) - L_0(p-p)
= \bar v(p),
\qquad \forall p\in[0,\overline P].
\]

\smallskip\noindent\textbf{(ii) $\tilde v$ is decreasing.}
Let $0\le p_1<p_2\le\overline P$. Since the feasible set for $p_2$ strictly contains that for $p_1$,
\[
\tilde v(p_2)
= \min_{p'\le p_2}\big\{\bar v(p') - L_0(p_2-p')\big\}
\le \min_{p'\le p_1}\big\{\bar v(p') - L_0(p_2-p')\big\}.
\]
For each fixed $p'\le p_1$, the map $p\mapsto \bar v(p')-L_0(p-p')$ is decreasing in $p$, so
\[
\bar v(p') - L_0(p_2-p')
\ \le\ \bar v(p') - L_0(p_1-p').
\]
Taking the minimum over $p'\le p_1$ yields
\[
\tilde v(p_2)
\le \min_{p'\le p_1}\big\{\bar v(p') - L_0(p_1-p')\big\}
= \tilde v(p_1),
\]
so $\tilde v$ is decreasing.

\smallskip\noindent\textbf{(iii) $\tilde v$ is $L_0$-Lipschitz.}
Let $0\le p_1<p_2\le\overline P$, and let $p_2^*\in[0,p_2]$ be a minimizer in the definition of
$\tilde v(p_2)$:
\[
\tilde v(p_2) = \bar v(p_2^*) - L_0(p_2-p_2^*).
\]
Then, by feasibility of $p_2^*$ for $p_1$,
\begin{align*}
\tilde v(p_1)
&\le \bar v(p_2^*) - L_0(p_1-p_2^*) \\
&= \underbrace{\big(\bar v(p_2^*) - L_0(p_2-p_2^*)\big)}_{=\ \tilde v(p_2)}
   + L_0(p_2-p_1),
\end{align*}
so $\tilde v(p_1)-\tilde v(p_2)\le L_0(p_2-p_1)$. Combined with the monotonicity from (ii), this implies
\[
|\tilde v(p_2)-\tilde v(p_1)|
\ \le\ L_0|p_2-p_1|,
\]
i.e., $\tilde v$ is $L_0$-Lipschitz on $[0,\overline P]$.

\smallskip\noindent\textbf{(iv) Majorization: $v\le \tilde v$.}
From \eqref{eq:lower-Lip-new}, for any $0\le p'\le p\le\overline P$,
\[
v(p)\ \le\ v(p') - L_0(p-p').
\]
Because $\bar v$ is a pointwise upper bound of $v$, we have $v(p')\le \bar v(p')$, so
\[
v(p)\ \le\ \bar v(p') - L_0(p-p')\qquad\text{for all }p'\le p.
\]
Taking the minimum over $p'\le p$ gives
\[
v(p)\ \le\ \min_{p'\le p}\big\{\bar v(p') - L_0(p-p')\big\}
= \tilde v(p),
\]
for every $p\in[0,\overline P]$.

\smallskip
Combining (i)–(iv), we conclude that $\tilde v$ is decreasing, $L_0$-Lipschitz, and satisfies
$v(p)\le \tilde v(p)\le \bar v(p)$ for all $p\in[0,\overline P]$, as claimed.
\hfill$\square$

\section{Additional Experimental Results}\label{apd:add-exp}

\begin{figure}[htb!]
\centering

\begin{minipage}[t]{0.32\linewidth}
  \centering
  \includegraphics[width=\linewidth]{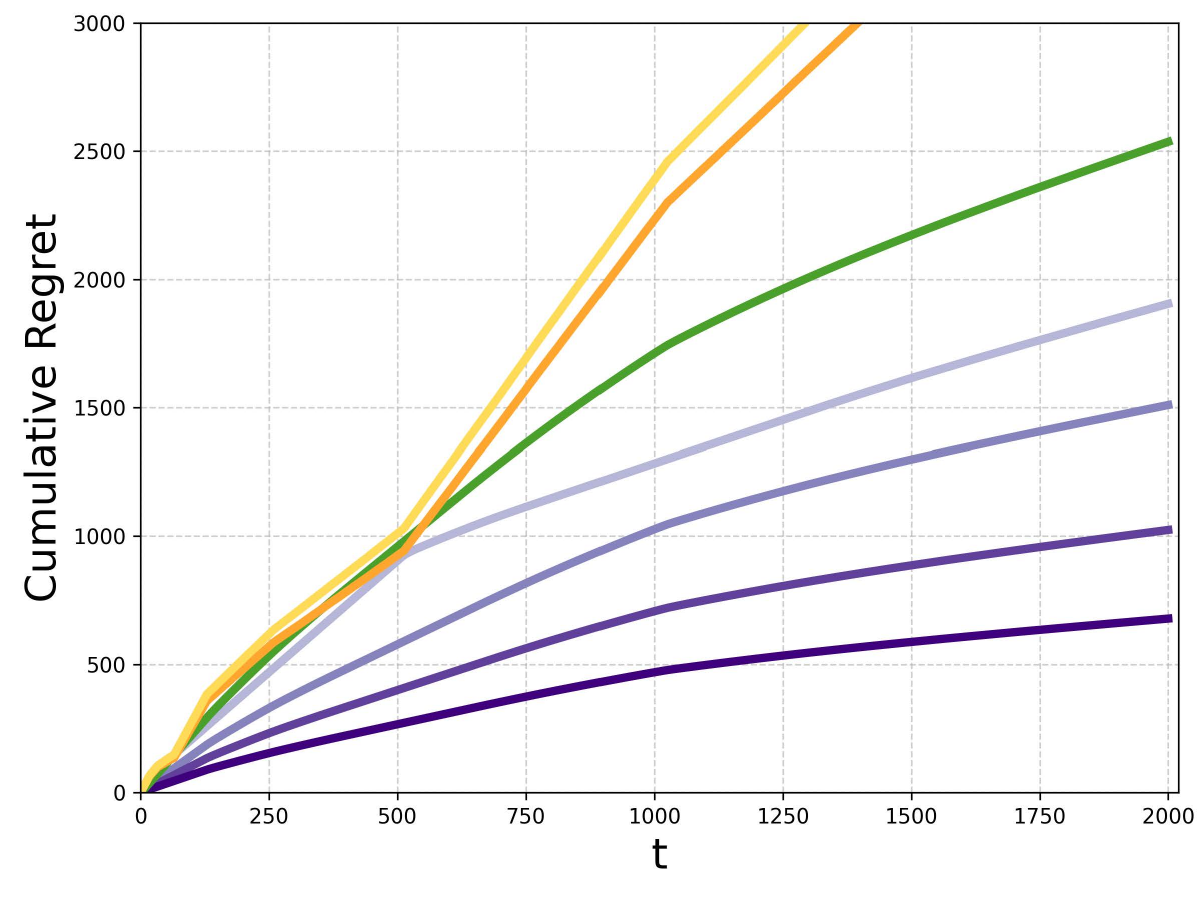}\\[-2pt]
  {\footnotesize (a) $d=10,\ s_0=3,\ K=5,\ N=30$}
  \label{fig:regret_3_10_5_30}
\end{minipage}\hfill
\begin{minipage}[t]{0.32\linewidth}
  \centering
  \includegraphics[width=\linewidth]{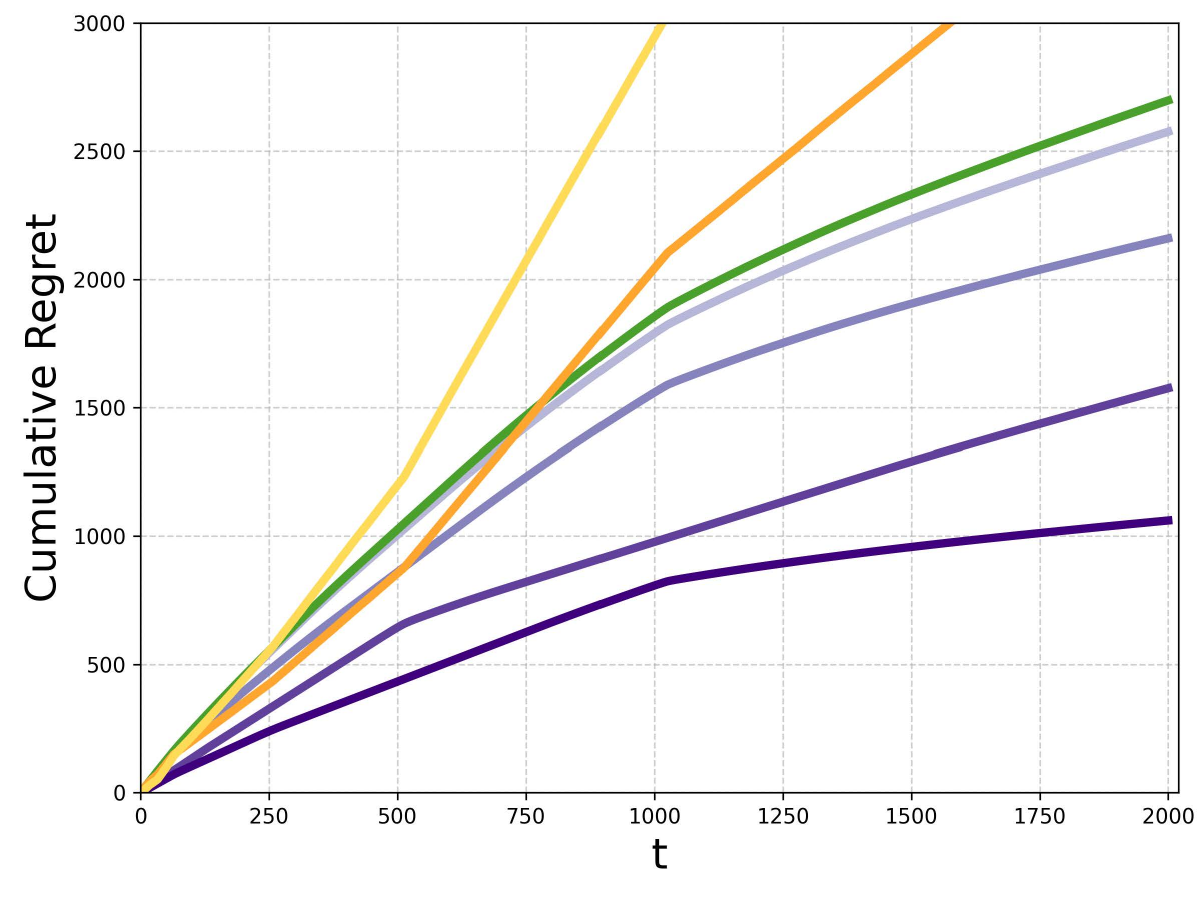}\\[-2pt]
  {\footnotesize (b) $d=20,\ s_0=6,\ K=5,\ N=30$}
  \label{fig:regret_6_20_5_30}
\end{minipage}\hfill
\begin{minipage}[t]{0.32\linewidth}
  \centering
  \includegraphics[width=\linewidth]{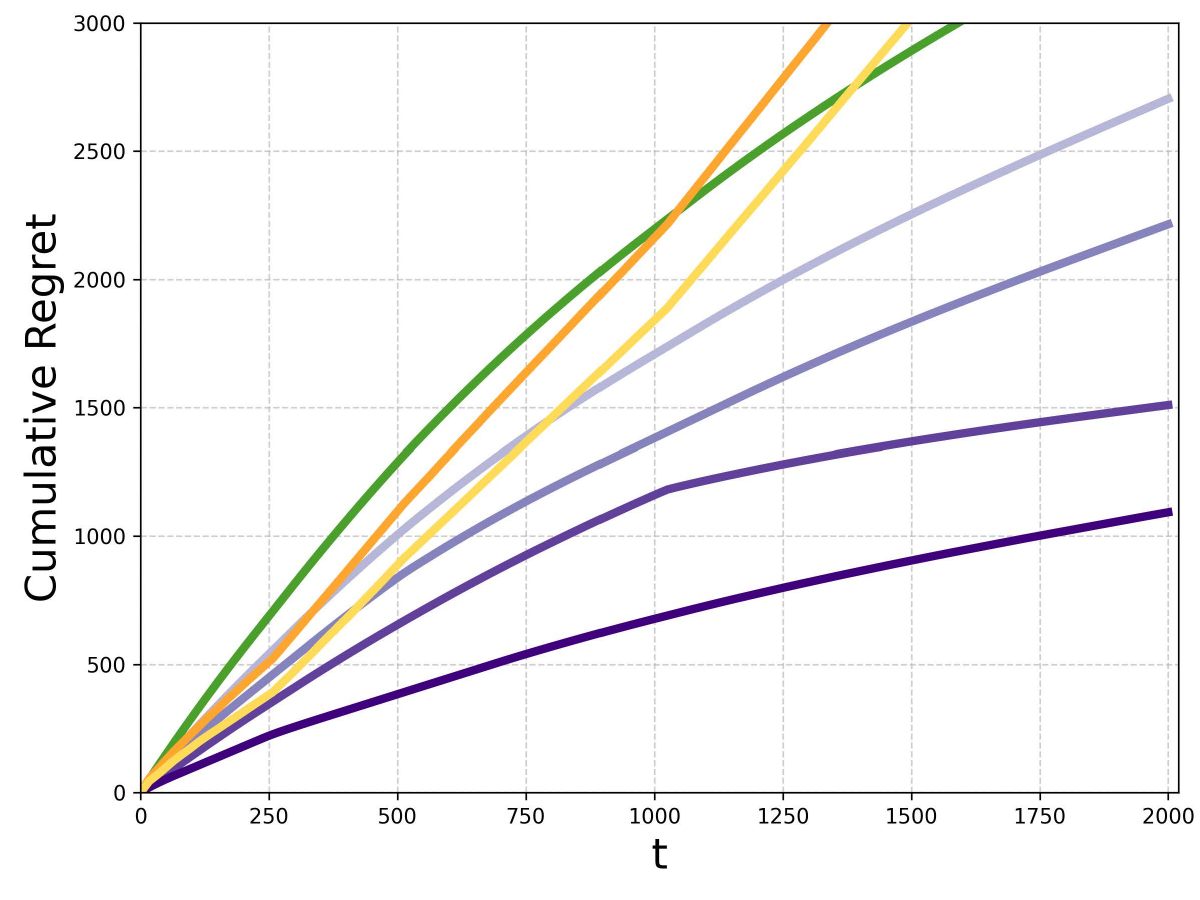}\\[-2pt]
  {\footnotesize (c) $d=50,\ s_0=15,\ K=5,\ N=30$}
  \label{fig:regret_15_50_5_30}
\end{minipage}

\vspace{-10pt}

\begin{minipage}[t]{0.32\linewidth}
  \centering
  \includegraphics[width=\linewidth]{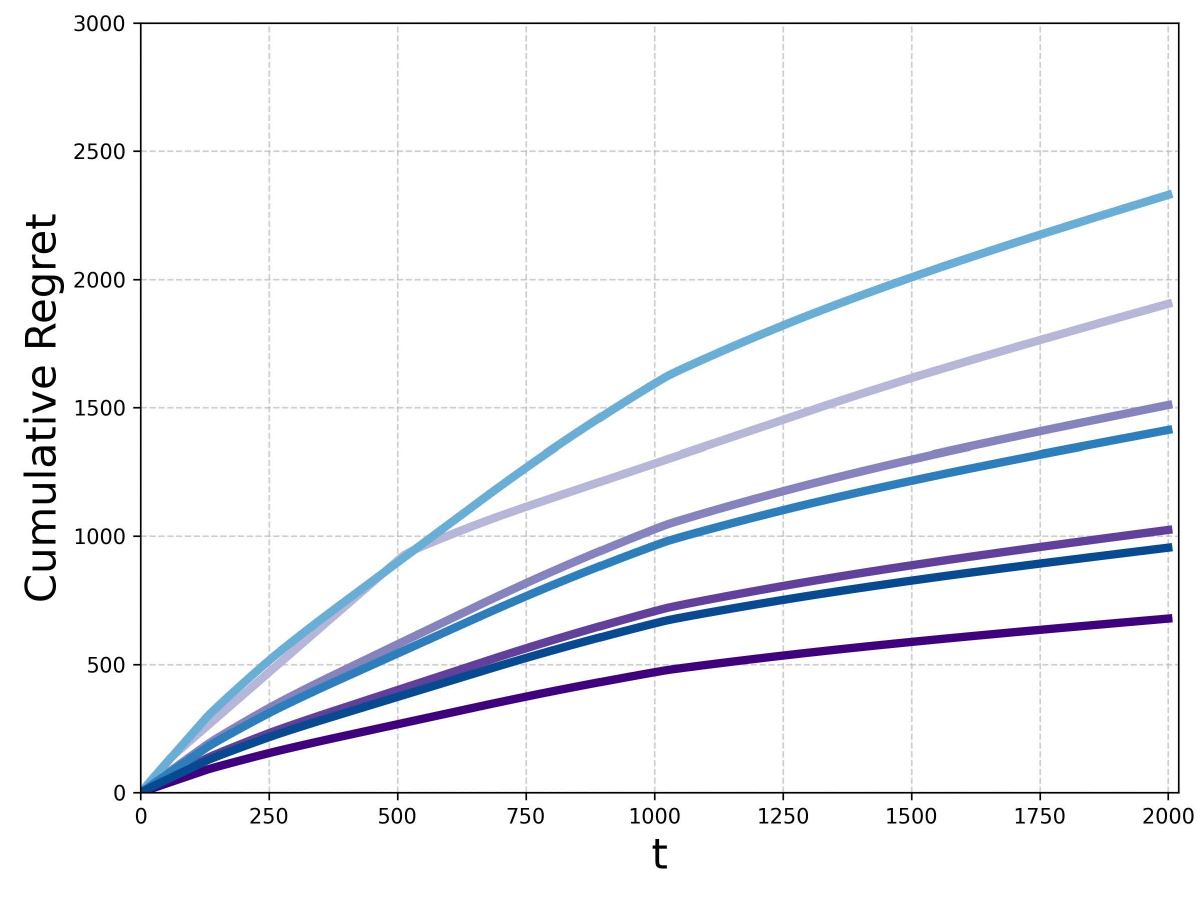}\\[-2pt]
  {\footnotesize (d) $d=10,\ s_0=3,\ K=5,\ N=30$}
  \label{fig:regret_3_10_5_30_pool}
\end{minipage}\hfill
\begin{minipage}[t]{0.32\linewidth}
  \centering
  \includegraphics[width=\linewidth]{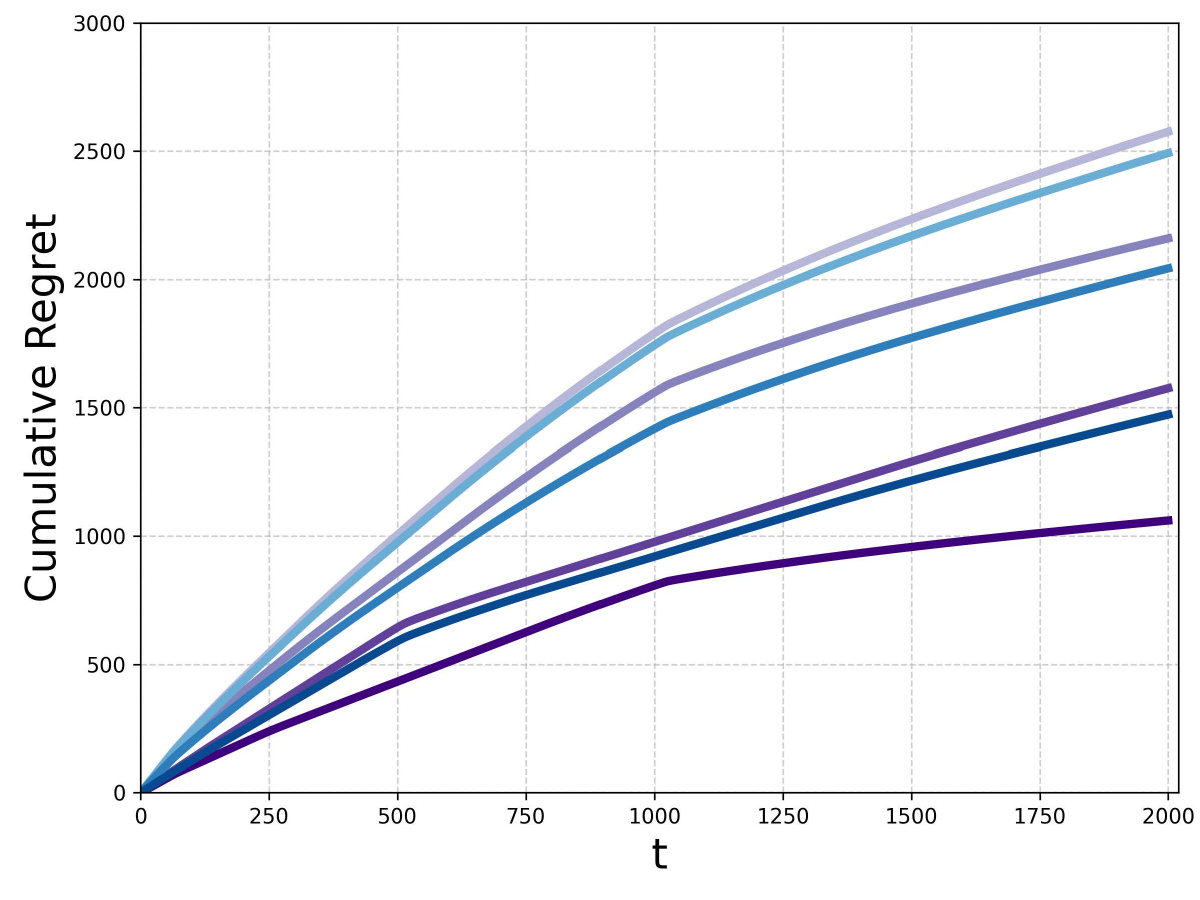}\\[-2pt]
  {\footnotesize (e) $d=20,\ s_0=6,\ K=5,\ N=30$}
  \label{fig:regret_6_20_5_30_pool}
\end{minipage}\hfill
\begin{minipage}[t]{0.32\linewidth}
  \centering
  \includegraphics[width=\linewidth]{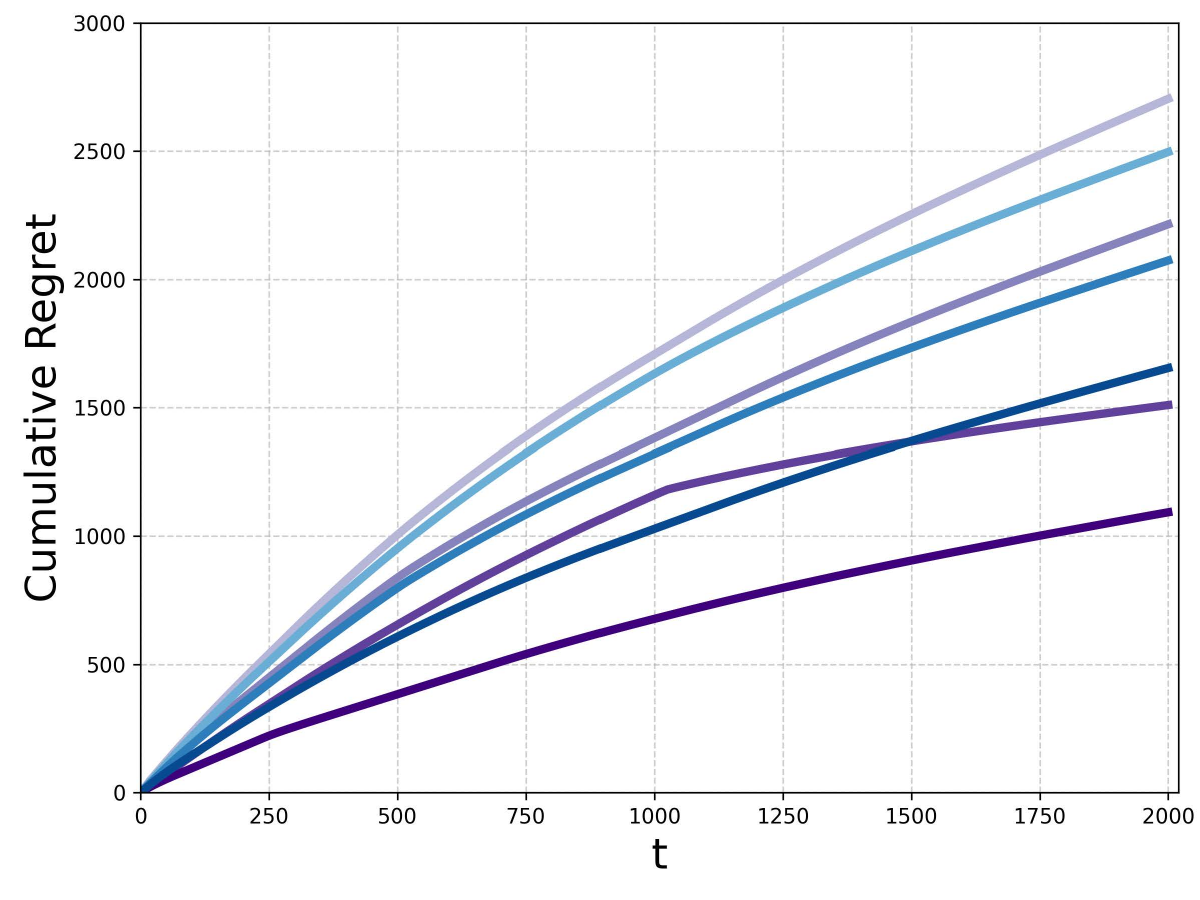}\\[-2pt]
  {\footnotesize (f) $d=50,\ s_0=15,\ K=5,\ N=30$}
  \label{fig:regret_15_50_5_30_pool}
\end{minipage}

\vspace{6pt}

\begin{minipage}[t]{\linewidth}
  \centering
  \includegraphics[width=0.9\linewidth]{exp/legend.pdf}
\end{minipage}

\caption{Cumulative regret on synthetic instances under varying feature dimension $d$, sparsity level $s_0$, and catalog size $N$. \textbf{Top row:} TJAP with $H\in\{0,1,3,5\}$ compared against CAP, M3P, and ONS--MPP. \textbf{Bottom row:} TJAP with $H\in\{0,1,3,5\}$ compared against the pooled estimator \textsc{Pool}$(H)$ for $H\in\{1,3,5\}$. Each curve is averaged over $10$ independent runs; all methods share the same price range $[0,\overline P]$ and observe identical contexts.}
\label{fig:regret-2}
\end{figure}

Figures~\ref{fig:regret-2} present the cumulative regret under other parameter configurations. The findings here echo the same trend: regret decreases with more source markets, and TJAP consistently outperforms the pricing-only baselines. Taken together, the experiments demonstrate that our algorithm is robust and effective under different sparsity levels.

\begin{figure}[htb!]
\centering

\begin{minipage}[t]{0.32\linewidth}
  \centering
  \includegraphics[width=\linewidth]{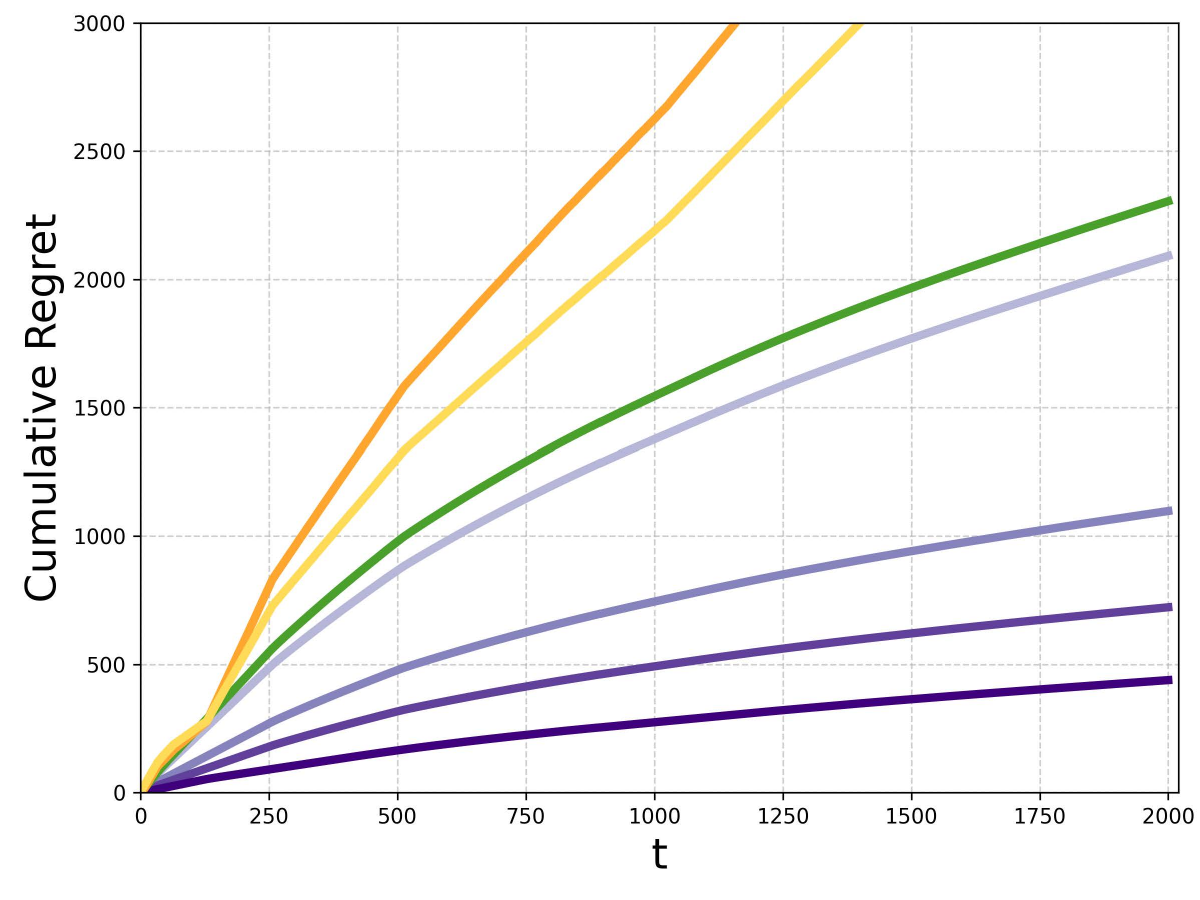}\\[-2pt]
  {\footnotesize (a) $d=10,\ s_0=3,\ K=5,\ N=100$}
  \label{fig:regret_3_10_5_100}
\end{minipage}\hfill
\begin{minipage}[t]{0.32\linewidth}
  \centering
  \includegraphics[width=\linewidth]{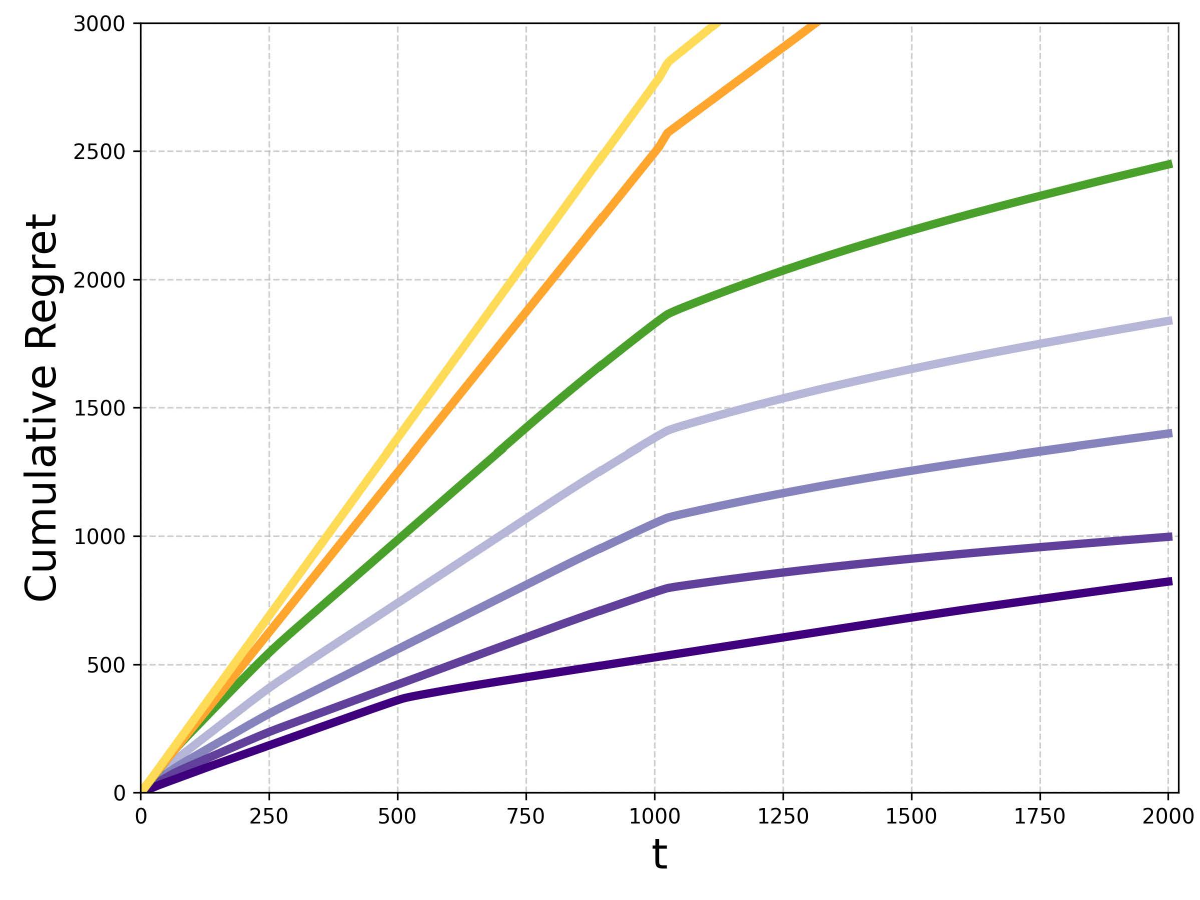}\\[-2pt]
  {\footnotesize (b) $d=20,\ s_0=4,\ K=5,\ N=30$}
  \label{fig:regret_4_20_5_30}
\end{minipage}\hfill
\begin{minipage}[t]{0.32\linewidth}
  \centering
  \includegraphics[width=\linewidth]{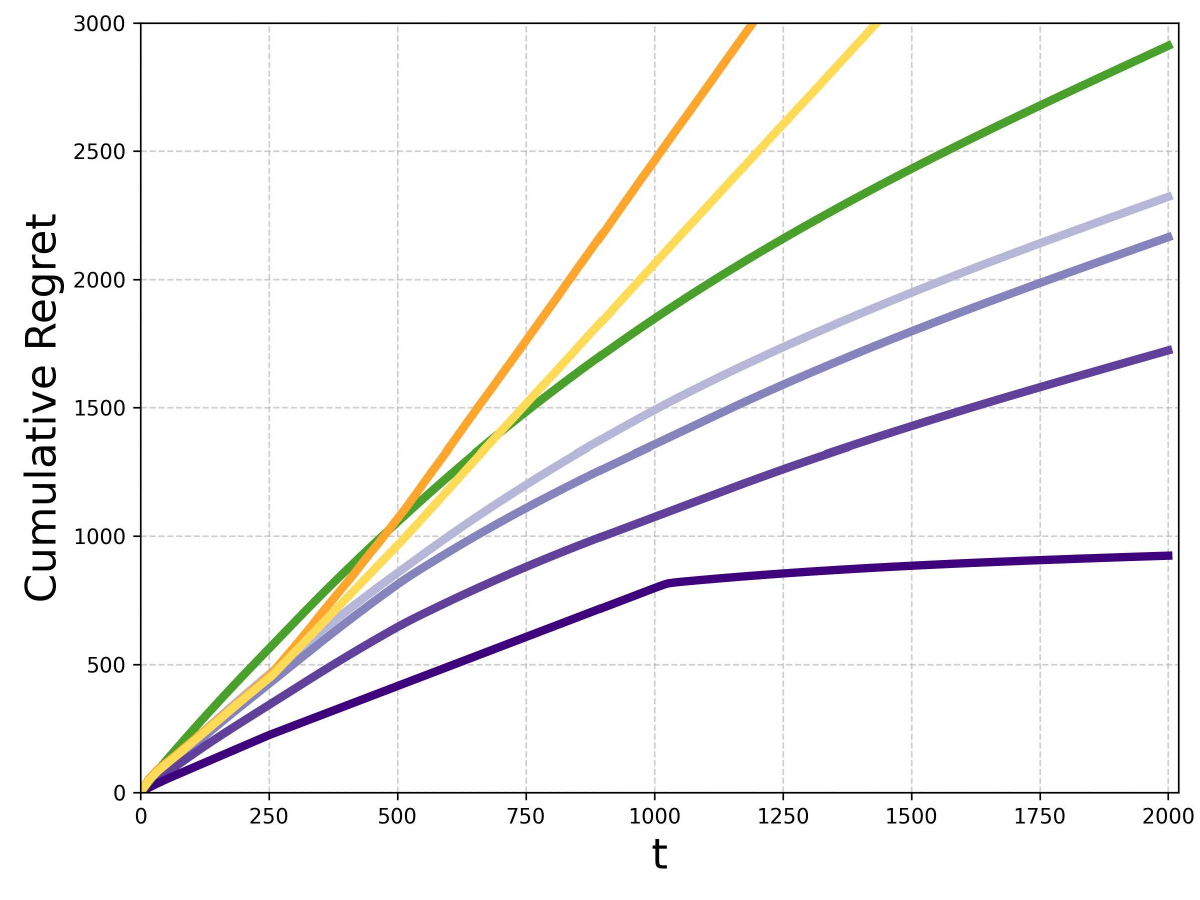}\\[-2pt]
  {\footnotesize (c) $d=50,\ s_0=10,\ K=5,\ N=30$}
  \label{fig:regret_10_50_5_30}
\end{minipage}

\vspace{-10pt}

\begin{minipage}[t]{0.32\linewidth}
  \centering
  \includegraphics[width=\linewidth]{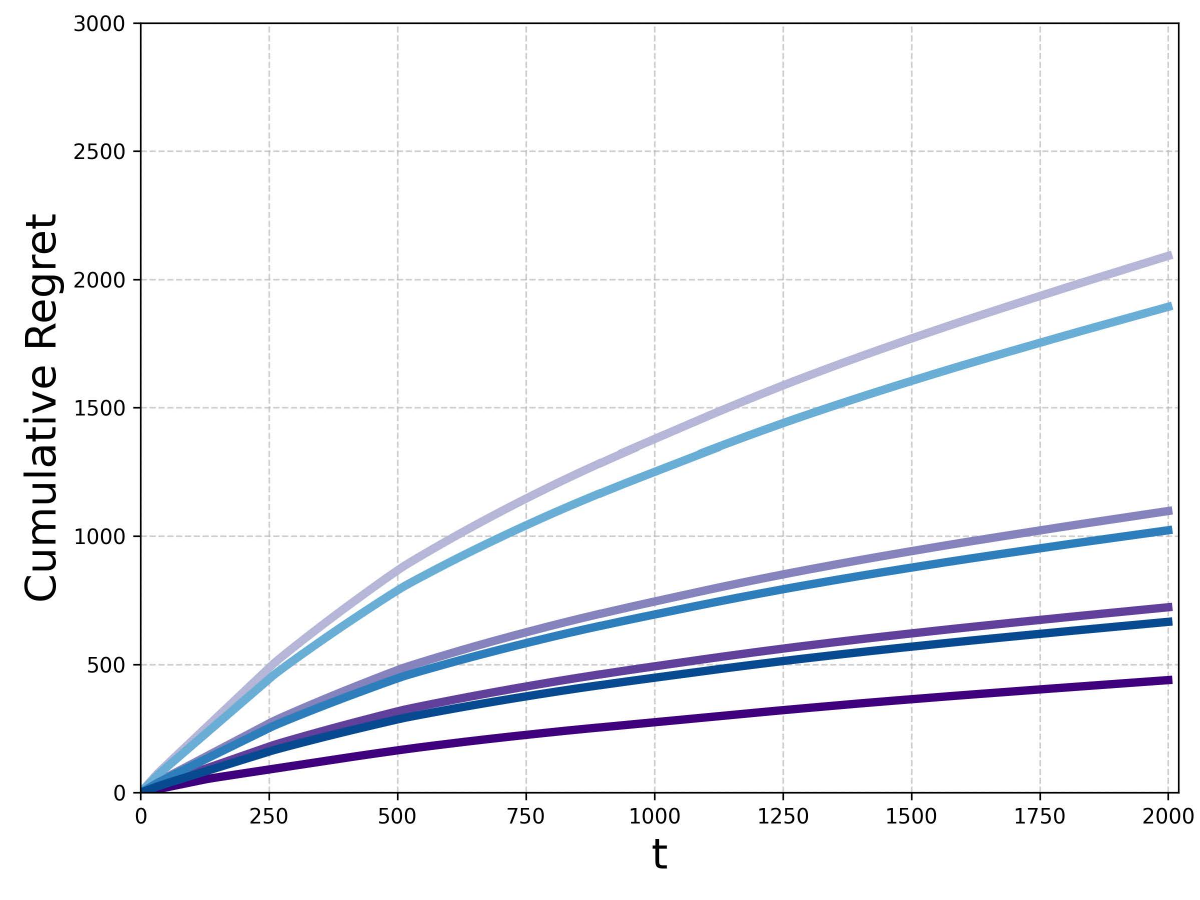}\\[-2pt]
  {\footnotesize (d) $d=10,\ s_0=3,\ K=5,\ N=100$}
  \label{fig:regret_3_10_5_100_pool}
\end{minipage}\hfill
\begin{minipage}[t]{0.32\linewidth}
  \centering
  \includegraphics[width=\linewidth]{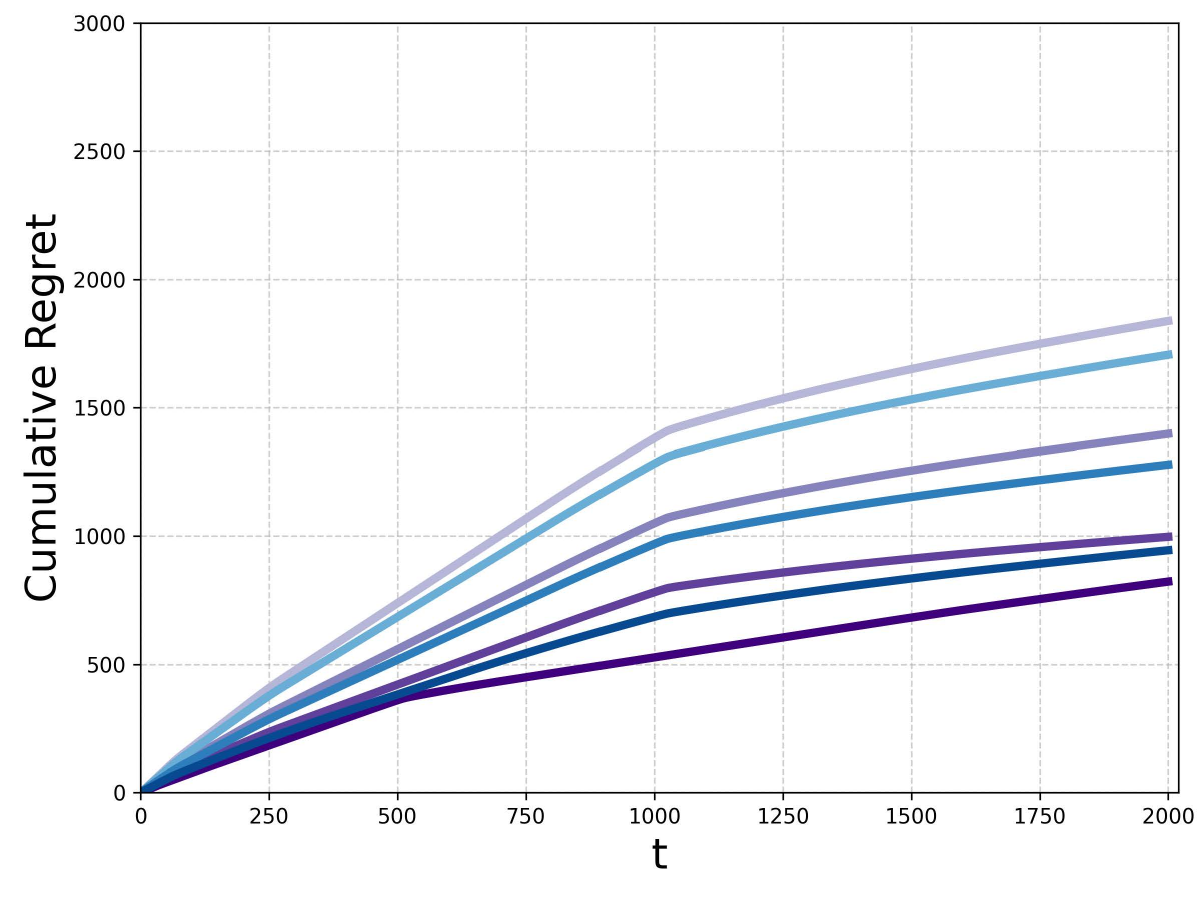}\\[-2pt]
  {\footnotesize (e) $d=20,\ s_0=4,\ K=5,\ N=30$}
  \label{fig:regret_4_20_5_30_pool}
\end{minipage}\hfill
\begin{minipage}[t]{0.32\linewidth}
  \centering
  \includegraphics[width=\linewidth]{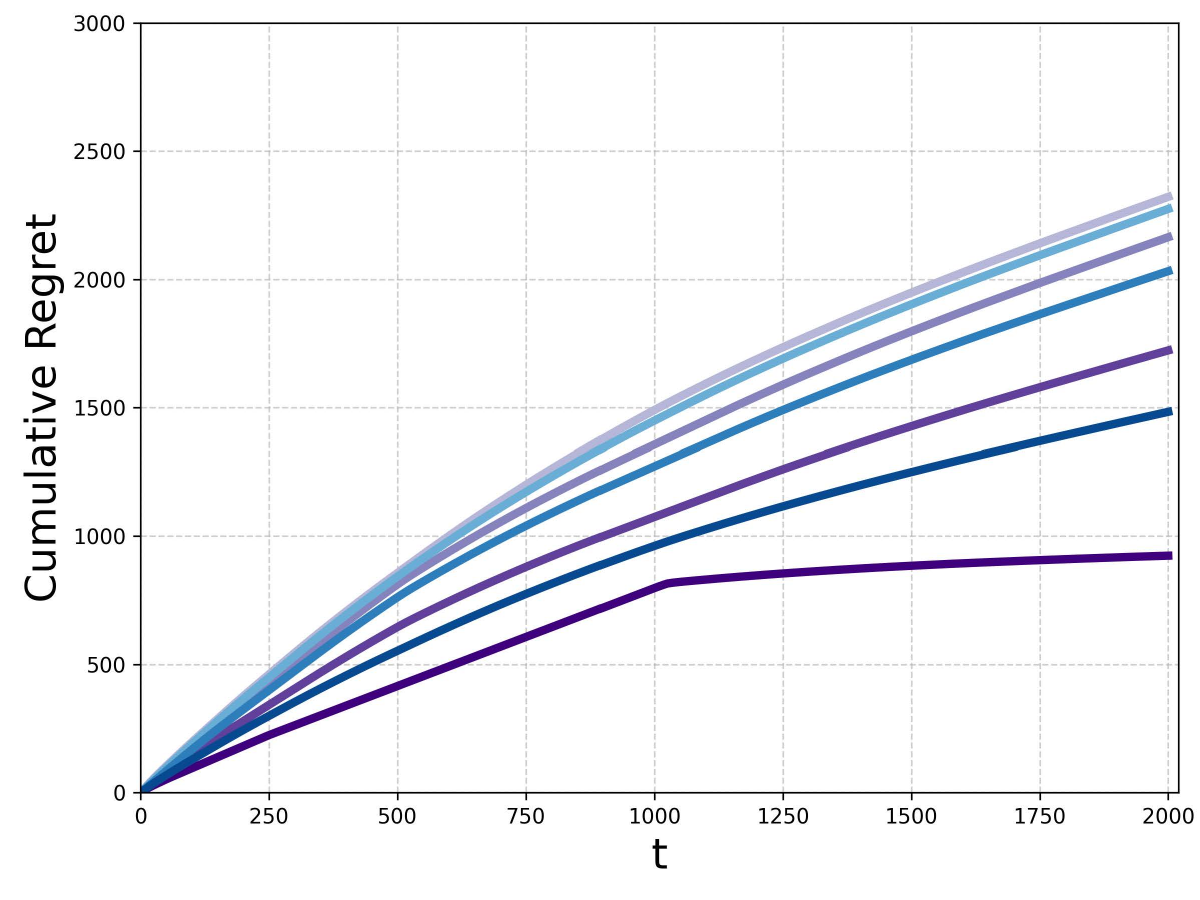}\\[-2pt]
  {\footnotesize (f) $d=50,\ s_0=10,\ K=5,\ N=30$}
  \label{fig:regret_10_50_5_30_pool}
\end{minipage}

\vspace{14pt}

\begin{minipage}[t]{0.32\linewidth}
  \centering
  \includegraphics[width=\linewidth]{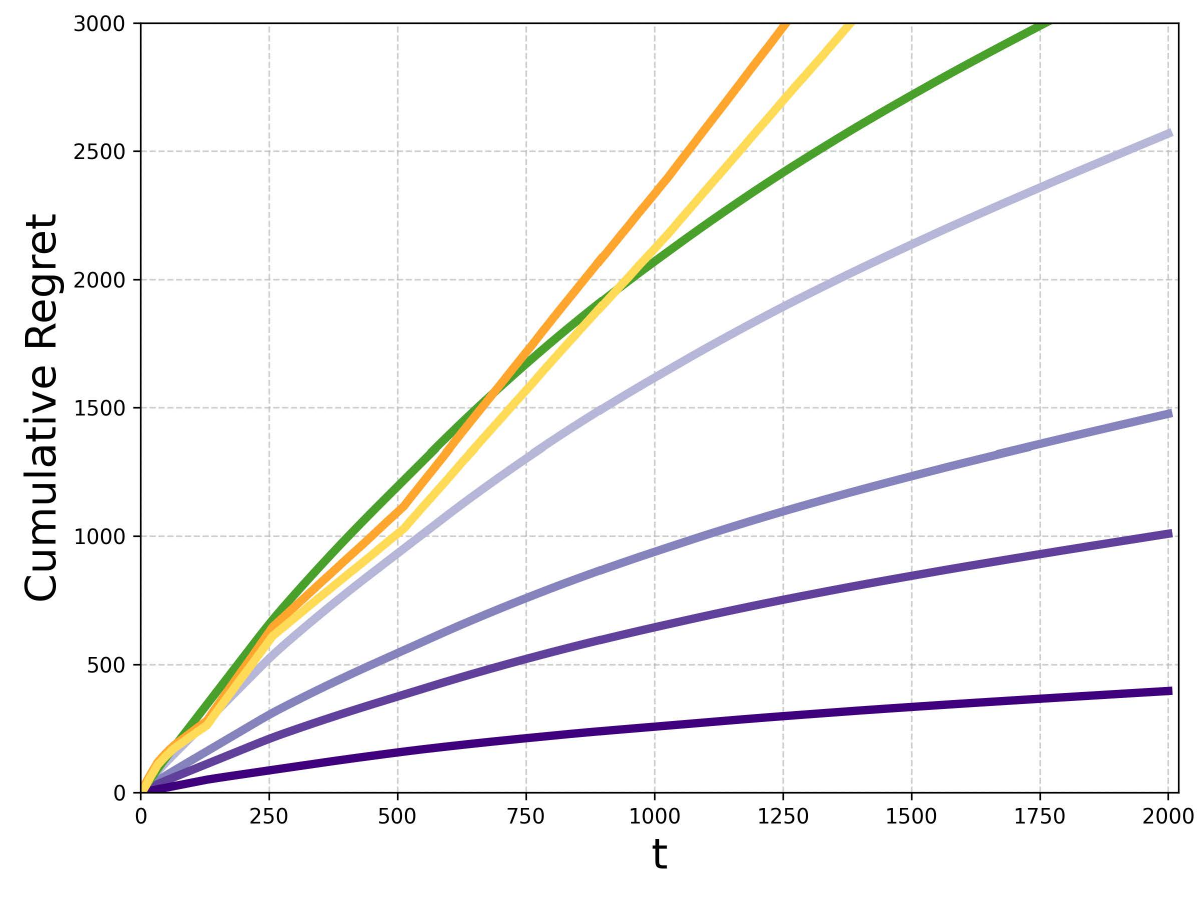}\\[-2pt]
  {\footnotesize (g) $d=10,\ s_0=2,\ K=5,\ N=100$}
  \label{fig:regret_2_10_5_100}
\end{minipage}\hfill
\begin{minipage}[t]{0.32\linewidth}
  \centering
  \includegraphics[width=\linewidth]{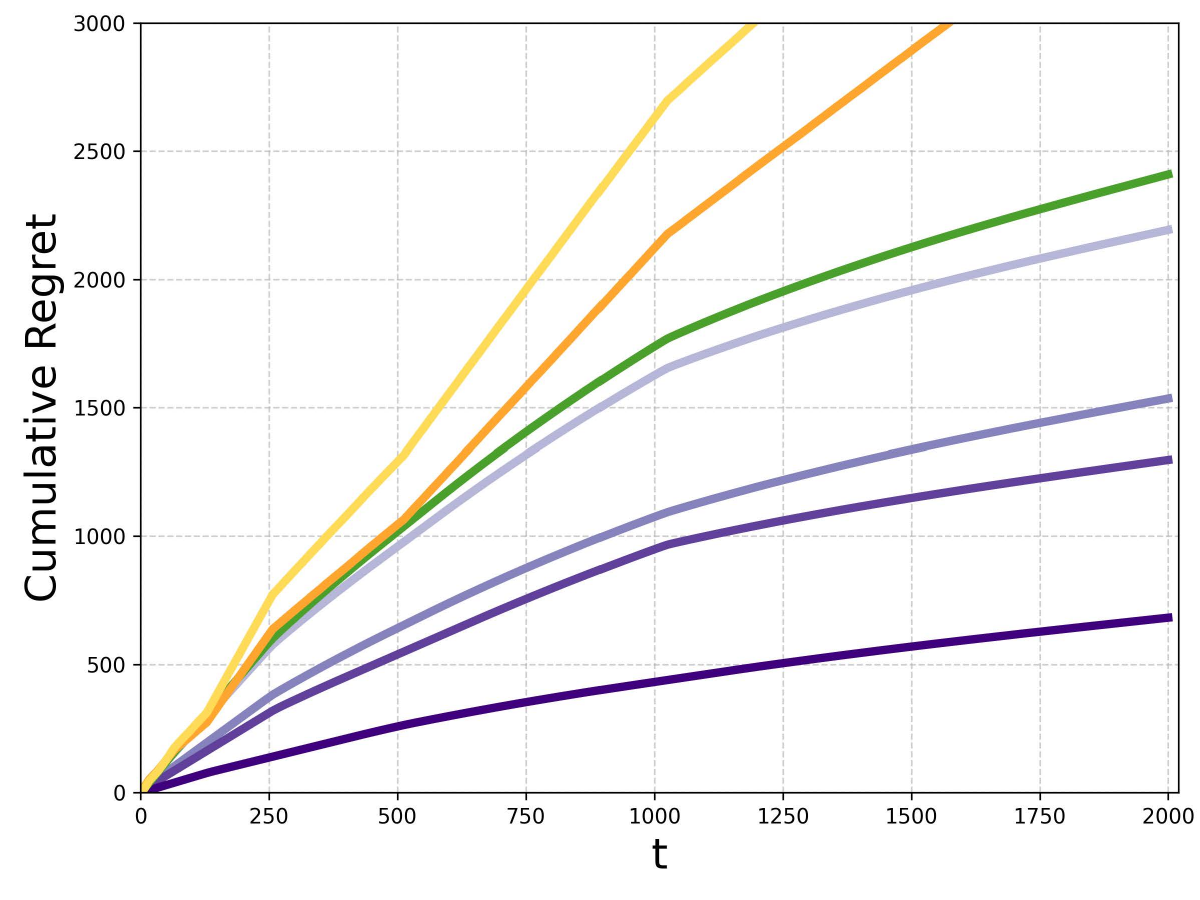}\\[-2pt]
  {\footnotesize (h) $d=20,\ s_0=4,\ K=5,\ N=100$}
  \label{fig:regret_4_20_5_100}
\end{minipage}\hfill
\begin{minipage}[t]{0.32\linewidth}
  \centering
  \includegraphics[width=\linewidth]{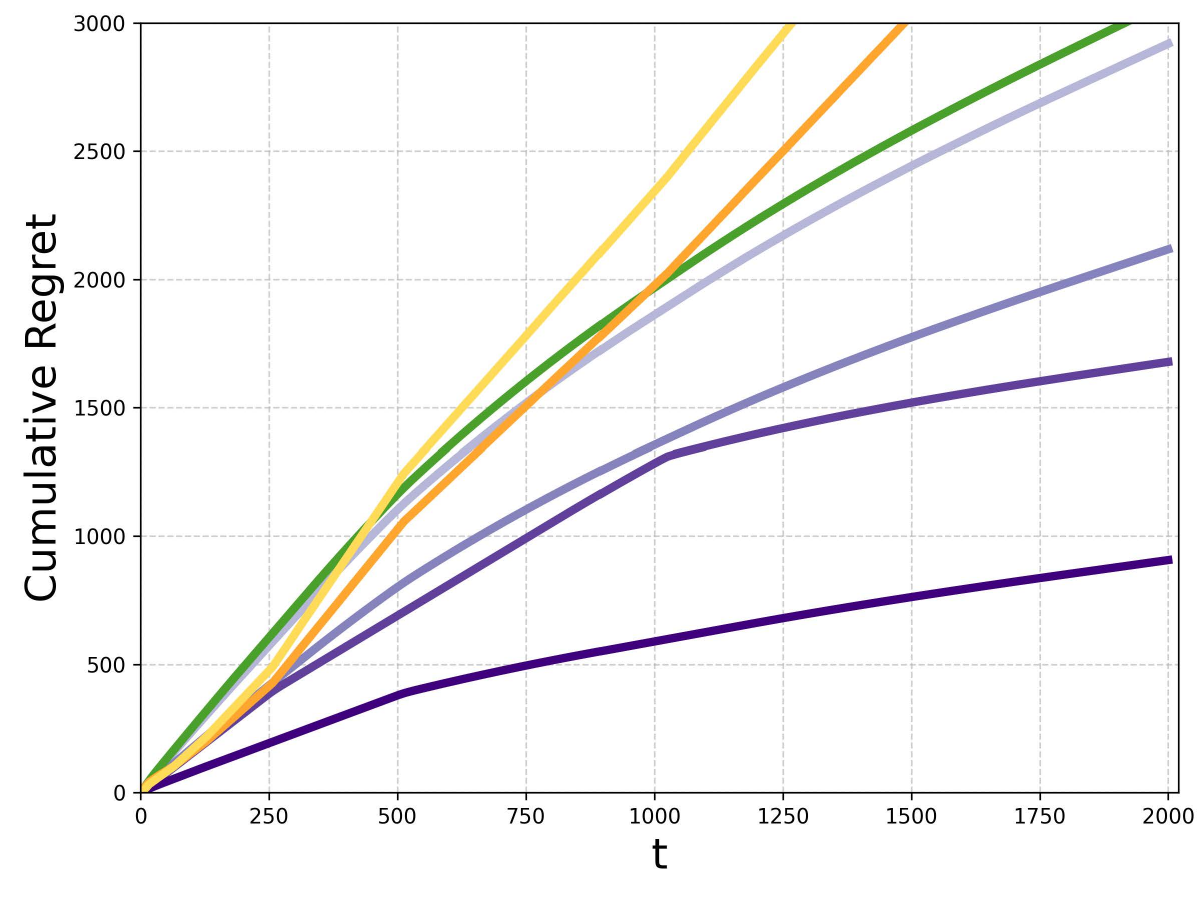}\\[-2pt]
  {\footnotesize (i) $d=50,\ s_0=10,\ K=5,\ N=100$}
  \label{fig:regret_10_50_5_100}
\end{minipage}

\vspace{-10pt}

\begin{minipage}[t]{0.32\linewidth}
  \centering
  \includegraphics[width=\linewidth]{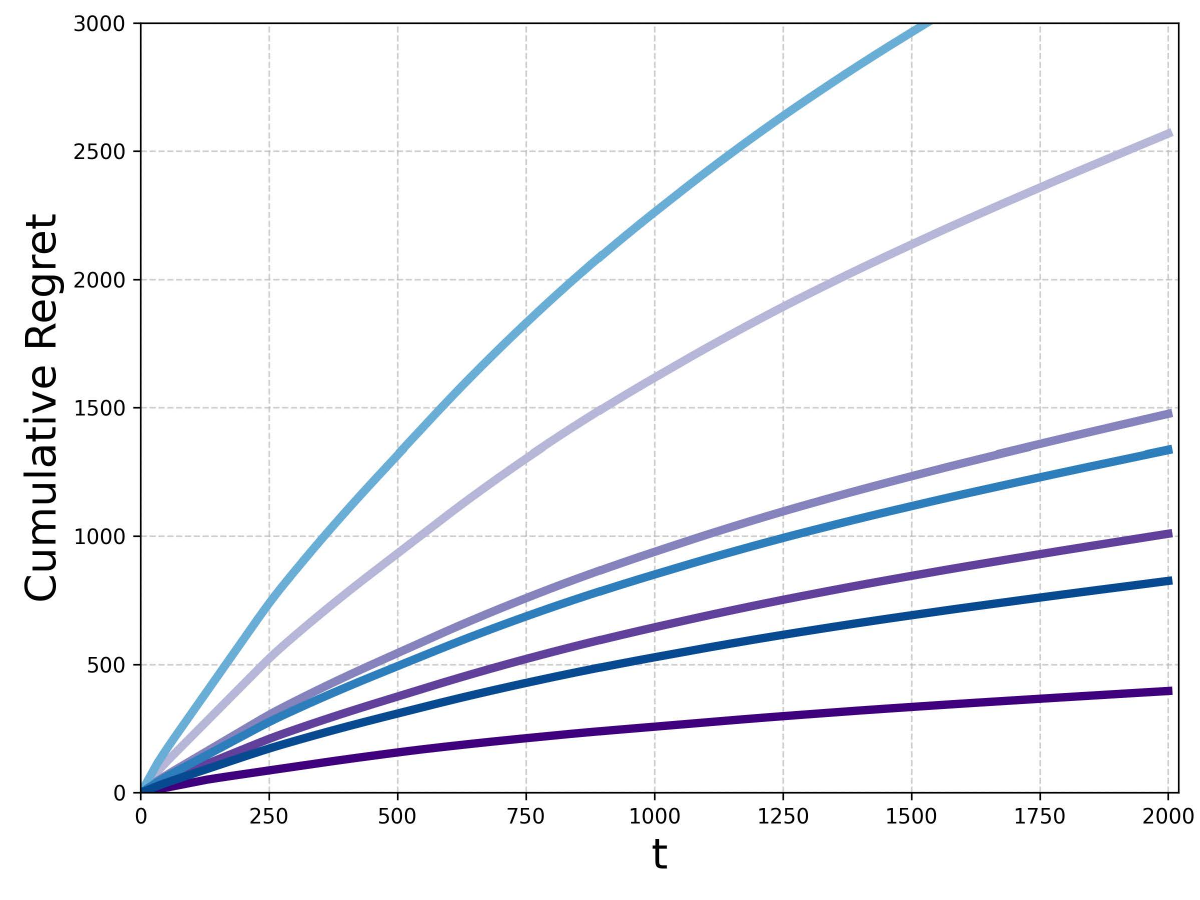}\\[-2pt]
  {\footnotesize (j) $d=10,\ s_0=2,\ K=5,\ N=100$}
  \label{fig:regret_2_10_5_100_pool}
\end{minipage}\hfill
\begin{minipage}[t]{0.32\linewidth}
  \centering
  \includegraphics[width=\linewidth]{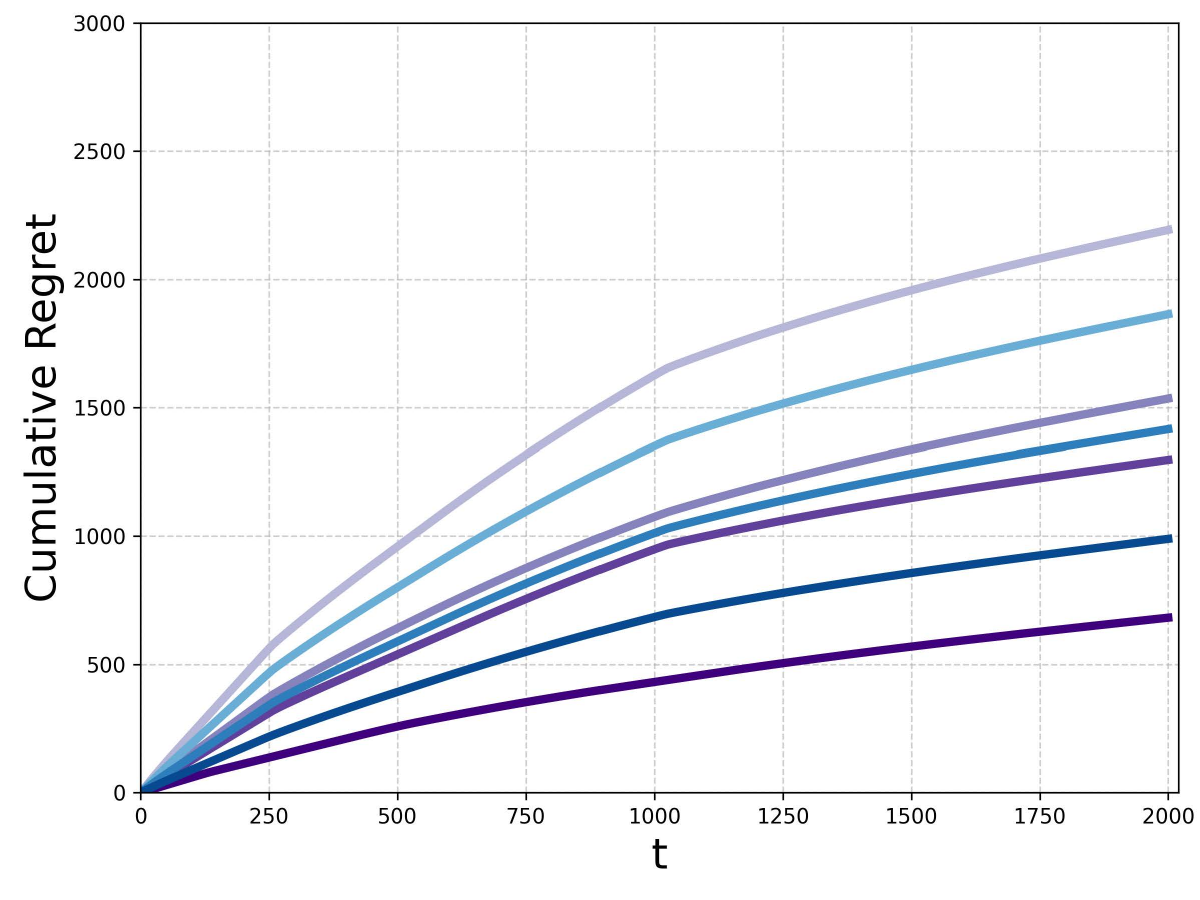}\\[-2pt]
  {\footnotesize (k) $d=20,\ s_0=4,\ K=5,\ N=100$}
  \label{fig:regret_4_20_5_100_pool}
\end{minipage}\hfill
\begin{minipage}[t]{0.32\linewidth}
  \centering
  \includegraphics[width=\linewidth]{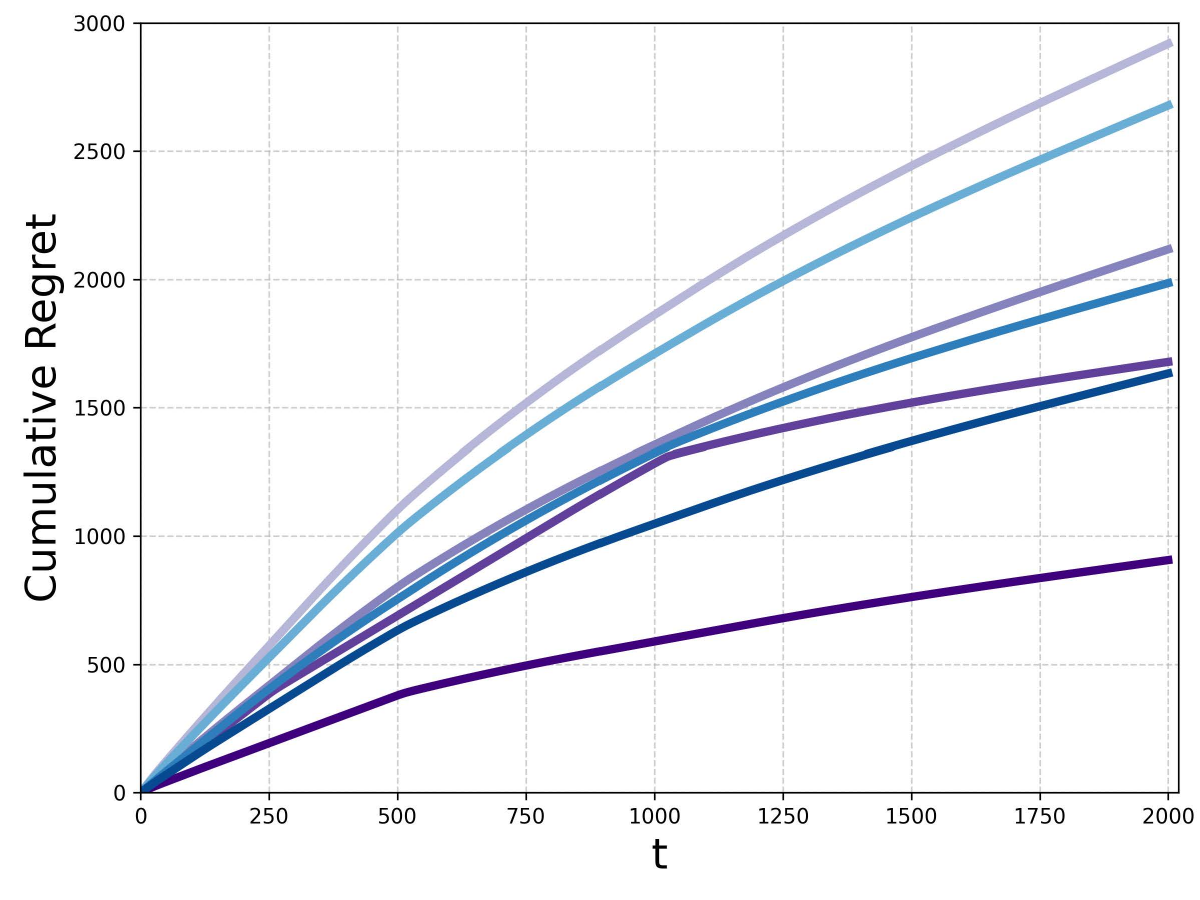}\\[-2pt]
  {\footnotesize (l) $d=50,\ s_0=10,\ K=5,\ N=100$}
  \label{fig:regret_10_50_5_100_pool}
\end{minipage}

\end{figure}

\end{appendices}

\end{document}